\documentclass[traditabstract]{aa}

\usepackage{amsmath,amssymb}
\usepackage{subfigure}
\usepackage{graphicx}
\usepackage{color}
\usepackage{hyperref}
\usepackage{natbib} \bibpunct{(}{)}{;}{a}{}{,}

\newcommand{\ds}{\displaystyle}
\newcommand{\ftns}{\footnotesize}
\newcommand{\Msun}{M_{\odot}}
\newcommand{\Halo}{\mathcal{H}}
\newcommand{\DM}{\mathcal{DM}}
\newcommand{\Gal}{\mathcal{G}}
\newcommand{\Disc}{\mathcal{D}}
\newcommand{\Bulge}{\mathcal{B}}

\newcommand{\Clumps}{\mathcal{C}}

\newcommand{\sfg}{g^{\star}}

\newcommand{\Torus}{\mathcal{T}}
\newcommand{\BH}{\bullet}

\begin{document}

\title{Towards a new modelling of gas flows in a semi-analytical model of galaxy formation and evolution} 

\authorrunning{M. Cousin et al }

\titlerunning{Towards a new modelling of gas flow in SAM}

\author{M. Cousin \inst{1} \and G. Lagache \inst{1}\and M. Bethermin \inst{3} \and B. Guiderdoni \inst{2}}

\institute{Institut d'Astrophysique Spatiale (IAS), B\^atiment 121, F- 91405 Orsay, France; Universit\'e Paris-Sud 11 and CNRS (UMR 8617) e-mail : morgane.cousin@ias.u-psud.fr \and Universit\'ee Lyon 1, Observatoire de Lyon, 9 avenue Charles Andr\'ee, Saint-Genis Laval, F-69230, France CNRS (UMR 5574), Centre de Recherche Astrophysique de Lyon, Ecole Normale Superieure de Lyon, Lyon, 69007, France \and European Southern Observatory, Karl-Schwarzschild-Str. 2, 85748 Garching, Germany \and Aix Marseille Universit\'e, CNRS, LAM (Laboratoire d'Astrophysique de Marseille) UMR 7326, 13388 Marseille, France}

\date{Received ? / Accepted ?}

\abstract{We present an extended version of the semi-analytical model, GalICS. Like its predecessor, eGalICS applies a post-treatment of the baryonic physics on pre-computed dark-matter merger trees extracted from an N-body simulation. We review all the mechanisms that affect, at any given time, the formation and evolution of a galaxy in its host dark-matter halo. We mainly focus on the gas cycle from the smooth cosmological accretion to feedback processes. To follow this cycle with a high accuracy, we introduce some novel prescriptions: i) a smooth baryonic accretion with two phases: a cold mode and a hot mode built on the continuous dark-matter accretion. In parallel to this smooth accretion, we implement the standard photoionisation modelling to reduce the input gas flow on the smallest structures. ii) a complete monitoring of the hot gas phase. We compute the evolution of the core density, the mean temperature and the instantaneous escape fraction of the hot atmosphere by considering that the hot gas is in hydrostatic equilibrium in the dark-matter potential well, and by applying a principle of conservation of energy on the treatment of gas accretion, SNe and super massive black hole feedback iii) a new treatment for disc instabilities based on the formation, the migration and the disruption of giant clumps. The migration of such clumps in gas-rich galaxies allows to form pseudo-bulges. The different processes in the gas cycle act on different time scales, and we thus build an adaptive time-step scheme to solve the evolution equations. The model presented here is compared in detail to the observations of stellar-mass functions, star formation rates, and luminosity functions, in a companion paper. Model outputs are available online.}


\maketitle

%
%
\section{Introduction}
After more than twenty years, (hybrid)-semi analytical models (SAMs) are the best tool for the physical interpretation of large surveys. Under the hypothesis that baryonic processes cannot strongly influence the structuration of the dark-matter, the dark-matter evolution is decoupled from the computation of the baryonic component. Even if this decoupling is a strong assumption (see \cite{van_Daalen_2011}), it allows a wide range of physical prescriptions to be explored in a realistic cosmological context and in a short computational time. 

Originally proposed by \cite{White_1991}, SAMs are still being developed \cite[e.g.][]{Cole_1991, Cole_2000, Hatton_2003, Baugh_2006, Croton_2006, Cattaneo_2006, Somerville_2008, Guo_2011, Henriques_2013}. After a first version described in a dedicated paper, the updates of these models are often fragmented into a large number of publications. In this context, we found it necessary to clarify all the steps of the modelling and build an up-to-date model based on the latest results obtained by hydrodynamic simulations and the comparison of empirical models with observations. Inspired by the previous GalICS model \citep{Hatton_2003}, we revisit all the standard mechanisms acting on the formation and the evolution of galaxies in our new SAM, step by step, and in a single paper. 

We mainly focus on the gas cycle, which is the main actor in galaxy stellar mass assembly. It is thus described from the cosmological smooth accretion to the complex feedback mechanisms. Obviously, all physical processes act in the gas cycle with different time scales. Indeed, gas-accretion, star formation, and clump migration time scales are all different. For this reason, we propose an adaptive time-step scheme to solve the gas evolution equations.

Like its predecessor \citep{Hatton_2003}, the new model is based on a dark-matter N-body simulation computed in the $\Lambda$ cold dark-matter paradigm ($\Lambda-CDM$). Indeed, even if the growth of dark-matter haloes can be followed using the \cite{Press_1974} formalism and their extensions \citep[e.g.][]{Cole_1991, Kauffmann_1993, Cole_1994, Somerville_1999}, the evolution of the dark-matter component is now commonly extracted from huge dark-matter simulations \citep[e.g.][]{Kauffmann_1999,Helly_2003,Hatton_2003}. In such high-resolution simulations, the dark-matter halo population covers a wide range of mass ($M_h \in [10^7 : 10^{13}]~\Msun$) and scale, from dwarf haloes to groups. To follow its time evolution, the hierarchical growth of dark-matter structures is commonly represented by merger trees in which each branch represents the growth of a given halo. The analysis of dark-matter simulations have set the current paradigm in which the dark-matter halo growth follows two modes \citep[e.g.][]{Lacey_1994,Tormen_1997,Wechsler_2002,Van_den_Bosch_2002,Zhao_2009,Fakhouri_2010a, Fakhouri_2010b, Genel_2010}:
\begin{itemize}
  \item{A smooth accretion that is difficult to define with accuracy. Indeed, it is closely linked to the mass resolution. For example, a $10^6\Msun$ dark-matter particle can be seen (or defined) as an unresolved dark-matter halo. However \cite{Genel_2010} show that the smooth accretion process converges to non-zero values when resolution increases. More precisely, they demonstrate that the smooth accretion can account for up to 40\% of the total mass of a dark-matter halo. These results prove that the naive expectation that all accretion would come through mergers when the resolution increases is wrong. The smooth dark-matter accretion is therefore a major vector of the growth of dark-matter structures. It is subject of a detailed description in this paper.}
    \item{The second mode is obviously linked to the merging of pre-existing structures.}
\end{itemize}

Applied to the evolution of the dark-matter component, the baryonic post-treatment is articulated around three new prescriptions:
\begin{itemize}
\item{A two-phase smooth baryonic accretion, with a cold and a hot mode that are built on the smooth dark-matter accretion.}
\item{A complete monitoring of the hot phase. The evolution of the mean temperature and density profile parameters are followed using explicit energy conservation laws, under the hydrostatic equilibrium hypothesis.}
\item{A new approach for describing of the disc gravitational instabilities. Based on the formation, migration and disruption of giant-clumps, this new prescription adapted for SAMs relies on the work by \cite{Dekel_2009b}.}
\end{itemize}
Beyond these three new prescriptions, the model examines the standard mechanisms of star formation regulation. For the low-mass structures two processes are in play:
\begin{itemize}
  \item{SNe (SNe) feedback. Originally proposed by \cite{Larson_1974} and \cite{White_1978}, this process is based on the injection of SN energy in the ISM, which generates a wind that expels a fraction of the gas from the disc structure. This mechanism has been widely used in SAM during the past 15 years \citep[e.g.][]{Kauffmann_1993, Cole_1994, Cole_2000, Silk_2003, Benson_2003,Hatton_2003, Croton_2006, Cattaneo_2006, Somerville_2008, Guo_2011, Henriques_2013}.}
    \item{Photoionisation. Originally proposed by \cite{Doroshkevich_1967}, photoionisation has been introduced in the context of the $CDM$ paradigm by \cite{Couchman_1986}, \cite{Ikeuchi_1986}, and \cite{Rees_1986}. The ultraviolet background, created by the first generations of stars, heats the gas surrounding the halo. In the smallest structures ($M_h<10^{10}~\Msun$), the temperature reached by the gas is high enough to prevent it from collapsing. The gas accretion on galaxies, hence the star formation, hosted by these small structures, are thus reduced. In the context of SAM, the impact of this mechanism has been previously explored by \cite{Benson_2002} \cite{Benson_2003} \cite{Somerville_2002} \cite{Croton_2006} and \cite{Guo_2011}, amoung others.}
\end{itemize}
The galaxy/dark-matter potential well is too deep for high-mass structures ($M_h>10^{12}~\Msun$). In this case, SNe and photoionisation processes are not powerful enough to eject some gas or to prevent the infall of baryons. SNe still produce winds but ejecta velocities are not sufficient to expel the gas. It is thus rapidly re-accreted in the disc. A popular interpretation that can explain the low star formation efficiency in $M_h>10^{12}~\Msun$ haloes is based on the activity of a super massive black hole (SMBH) hosted by the galaxy \cite[e.g.][]{Croton_2006,Cattaneo_2006,Bower_2006,Malbon_2007,Somerville_2008,Ostriker_2010,Bower_2012}. If a fraction of the energy produced by the accretion on the SMBH is used to heat the hot gas phase, the efficiency of the cooling process is decreased. The accretion of the gas in the galaxy, and thus the star formation, are then reduced.\\

With these standard recipes some profound disagreements between model predictions and observations are still present, even if observed local ($z=0$) galaxy properties are generally reproduced well. For example, one of the most important problems is the strong excess of low-mass galaxies predicted at high $z$. Indeed the stellar mass functions measured at $1<z<4$, for example by \cite{Ilbert_2010}, \cite{Ilbert_2013}, or \cite{Caputi_2011}, demonstrate that SAMs predict a number of low-mass galaxies roughly ten times larger than what observed. Models cannot also easily explain the cosmic star formation history in detail and the bulk of star formation activity observed at $1<z<3$. In addition, recent studies show that the link between dark-matter haloes and their host galaxies is not well understood \citep[e.g.][]{Leauthaud_2012}. Some empirical models \citep[e.g.][]{Bethermin_2012a,Behroozi_2012b,Behroozi_2012a} suggest a non-monotonically growth of the stellar mass with the dark-matter halo mass. Indeed, the star formation activity derived from observations seems to be very inefficient both in low-mass ($M_h< 10^{10}~\Msun$) and high-mass haloes ($M_h> 10^{13}~\Msun$). The star formation activity seems to peak for galaxies evolving at intermediate halo mass ($M_h\simeq 10^{12}~\Msun$) \citep{Eke_2004,Eke_2005,Guo_2010}.\\

While the current paper describes all the standard prescriptions used in the new model in detail, the reasons for the tensions between models and observations are discussed in a companion paper where we compare the predictions with fundamental observations, such as the stellar mass function, $M_h$ versus $M_{\star}$ relation, IR luminosity function, and specific star formation rate (Cousin et al 2014). The companion paper describes, in particular, the impact of SNe feedback and photoionisation process on the stellar mass assembly. It emphasises the difficulty matching observations by using only the standard star formation regulation processes and proposes, in this context, an ad hoc modification of the standard star formation cycle.\\ 

The paper is organised as followed. In Sect. \ref{dark_matter_section} we describe the dark-matter simulation used to process the dark-matter evolution. We then present an explicit description of the smooth dark-matter process based on background particles. In Sect. \ref{baryonic_accretion}, we describe the bimodal baryonic accretion prescription adopted in the model. It is based on a photoionisation model and on two gas reservoirs dedicated to cold and hot phases. Section~\ref{hot_phase} presents a complete modelling of the monitoring of the hot gas phase. Ejecta processes coming from SNe and/or SMBH activity are also presented, together with the prescription for computating the gas escape fraction and the cooling rate. In Sect.~\ref{galaxy}, we present all the properties and evolution laws of discs and bulges. The new clumpy component that derives from disc instabilities is also fully described in this section. In Sect.~\ref{merger_events} we describe the effect of merger events onto galaxy evolution. Section~\ref{Overview_perspectives} closes the paper by giving an overview and some research perspectives. In addition, we have added in Appendix \ref{adaptive_time_step}, a description of the adaptive time-step scheme used to follow the different dynamical time scales of the gas cycle.

%
%

\section{dark-matter}
\label{dark_matter_section}

In this section, we first describe the dark-matter merger-tree structure and then define the smooth accretion process. We also give the useful dark-matter halo properties. All symbols and their definition are listed in Table~\ref{dm_symbol}.

\subsection{Dark-matter evolution}

\begin{figure*}[t]
  \begin{center}
    \includegraphics[scale = 0.58]{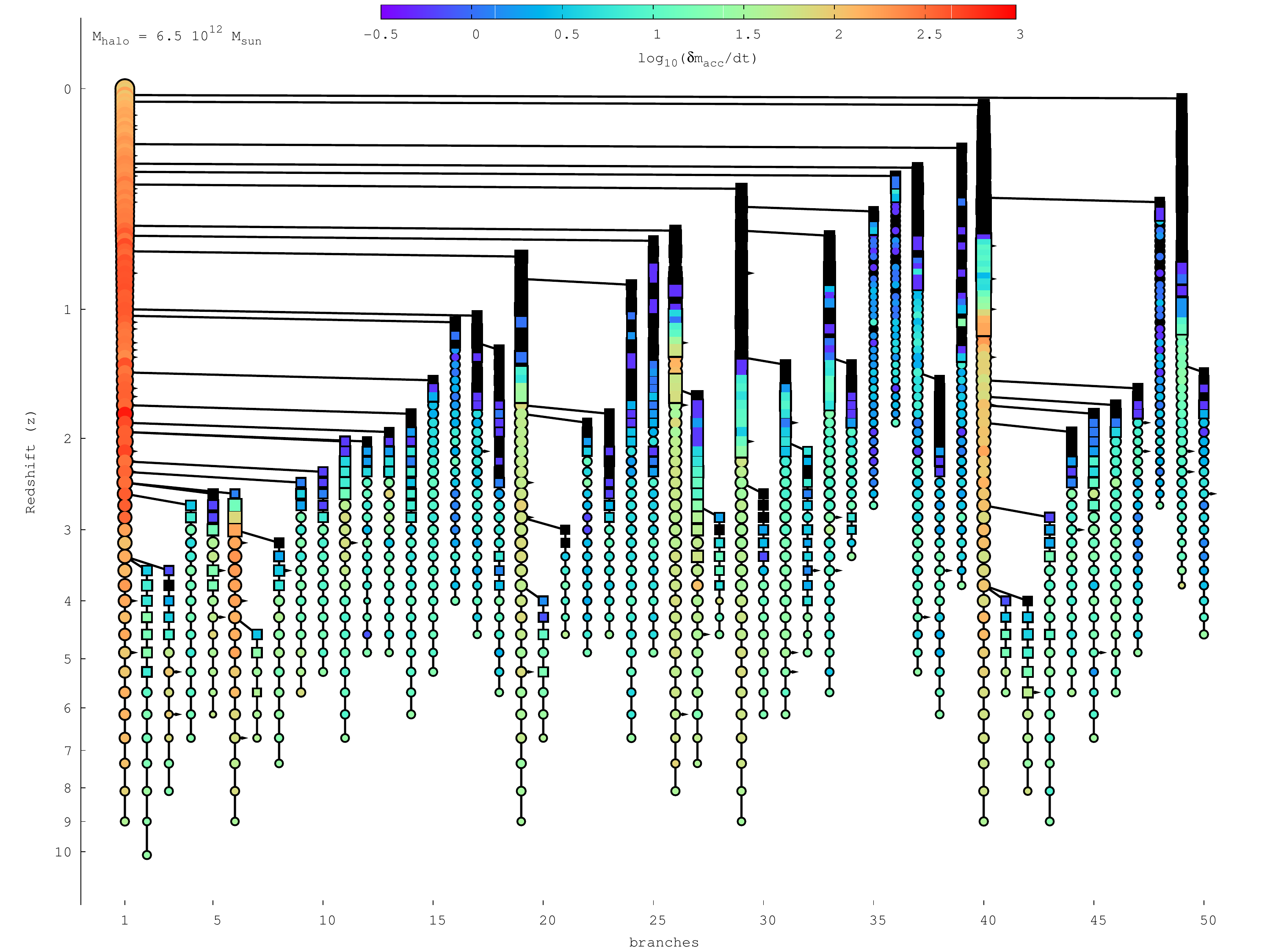}
  \caption{\tiny{Merger tree showing the evolution of the dark-matter smooth accretion. We only show the fifty most massive branches. The junctions with less massive branches are indicated with small arrowheads. The main branch (on the left) is built up through a smooth accretion process $\dot{M}_{acc}$ (Eq. \ref{dm_acc_rate}) and multiple mergers with other haloes. The evolution of the smooth dark-matter accretion through the merger tree is shown with colour. A halo without any dark-matter accretion is shown with a black symbol. Circles are used for main haloes (M-H), while squares are used for subhaloes (S-H). The size of the symbols is proportional to the dark-matter halo virial mass. It is clearly visible that before a merger event with a larger structure, a halo is systematically identified as a substructure of this more massive halo. During this subhalo phase, the dark-matter accretion rate strongly decreases and may even stop.}}
  \label{tree_dm_acc}
  \end{center}
\end{figure*}

Like its predecessor, the new version \textit{e}GalICS uses a pure N-body simulation to follow the dark-matter evolution. We use an N-body cosmological simulation based on the WMAP-3yr cosmology ($\Omega_m = 0.24$, $\Omega_{\Lambda} = 0.76$, $f_b = 0.16$, $h = 0.73$). In a volume of $(100h^{-1})^3 \simeq 150~Mpc^3$, $1024^3$ particles evolve with an elementary mass of $m_p = 8.536~10^7~\Msun$.
Dark-matter merger trees (see Fig.~\ref{tree_dm_acc}) are extracted with the halo-finder program \verb?AdaptaHOP? proposed by \cite{Aubert_2004} and extended by \cite{Tweed_2009}. The dark-matter assembly is therefore pre-computed for the different time steps. Its evolution cannot be modified by baryonic processes.\\ 

The halo-finder version used here \citep{Tweed_2009} allows us to extract and separate different levels of structures: main haloes (M-H) and subhaloes (S-H or satellites). In the two cases, we only consider dark-matter structures containing at least 20 dark-matter particles. This limit gives a minimum dark-matter halo mass $M_h^{min} = 1.707\times10^9~\Msun$. Subhaloes are extracted and identified as overdensities in a pre-existing structure that can be a M-H or another S-H. Currently, it is possible to extract up to six different levels of S-H. The halo finder generates all the links between all haloes in time and in structuration level. 

\begin{table}[t]
  \begin{center}
    \tiny{
      \begin{tabular*}{0.5\textwidth}{@{\extracolsep{\fill}}cr}
        \hline
        Symbol & Definition \\
        \hline
        $m_p^{\star}$     & Smooth accreted particule mass [Eq. \ref{dm_acc_rate}] \\
        $\dot{M}_{acc}$  & Smooth dark-matter accretion rate [Eq. \ref{dm_acc_rate}]\\
        $M_{acc}$           & Integrated background mass [Eq. \ref{dm_acc_mass}]\\
        $M_{fof}$           & Halo finder mass\\
        $M_{vir}$           & Virial mass\\
        $\rho_{dm}(r)$   & Radial density profile [Eq. \ref{rho_dm}]\\
        $r_{dm}$            & Core radius [Eq. \ref{rho_dm}]\\
        $\rho_{dm,0}$    & Core density [Eq. \ref{rho_dm}]\\
        $M_h(<r)$        & Halo mass enclosed in the radius $r$ [Eq. \ref{dm_mass_r}]\\
        $\phi(x)$         & Geometrical function (dark-matter mass) [Eqs. \ref{dm_mass_r}, \ref{phix}]\\
        $r_{vir}$            & Virial radius [Eqs. \ref{dm_mass_rvir}, \ref{T_vir}, \ref{Vesc_dm}]\\ 
        $c_{dm}$           & Halo concentration [Eq. \ref{dm_mass_rvir}]\\
        $T_{vir}$            & Virial temperature [Eqs. \ref{T_vir}, \ref{Ethacc}]\\ 
        $t_{dyn}$            & Dynamical time [Eqs. \ref{t_dyn}, \ref{free-fall-rate}]\\
        $V_{esc,\DM}$     & Escape velocity [Eqs. \ref{Vesc_dm} and \ref{f_esc}]\\
        $V_{circ,dm} $     & Circular velocity [Eq. \ref{Vcirc_dm}]\\
        $\lambda$       & Halo spin parameter  [Eq. \ref{rd_evolution}]\\
        \hline
      \end{tabular*}}
  \end{center}  
  \caption{\footnotesize{Dark-matter symbols and their definition. Symbols are listed in order of appearance.}}
  \label{dm_symbol}
\end{table}

\subsection{Dark-matter background accretion}

To follow the dark-matter halo growth and, later on, the baryonic accretion process, we identify at a given time step $n$, all particles that are detected in a halo $\Halo$, and that have never been identified in an other halo $\Halo$' before. These background particles ($m_p^{\star}$) play an important role. Indeed, we consider them as the only source of new baryonic mass. In this framework, the accretion rate on a given structure at a given time step is defined as
\begin{equation}
  \dot{M}_{acc} = \dfrac{\sum_{\Halo} m_p^{\star}}{t^{n}-t^{n-1}}.
  \label{dm_acc_rate}
\end{equation}

Figure \ref{tree_dm_acc} shows the evolution of the dark-matter accretion rate $\dot{M}_{acc}$. We can see that before merging with a larger structure (M-H), a halo is systematically identified as a S-H of this more massive halo. During this pre-merger period, it is clear that the dark-matter accretion rate ($\dot{M}_{acc}$) decreases strongly and may even stop. In these conditions, the diffuse accretion is then supported by the most massive structure. The decrease in background accretion is also visible during fly-by events, as identified in the $18^{th}$ branch of the merger tree shown in Fig.~\ref{tree_dm_acc}.\\

By following with time the background accretion process for a given structure $i$, we can define an integrated accretion mass:
\begin{equation}
  M_{acc,i}^{n} = \underbrace{{M}_{acc,i}^{n-1}}_{I} + \underbrace{\sum_{\Halo} m_p^{\star}}_{II} + \underbrace{\sum_{j \neq i} M_{acc,j}^{n-1}}_{III}.
  \label{dm_acc_mass}
\end{equation}
In this equation,
\begin{itemize}
  \item{$I~~$ is the integrated background mass accreted by the halo $i$ up to the previous time step $n-1$,}
  \item{$II~$ is the background mass accreted during the time step [$n-1$, $n$]. This mass is computed using all particles identified in the halo $i$ at the time step $n$ and that have never been identified in an other halo before.}
  \item{$III$ corresponds to the integrated background mass accreted by all the progenitors $j$ of the halo $i$.}
\end{itemize}

\subsection{Remark on the dark-matter background accretion}
For M-H, the integrated accretion mass ($M_{acc}$) is closed to the well known halo finder mass $M_{fof}$, which is defined as the mass of all particles instantaneously detected in a structure. In contrast, the integrated background mass $M_{acc}$ for S-H is often higher than $M_{fof}$. Indeed, S-H are sensitive to the stripping process, and a fraction of the halo mass ($M_{fof}$) can be ripped by gravitational interactions acting before a merger event. In the context of our dark-matter background accretion process, one of the key points of the model resides in the fact that the stripped particles that have already been identified in a structure cannot be considered in any other structure.\\

The mass estimator $M_{acc}$ summarises the accretion history of a given halo. When the instantaneous mass is needed, we use the virial mass $M_{vir}$, as in other SAMs. This mass is computed using all particles that follow the virial theorem for a given halo.  

\subsection{Dark-matter halo properties}

\subsubsection{Density profile}

The spatial distribution of a dark-matter halo mass is described by a Navarro Frenk and White (NFW) density profile \citep{Navarro_1995, Navarro_1996, Navarro_1997} defined as:
\begin{equation}
  \rho(r) = \dfrac{\rho_{dm}}{\frac{r}{r_{dm}}\left(1 + \frac{r}{r_{dm}}\right)^2}
  \label{rho_dm}
\end{equation}
where $r_{dm}$ is the characteristic scale radius of the dark-matter halo and $\rho_{dm}/4$ is the density at $r=r_{dm}$. 
With this density profile, the dark-matter halo mass $M_h(<r)$ contained in the radius $r$ is given by \citep[e.g.][]{Suto_1998, Makino_1998, Komatsu_2001} 
  \begin{equation}
    M_h(<r) = 4\pi\rho_{dm}r_{dm}^3\phi\left(x\right)
    \label{dm_mass_r}
  \end{equation}
where $\phi(x)$ is a geometrical function defined as
\begin{equation}
  \phi(x) = ln(1 + x) - \dfrac{x}{1 + x}~~\mbox{with}~~x = \frac{r}{r_{dm}}.
  \label{phix}
\end{equation}
Obviously, if $r_{vir}$ is the virial radius of the dark-matter halo extracted from the dark-matter simulation (see Sect. 2.3 of \cite{Hatton_2003} for more information):
  \begin{equation}
    M_h(<r_{vir}) = M_{vir} = 4\pi\rho_{dm}r_{dm}^3\phi\left(c_{dm}\right)
    \label{dm_mass_rvir}
  \end{equation}
where $c_{dm}$ is the dark-matter concentration parameter defined as $c_{dm} = r_{vir}/r_{dm}$. With the dark-matter virial mass and the virial radius, it is possible to define the dynamical time of the dark-matter structure:
  \begin{equation}
    t_{dyn} = \sqrt{\dfrac{r_{vir}^3}{GM_{vir}}}.
    \label{t_dyn}
  \end{equation}
This time will be used in the next sections to define the baryonic accretion rate.

\subsubsection{Virial temperature and escape velocity}

In the hot atmosphere evolution monitoring, we use the virial temperature $T_{vir}$ linked to the dark-matter halo: 
  \begin{equation}
    T_{vir} = 35.9\dfrac{G~M_{vir}}{r_{vir}}~~[K].
    \label{T_vir}
  \end{equation}
We also use the escape velocity of the dark-matter halo defined as 
  \begin{equation}
    V_{esc,dm} = \sqrt{8\pi G \rho_{dm} r_{dm}^2\dfrac{ln\left(1 + x\right)}{x}}~~\mbox{with}~~x = \frac{r}{r_{dm}}
    \label{Vesc_dm}
  \end{equation}
and the circular velocity of the dark-matter halo defined as
  \begin{equation}
    V_{circ,dm} = \sqrt{4\pi G \rho_{dm} r_{dm}^2\dfrac{\phi(x)}{x}}~~\mbox{with}~~x = \frac{r}{r_{dm}}.
    \label{Vcirc_dm}
  \end{equation}

%
%

\section{Baryonic accretion}
\label{baryonic_accretion}

The evolution of the dark-matter component is therefore known using a large set of merger trees. The dark-matter halo properties are listed and defined in the previous section. We now have to add the baryonic content on the hierarchical structure of the dark-matter. This section describes the link between the dark-matter smooth accretion and the baryons that will feed the galaxy. All parameters used in this section and their definition are listed in Table~\ref{halo_symbol}.

As explained previously, we consider that the dark-matter background-accreted mass ($\dot{M}_{acc}$) is the only vector of baryonic accretion. The link between the dark-matter and baryonic accretion is defined through the effective accreted baryonic fraction, $f_b$ (Eq. \ref{baryonic_fraction}). 

\subsection{Photoionisation}
\label{photoionisation}

The value taken by this parameter $f_b$ follows two distinct regimes. Before reionisation ($z_{reion}$), we consider that the baryons feed the dark-matter structures following the universal baryonic fraction $\left<f_b\right>=\frac{\Omega_b}{\Omega_m}$. After reionisation, the heating of the gas generated by the first generation of stars limits the baryonic gas accretion in the smallest structures \citep{Kauffmann_1993}. Originally proposed by \cite{Doroshkevich_1967}, the photoionisation mechanism has been developed in the CDM paradigm by \cite{Couchman_1986}, \cite{Ikeuchi_1986} and \cite{Rees_1986}. This process is widely used in the context of galaxy evolution \cite{Efstathiou_1992, Babul_1992, Shapiro_1994, Quinn_1996, Thoul_1996, Bullock_2000, Gnedin_2000, Benson_2002, Somerville_2002, Croton_2006, Hoeft_2006, Okamoto_2008, Somerville_2008, Guo_2011, Somerville_2012, Henriques_2013}. 

The effective baryonic fraction $f_b$ therefore depends on both the redshift $z$ and dark-matter halo mass $M_{vir}$. The most commonly used formulation is that proposed by \cite{Gnedin_2000} and \cite{Kravtsov_2004}:

\begin{equation}
\begin{small}
f_b = \left<f_b\right>\left\{
  \begin{array}{ll}
   \left[1 + (2^{\alpha/3}-1)\left(\dfrac{M_{vir}}{M_c(z)}\right)^{-\alpha}\right]^{-3/\alpha}  & \mbox{: if } z<z_{reion} \\
    & \\
    1 & \mbox{: otherwise}.
  \end{array}\right.
\end{small}
\label{baryonic_fraction}
\end{equation}
Following \cite{Okamoto_2008}, we use $\alpha=2$. The characteristic halo mass follows\footnote{We use here a fit of the relation given in Fig. 15of \cite{Okamoto_2008}}: 
\begin{equation} 
  M_c(z) = 8.22\times10^{9}\rm exp(-0.7z)~~[\Msun ].
  \label{filtering_mass}
\end{equation} 
According to this filtering mass, haloes with $M_{vir} = M_c$ really collect only half of the baryons coupled with the accreted dark-matter. 

\begin{table}[t]
  \begin{center}
    \footnotesize{
      \begin{tabular*}{0.5\textwidth}{@{\extracolsep{\fill}}cr}
        \hline
        Symbol & Definition \\
        \hline
        $f_b$                      & Effective accreted baryonic fraction [Eq. \ref{baryonic_fraction}]\\
        $\left<f_b\right>$ & \textbf{Universal baryonic fraction = 0.18} [Eq. \ref{baryonic_fraction}]\\
        $M_c(z)$                 & Photoionisation characteristic halo mass [Eq. \ref{filtering_mass}]\\
        $\dot{M}_{bar}$       & Baryonic accretion rate  [Eq. \ref{bar_acc_rate}]\\
        $f_{sh}$                   & Shock-heated fraction (in accretion) [Eq. \ref{shock_heated_fraction},  Fig \ref{Fig_shock_heated_fraction}]\\
        $\dot{M}_{flt}$        & Filament formation rate [Eq. \ref{flt_formation_rate}, Fig. \ref{baryon_halo_transfer}]\\
        $\dot{M}_{ff}$         & Cold free-fall rate (galaxy input) [Eq. \ref{free-fall-rate}, Fig. \ref{baryon_halo_transfer}]\\
        $\varepsilon_{ff}$   & \textbf{Free-fall rate efficiency = 1.0} [Eq. \ref{free-fall-rate}]\\
        $M_{cold}$               & Cold accretion phase  [Eqs. \ref{free-fall-rate} and \ref{cold_mass_evolution}, Fig. \ref{baryon_halo_transfer}]\\
        $\dot{M}_{sh}$        &  Cosmological hot gas accretion rate  [Eq. \ref{hot_cosmo_acc}, Fig. \ref{baryon_halo_transfer}]\\
        $M_{hot}$                & Hot atmosphere mass [Eqs. \ref{hot_mass_evolution}, \ref{Mhot_r}, Fig. \ref{baryon_halo_transfer}]\\
        \hline
      \end{tabular*}}
  \end{center}  
  \caption{\footnotesize{Baryonic accretion symbols and their definition. Symbols are listed in order of appearance. Model parameters are in boldfaced text.}}
  \label{halo_symbol}
\end{table}

\begin{figure}[h]
  \begin{center}
    \includegraphics[scale = 0.9]{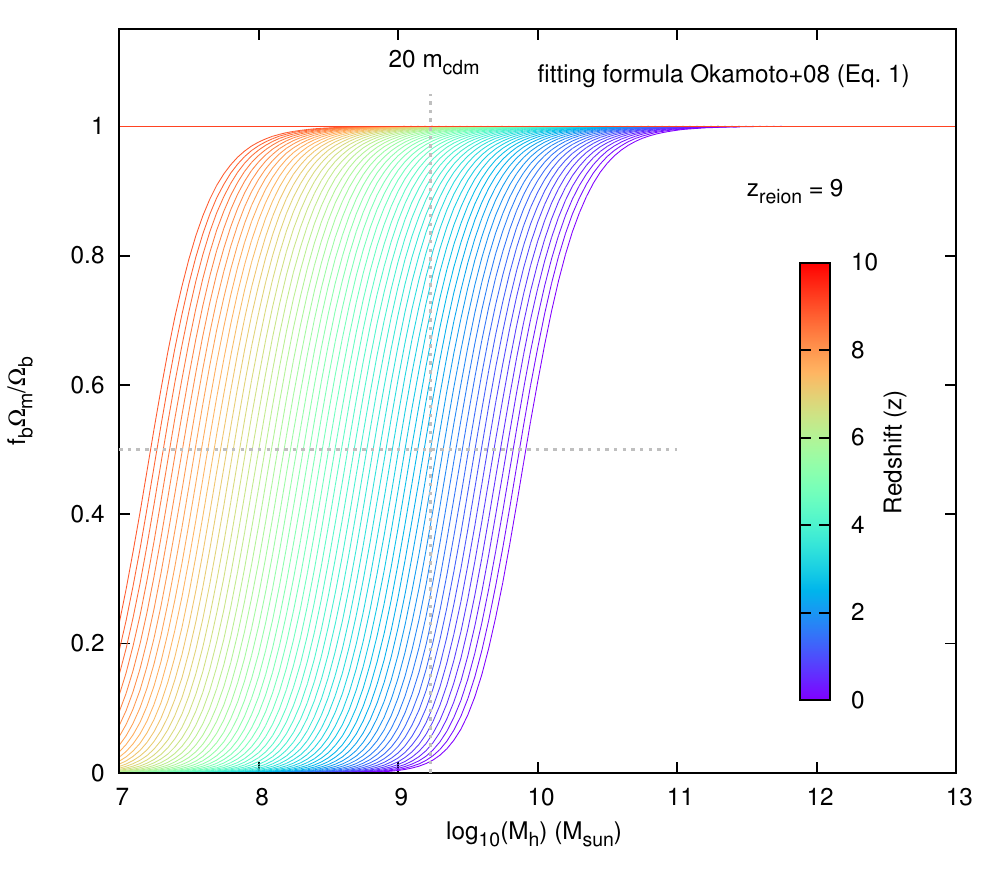}
  \caption{\ftns{Evolution of the effective baryonic fraction ($f_b$: Eq. \ref{baryonic_fraction}) with dark-matter halo mass and redshift. We use the \cite{Okamoto_2008} prescription with a reionisation redshift $z_{reion} = 9$ (see text for more details). The vertical grey line indicates our mass resolution limit (20 dm-particules). At this mass resolution and for a redshift of about 2, the effective baryonic fraction $f_b$ applied to the accretion flux is $\frac{\Omega_b}{\Omega_m}$.}}
  \label{Fig_photoionisation}
  \end{center}
\end{figure}

Figure \ref{Fig_photoionisation} shows the evolution of the effective accreted baryonic fraction as a function of both the redshift and the dark-matter halo mass. The mass resolution limit is 20 dm-particules. At this mass scale, the effective baryonic fraction is strongly reduced ($f_b < \left<f_b\right>/2$) only in the smallest structures evolving at $z<2$. As demonstrated in Cousin et al. (2014), the impact of photoionisation is limited. Even if the amount of accreted gas is reduced, it is not sufficient to significantly limit the star formation activity. 

\subsection{Bimodal accretion}

Following the definitions summarised by Eqs.~\ref{dm_acc_rate} and \ref{baryonic_fraction}, the baryonic accretion rate on a dark-matter halo is given by
\begin{equation} 
  \dot{M}_{bar} = f_b(z,M_{vir})\dot{M}_{acc}. 
  \label{bar_acc_rate}
\end{equation}

As proposed by \cite{Khochfar_2009} we adopt an explicit bimodal accretion composed of a cold and a hot gas flow. These two parallel accretion modes have been discussed in \citealt{Keres_2005, Dekel_2009a, Dekel_2009b, Khochfar_2009, Voort_2010} and, \cite{Faucher-Giguere_2011}.\\

\begin{figure}[h]
  \begin{center}
    \includegraphics[scale = 0.3]{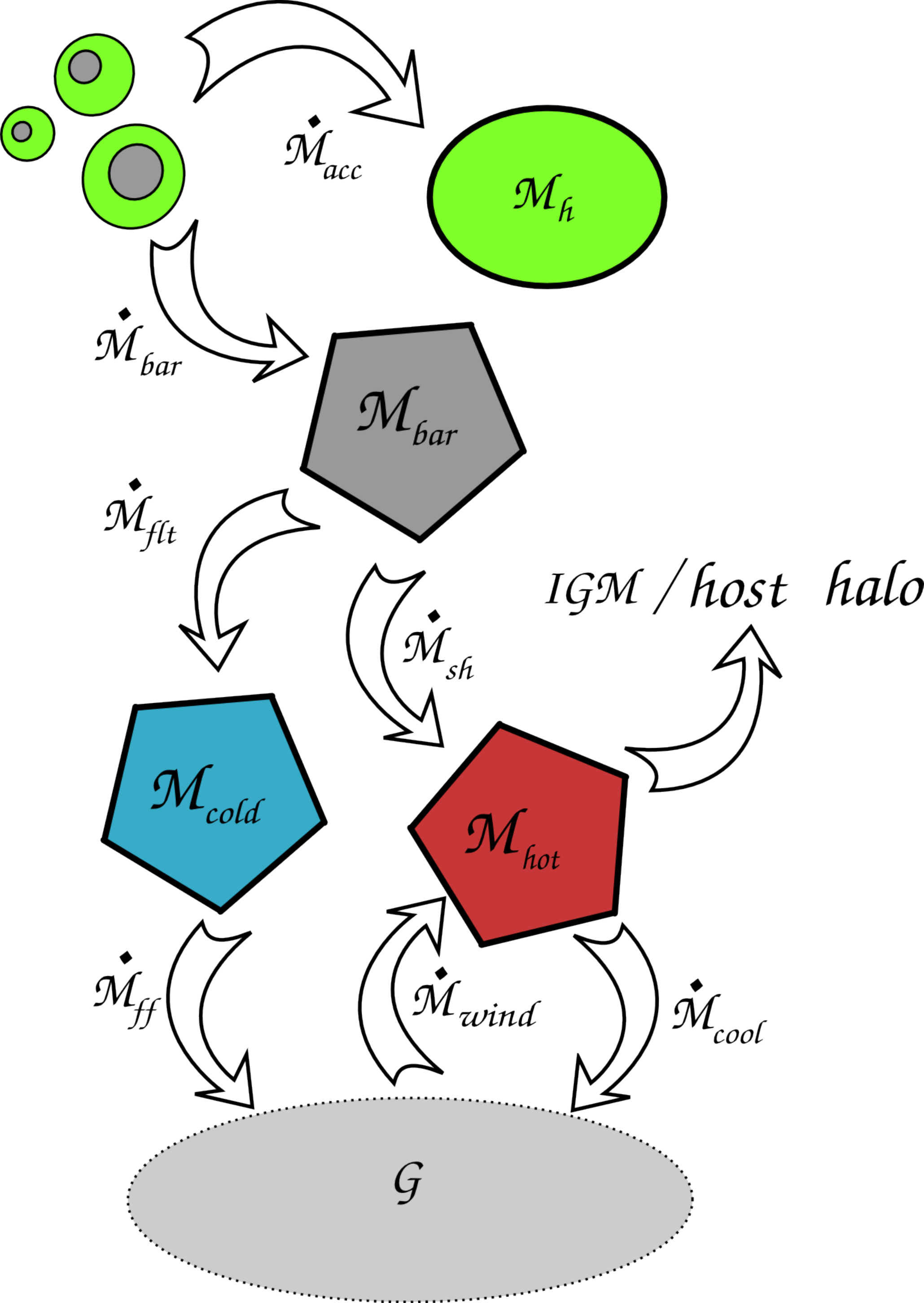}
    \caption{\ftns{\textbf{Large-scale exchanges}. On a large scale, dark-matter and baryons are coupled. During the accretion process and while the dark-matter smooth accretion ($\dot{M}_{acc}$) feeds the dark-matter halo ($M_h$), the associated baryons ($\dot{M}_{bar}$) feed a baryonic reservoir ($M_{bar}$). In the context of bimodal accretion, this baryonic reservoir is divided into two parts: the hot reservoir representing the hot atmosphere ($\dot{M}_{sh}$ and $M_{hot}$), and the cold reservoir representing filamentary streams ($\dot{M}_{flt}$ and $M_{cold}$). The cold gas feeds the galaxy ($\dot{M}_{ff}$) directly. The hot gas follows radiative cooling process and then also feeds the galaxy ($\dot{M}_{cool}$). As described in the text, a galaxy can eject material. The ejecta ($\dot{M}_{wind}$) are continuously added to the hot-gas reservoir ($M_{hot}$) or definitively lost in the $IGM$ according to their velocity distribution. Indeed, the gas stored in the $IGM$ reservoir will never be considered.}}
  \label{baryon_halo_transfer}
  \end{center}
\end{figure} 

On the first hand, we distinguish a cold mode, for which the gas is accreted through filamentary streams. This mode is very efficient. Even if the gas is previously shock-heated \citep{Nelson_2013}, it cools with a very high efficiency ($t_{cool}<<t_{dyn}$) and falls in the centre of the halo in an approximately free-fall time ($t_{dyn}$, see Eq. \ref{t_dyn}). The first galaxy discs are formed quickly with this accretion mode. This mode dominates the growth of galaxies at high redshifts ($z>3$) and the growth of galaxies in the lowest mass haloes ($M_{vir} < 10^{11}~\Msun$) at any times. \\

On the other hand, in more massive haloes ($M_{vir} > 10^{12}~\Msun$) and at late times, the accretion is dominated by a hot mode. For these haloes, the cosmological accreted gas falls into the dark-matter structure and is shock-heated to temperatures close to the virial temperature ($T_{vir}$). This gas participates in the development of a hot stable atmosphere ($T > 10^6~K$) around the central host galaxy.\\ 

Consequently, and as shown in Fig.~\ref{baryon_halo_transfer}, the baryonic cosmological accretion is divided into two phases ($M_{cold}$ and $M_{hot}$). At a given redshift and for a given halo, the hot mass fraction ($f_{sh}(M_{vir},z)$, shock heated) is computed following \cite{Lu_2011} prescription (their Eqs. 24 and 25):  
\begin{equation}
  \begin{split}
    f_{sh}(M_{vir},z) = \dfrac{1}{2}\left[0.1\times exp\left[-\left(\frac{z}{4}\right)^2\right]+ 0.9\right]~~\times \\
    ~~~\left[1+erf\left(\dfrac{logM_{vir}-11.4}{0.4}\right)\right].  
    \label{shock_heated_fraction}
  \end{split}
\end{equation}

\begin{figure}[h]
  \begin{center}
    \includegraphics[scale = 0.9]{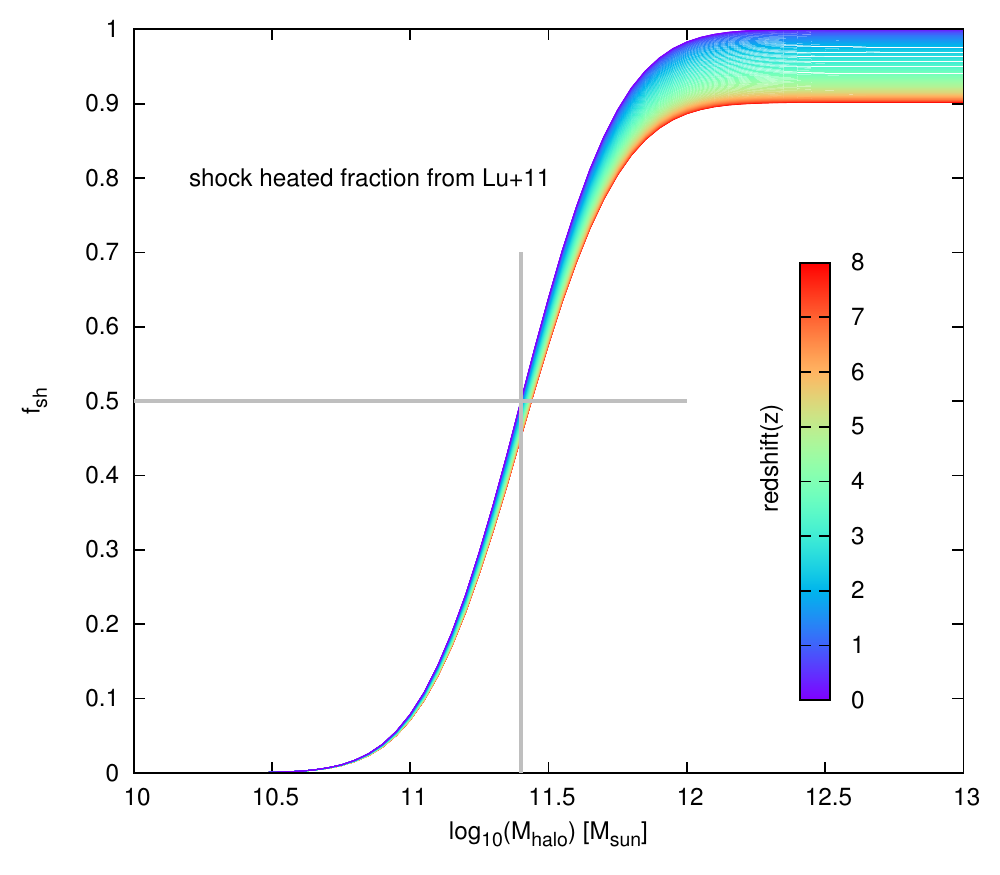}
  \caption{\ftns{Fraction of hot isotropic gas distributed in the baryonic accretion following \citep{Lu_2011}. This fraction depends on the dark-matter halo mass and redshift (colour code).}}
  \label{Fig_shock_heated_fraction}
  \end{center}
\end{figure}

Figure~\ref{Fig_shock_heated_fraction} gives the evolution of $f_{sh}$ with redshift and mass. As explained previously, the cold mode is dominated in the smallest structures ($M_{vir} < 10^{11}~\Msun$), and thus $f_{sh}(M_{vir},z)<<1$. Unlike for these small objects, the cosmological accretion onto the largest haloes is dominated by the hot mode, and $f_{sh}(M_{vir},z) \simeq 1$. It is interesting to note that, at high redshift ($z>4$), even for the largest haloes ($M_{vir} > 10^{12}~\Msun$), a small fraction ($\simeq 10\%$) of the accreted gas is always attributed to the cold mode. This residual cold fraction decreases with time and tends to zero in the local Universe.

\subsubsection{Cold gas accretion, the filamentary part}

Before feeding the galaxy, the cold gas falls in the dark-matter potential well and forms some filamentary structures. This cold gas follows a very collimated distribution. The formation rate of these filamentary structures ($\dot{M}_{flt}$) is deduced from the dark-matter accretion rate ($\dot{M}_{acc}$, Eq.~\ref{dm_acc_rate}) and from the fraction of shock-heated gas ($f_{sh}$, Eq.\ref{shock_heated_fraction}). We define  
\begin{equation}
  \dot{M}_{flt} = (1 - f_{sh})\dot{M}_{bar}\, .
  \label{flt_formation_rate}
\end{equation}
The cold gas is represented by the cold reservoir $M_{cold}$ (Fig.~\ref{baryon_halo_transfer}). Once collimated in the filamentary structure, the cold gas falls into the centre of the dark-matter halo, and feeds a galaxy disc with a rate $\dot{M}_{ff}$ (see Fig. \ref{baryon_halo_transfer}) close to the free-fall  rate ($ff$). We define 
\begin{equation}
  \dot{M}_{ff} = \varepsilon_{ff}\dfrac{M_{cold}}{2t_{dyn}}\, ,
  \label{free-fall-rate}
\end{equation}
where $\varepsilon_{ff}=1.0$ is an efficiency parameter.

The evolution terms of the cold gas reservoir are thus completely known. They follow 
\begin{equation}
  \dfrac{dM_{cold}}{dt} = \dot{M}_{cold} = \dot{M}_{flt} - \dot{M}_{ff}\, . 
  \label{cold_mass_evolution}
\end{equation}
 
\subsubsection{Hot gas accretion, the shock-heated part}
\label{shock_heated_part}

The other part of the baryonic cosmological accretion on a given dark-matter halo is an isotropically distributed gas that is shock-heated when its falls in the potential well. The shock is localised around the virial radius ($r_{vir}$) of the dark-matter structure. The accretion rate ($\dot{M}_{sh}$) of this hot isotropic gas phase is given by 
\begin{equation} 
  \dot{M}_{sh} = f_{sh}\dot{M}_{bar}\, .
  \label{hot_cosmo_acc}
\end{equation}
At any given time, the total hot gas mass ($M_{hot}$ in Fig.~\ref{baryon_halo_transfer}) located around the galaxy is the sum of this cosmological shock-heated gas and the ejected gas produced by the galaxy wind ($\dot{M}_{wind}$ in Fig. \ref{baryon_halo_transfer} and Sect. \ref{ejecta}). \\

In summary, the time evolution of the hot atmosphere follows the differential equation
\begin{equation}
  \dfrac{dM_{hot}}{dt} = \dot{M}_{hot} = \dot{M}_{sh} + \dot{M}_{wind} - \dot{M}_{cool} - \dot{M}_{IGM}\, ,
  \label{hot_mass_evolution}
\end{equation}
where $\dot{M}_{sh}$ is given by Eq. \ref{hot_cosmo_acc}, $\dot{M}_{wind}$ is the ejecta rate (wind) produced by the galaxy (see Sect. \ref{ejecta}), and $\dot{M}_{cool}$ is the cooling rate computed by taking the radiative cooling of the hot gas into account (see Sect.~\ref{cooling}). Finally, $\dot{M}_{IGM}$ is an output rate corresponding to the mass fraction (wind + pre-existing hot gas phase), which leaves the hot atmosphere due to its high velocity (see Sect. \ref{escape_fraction} for more details). Each term of this equation is shown in Fig.~\ref{baryon_halo_transfer}. 

\subsubsection{Total baryonic accretion on the galaxy}

\begin{figure*}[t]
  \begin{center}
    \includegraphics[scale = 0.9]{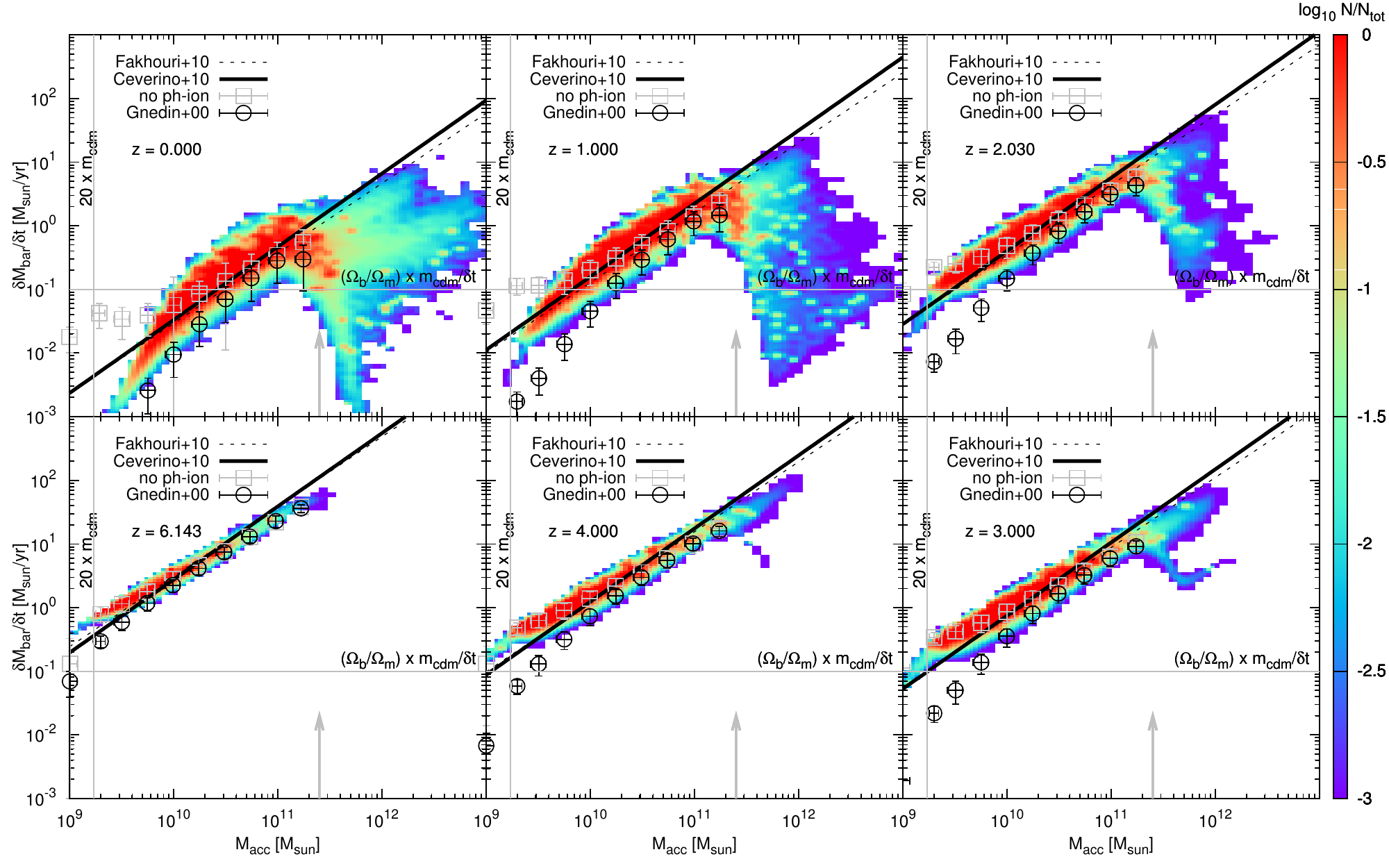}
  \caption{\ftns{The total smooth baryonic accretion onto the galaxy. The colour shading shows the total baryonic accretion rate (free-fall coming from filaments and cooling flows) from our reference model, as a function of dark-matter halo mass, and for a set of redshifts. The colour scale indicates the normalised logarithmic density of objects.} We add for information the transition mass ($10^{11.4}~\Msun$, grey vertical arrow) used  in the accretion mode repartition (Eq.~\ref{shock_heated_fraction}). The dotted and solid black lines show the average baryonic accretion rate deduced from hydrodynamic simulations by \cite{Fakhouri_2010b} and \cite{Ceverino_2010}, respectively. Black circles are for a model with photoionisation prescription by \cite{Gnedin_2000}  instead of our \cite{Okamoto_2008} standard prescription. Grey squares show the accretion for a model without any photoionisation process.}
  \label{Fig_bar_acc}
  \end{center}
\end{figure*}

Figure~\ref{Fig_bar_acc} shows the total baryonic accretion on the galaxy ($\dot{M}_{cool} + \dot{M}_{ff}$) for a set of redshifts ($z\in [0 : 6]$). The grey vertical arrow indicates the transition mass ($10^{11.4}~\Msun$) used in \cite{Lu_2011} (Eq. \ref{shock_heated_fraction}) for the accretion mode repartition. For haloes with mass lower (higher) than this threshold the accretion is dominated by cold mode (cooling process). The colour histogram is built with our reference model. This reference model uses \cite{Okamoto_2008} photoionisation prescription (see Sect.~\ref{photoionisation}). The decrease in accretion efficiency for massive haloes is clearly visible. For comparison, we plot for haloes with mass lower than $M_{vir}=10^{11.4}~\Msun$, the average accretion on the galaxy computed in the case of \cite{Gnedin_2000} photoionisation prescription. As explained in Cousin et al. (2014) the \cite{Gnedin_2000} photoionisation prescription is more efficient, and in this case, the average accretion is lower than with our reference model. We also show in the figure the mean accretion on the galaxy in a model without any photoionisation. At high redshift ($z>2$), the difference between our reference model and the model without any photoionisation is very small. For $z<2$, the differences are larger, especially for the smallest structures (by a factor $10$ at $z = 1$). Again, this confirms that the \cite{Okamoto_2008} photoionisation prescription only has a real impact on the smallest structures ($M_{vir} <10^{10}~\Msun$) and at low redshift. Finally, we plot the mean baryonic accretion computed in hydrodynamic simulations by \cite{Ceverino_2010} (their Eq. 7, $\dot{M}$) and \cite{Fakhouri_2010b} (their Eq. 2, $\frac{\Omega_b}{\Omega_m}<\dot{M}>_{median}$) for comparison. We see that our model is in good agreement with these two simulations. 
  
\subsubsection{Some remarks concerning the bimodal accretion prescription.}

Our accretion model is based on an explicit bimodal baryonic accretion. This prescription allows us to follow the mass of cold and/or hot gas evolving around the galaxy. Indeed, even if the original calculation proposed by \cite{White_1991} included the cooling and the free-fall timescales \citep{Benson_2011}, it was difficult to follow the mass of gas associated with each of these modes. In the context of our explicit two-phase accretion, it is easier to identify the different galaxy populations associated with a given accretion mode. The total fresh gas accreted on the galaxy ($\dot{M}_{ff} +  \dot{M_{cool}}$) in our model has been compared with the prediction from the original calculation proposed by \cite{White_1991}.  It differs by less than 10\%.\\

Some recent works based on a new computational hydrodynamic technique (e.g. \citealt{Nelson_2013}), are questioning the existence of cold streams. It seems that the different computational methods (SPH, AMR, or moving mesh) do not give the same results. In parallel, the impact of the numerical resolution is also very important.  At the time of writing, the debate is still open.   

%
%

\section{The hot halo phase}
\label{hot_phase}

We have presented the hot gas accretion mode that is the first input term of the hot gas phase. In a classical galaxy evolution model, SNe and the SMBH activity heat a fraction of the gas. This gas can be ejected from the galaxy ($\dot{M}_{wind}$) and thus has to be added to the hot gas phase. It is considered as the second input term of the hot gas phase. This section describes the two distinct feedback processes, SN and SMBH. All symbols used in the equations of this section and their definition are summarised in Table~\ref{ejecta_symbol}.\\

\begin{table}[t]
  \begin{center}
    \footnotesize{
      \begin{tabular*}{0.5\textwidth}{@{\extracolsep{\fill}}cr}
        \hline
        Symbol & Definition \\
        \hline     
        $\dot{M}_{wind}$          & Total galaxy ejecta rate [Eqs. \ref{hot_mass_evolution} and \ref{gal_ejecta}, Fig. \ref{baryon_halo_transfer}]\\
        $\dot{M}_{wind,sn}$      & SN galaxy ejecta rate [Eqs. \ref{gal_ejecta} and \ref{sn_energy_conservation}]\\
        $\dot{M}_{wind,\BH}$    & SMBH galaxy ejecta rate [Eq. \ref{agn_tot_ejecta_rate}]\\
        $V_{wind}$                   & Wind velocity (due to SN) [Eqs. \ref{sn_energy_conservation} and \ref{Vwind}]\\
        $f_{Kin,sn}$                  & \textbf{SN kinetic energy fraction = 0.3} [Eq. \ref{sn_energy_conservation}]\\
        $E_{sn}$                      & \textbf{Total SN energy = $10^{44}$~Joules} [Eq. \ref{sn_energy_conservation}]\\
        $\eta_{\BH}$                & \textbf{Outflow/effective accretion = 0.6} [Eq. \ref{agn_out_acc}]\\ 
        $\dot{M}_{\BH,out}$      & SMBH ejecta rate [Eqs. \ref{agn_out_acc}, \ref{agn_mass_conservation}, \ref{agn_in_out_rate}, \ref{agn_energy_conservation}, Fig. \ref{bulge_agn_transfer}]\\
        $\dot{M}_{\BH,acc}$      & SMBH real acc rate [Eqs. \ref{agn_out_acc}, \ref{agn_mass_conservation}, \ref{agn_in_out_rate}, \ref{agn_energy_conservation}, Fig. \ref{bulge_agn_transfer}]\\
        $\dot{M}_{\BH,inf}$       & SMBH total infall rate [Eqs. \ref{agn_mass_conservation}, \ref{agn_tot_ejecta_rate}, \ref{agn_inf}, Fig. \ref{bulge_agn_transfer}]\\
        $f_{Kin,\BH}$                 & \textbf{SN kinetic energy fraction = $10^{-3}$} [Eq. \ref{agn_energy_conservation}]\\
        $\left<V_{wind}\right>$ & Mean velocity wind [Eq \ref{Vwind_gal}]\\
        $\dot{Q}_{sn}$              & SN non-kinetic power [Eq \ref{Qsn}]\\
        $\dot{Q}_{\BH}$            & AGN non-kinetic power [Eq \ref{Qagn}]\\
        $E_{Th,wind}$                 & Thermal energy (wind) [Eq. \ref{Ethwind}]\\
        $f_{Th,sn}$                    & \textbf{SN thermal energy fraction = 0.1} [Eq. \ref{Ethwind}]\\
        $f_{Th,\BH}$                  & \textbf{AGN thermal energy fraction = $10^{-3}$} [Eq. \ref{Ethwind}]\\
        $E_{Th,ath}$                  & Thermal energy (hot phase) [Eq. \ref{Ethatm}]\\
        $E_{Th,Acc}$                 & Thermal energy (accreted hot gas) [Eqs. \ref{Ethacc}]\\
        $\overline{T}_{atm}$    & Mean temperature of the hot phase [Eq. \ref{Tatm}]\\ 
        $\overline{T}_{wind}$   & Mean temperature of the wind [Eq. \ref{Twind}]\\ 
        $f_{esc} $                     & Escape gas mass fraction [Eq. \ref{f_esc}]\\
        \hline
      \end{tabular*}}
  \end{center}  
  \caption{\footnotesize{Ejecta and heating processes symbols and their definition. Symbols are listed in order of appearance. Model parameters are in boldfaced text.}}
  \label{ejecta_symbol}
\end{table}

We present in the following (Sects.~\ref{ejecta} to \ref{cooling}) a self-consistent model of the hot atmosphere. We begin (Sect. \ref{ejecta}) by a complete description of the ejecta rate due to SNe and/or AGN. We continue (Sect.~\ref{energy_transfers}) by a description of the kinetic and thermal energy carried by the ejected mass. We add to this energy calculation a model of energy transfer from the ejecta to the pre-existing hot gas phase. These energy considerations allow us to compute (Sect.~\ref{escape_fraction}) an evolving hot halo temperature and the fraction of mass that can stay in equilibrium in the dark-matter halo potential well. The other part of the mass, which leaves the hot atmosphere, is removed from the hot gas reservoir $M_{hot}$ and is transferred to (see Fig. \ref{baryon_halo_transfer}): 
\begin{itemize}
\item{A passive $IGM$ reservoir if the structure is identified as a main halo,}
\item{the hot gas reservoir of the host M-H if the structure is a S-H (satellite).}
\end{itemize}

We give the formalism for this hot atmosphere model for two different equations of state. In our reference model, we use the ideal gas case (Sect.~\ref{ideal_gas_case}). We also give the equations for the polytropic gas case in Appendix \ref{poly_gas_case}.\\ 

SNe and SMBH (see Figs. \ref{disc_transfer} and \ref{bulge_agn_transfer}) generate outflows ($\dot{M}_{wind}$ in Fig. \ref{baryon_halo_transfer}). The total galaxy ejecta rate is given by the following relations: 
  \begin{equation}
    \dot{M}_{wind}  = \dot{M}_{wind,sn}+ \dot{M}_{wind,\BH} 
    \label{gal_ejecta}
  \end{equation}
We develop these terms in the following section. 

\subsection{Ejecta processes}
\label{ejecta}

\subsubsection{SNe ejecta}
\label{sn_ejecta}

The instantaneous ejecta rate produced by SNe ($\dot{M}_{wind,sn}$) is deduced with the principle of energy conservation. This has already been used \citep[see][]{Dekel_1986, Kauffmann_1993} and \cite{Efstathiou_2000} for a review. We link the kinetic energy per time unit to the power produced by SNe following
  \begin{equation}
    \dfrac{1}{2}\dot{M}_{wind,sn,}V_{wind}^2  = \varepsilon_{sn}f_{Kin,sn}E_{sn}\eta_{sn}\dot{M}_{\star} \ ,
    \label{sn_energy_conservation}
  \end{equation}
where
\begin{itemize}
\item{$(1 - \varepsilon_{sn})$ is the fraction of SNe kinetic energy used to feed the disc velocity dispersion (see Sect.~\ref{disc_vel_disp});}
\item{$f_{Kin,sn}E_{sn}$ is the SNe energy converted in kinetic energy. We use $E_{sn} = 10^{44}$~Joules and $f_{Kin,sn} = 0.3$ \citep{Kahn_1975, Aguirre_2001};}
\item{$\eta_{sn}$ is the number of SNe produced per unit of stellar mass ($[\eta_{sn}] = \Msun^{-1}$). This parameter is linked to the initial mass function IMF;\footnote{The number of SNe increases with the number of massive stars in the initial mass function (IMF). Thus, top-heavy IMFs have a larger $\eta_{sn}$. In this context, the SNe feedback is also larger.}}.
\item{$\dot{M}_{\star}$ is the star formation rate.}
\end{itemize}

This formulation of the kinetic energy per unit of time leads to a degeneracy between the ejected mass rate ($\dot{M}_{wind,sn}$) and the wind velocity ($V_{wind}$). To break the degeneracy, we compute the wind velocity using a model extracted from \cite{Bertone_2005} and based on \cite{Efstathiou_2000} and \cite{Shu_2005}. The wind velocity can be linked to the star formation rate \citep{Martin_1999} and seems to be independent on the galaxy morphology \citep{Heckman_2000, Frye_2002}. We therefore use the following relation (\citealt{Bertone_2005}, their Eq. 9) 
  \begin{equation}
    V_{wind} = 623\left(\dfrac{\dot{M}_{\star}}{100~\Msun\cdot yr^{-1}}\right)^{0.145}~~[km\cdot s^{-1}]\, .
    \label{Vwind}
  \end{equation}

\subsubsection{Super-massive-black hole activity}
\label{agn_ejecta}

An accreting SMBH produces wind. The outflow rate ($\dot{M}_{\BH,out}$) is linked to the effective accretion rate on the SMBH ($\dot{M}_{\BH,acc}$) by the relation \citep{Ostriker_2010}
  \begin{equation}
    \eta_{\BH}=\dfrac{\dot{M}_{\BH,out}}{\dot{M}_{\BH,acc}}\, .
    \label{agn_out_acc}
  \end{equation}
This parameter $\eta_{\BH}$ is currently distributed in the range $[0.1 : 1]$ \citep{Ostriker_2010}. In our reference model we use $\eta_{\BH}=0.6$. 
If $\dot{M}_{\BH,inf}$ is the total infall rate (see Sect. \ref{agn_component} and Eq. \ref{agn_inf} for explicit prescriptions), the conservation of the mass gives obviously
  \begin{equation}
    \dot{M}_{\BH,inf} = \dot{M}_{\BH,acc} + \dot{M}_{\BH,out}\, ,
    \label{agn_mass_conservation}
  \end{equation}
and therefore
  \begin{equation}
    \dot{M}_{\BH,acc} = \dfrac{\dot{M}_{\BH,inf}}{1 + \eta_{\BH}}~~\mbox{and}~\dot{M}_{\BH,out} = \dfrac{\dot{M}_{\BH,inf}\eta_{\BH}}{1 + \eta_{\BH}}\, .
    \label{agn_in_out_rate}
  \end{equation}
Using the same energy conservation criterion as that used for SNe ejecta (see Sect. \ref{sn_ejecta} and Eq. \ref{sn_energy_conservation}), we obtain 
  \begin{equation}
    \dfrac{1}{2}\dot{M}_{\BH,out} V_{jet}^2 = f_{Kin,\BH}\dot{M}_{\BH,acc}c^2\, ,
    \label{agn_energy_conservation}
  \end{equation}
where
\begin{itemize}
\item{$f_{Kin,\BH}$ is the fraction of energy produced by the accretion on the SMBH that is converted into kinetic energy. This parameter is not well known but observations and numerical simulations give a range of plausible values, $f_{Kin,\BH} \in \left[10^{-4} : 10^{-3}\right]$ \citep{Ostriker_2010,Proga_2000, Stoll_2009}. In our reference model we use $f_{Kin,\BH} = 10^{-3}$. This value is compatible with observed outflow velocities, as given for example by \cite{Emonts_2005} and \citealt{Morganti_2005a,Morganti_2005b}.}
\item{$V_{jet}$ is the outflow velocity.}
\end{itemize}

Observations of ionised and neutral gas outflows in radio galaxies show that the SMBH activities and, more precisely, the jet have an impact on the gas content \citep[e.g.][]{Morganti_2005a,Morganti_2005b,Emonts_2005,Nesvadba_2006,Nesvadba_2008,Lehnert_2011,Guillard_2012}. However, it is still unclear how the gas is affected.
To compute the galaxy ejecta rate linked to the SMBH activity ($\dot{M}_{wind,\BH}$), we use the momentum conservation between the SMBH jet and the gas in the galaxy. We assume that a fraction of the gas can reach a velocity equal to the galaxy escape velocity ($V_{esc,\Gal}$) thanks to momentum transfer. The total ejecta coming from the SMBH activity is therefore
  \begin{equation}
    \begin{tiny}
    \dot{M}_{wind,\BH}  = \dfrac{\dot{M}_{\BH,inf}\eta_{\BH}}{1 + \eta_{\BH}}\left[1 + \left(\dfrac{c}{\varepsilon_{\BH}V_{esc,\Gal}}\right)\sqrt{\dfrac{2 f_{Kin,\BH}}{\eta_{\BH}}}\right] 
    \label{agn_tot_ejecta_rate}
    \end{tiny}
  \end{equation}
where 
\begin{itemize}
  \item{$\varepsilon_{\BH}=0.6$ is the coupling factor between the jet momentum and the gas reservoir, This factor is adjusted to produce an outflow rate of a few solar masses per year during the secular evolution of galaxies and a few tens of solar masses per year during the merger induced activity \citep[e.g.][]{Morganti_2005a, Morganti_2005b, Emonts_2005}.}
  \item{$V_{esc,\Gal}$ is the escape velocity of the galaxy (Eq.~\ref{gal_escape_velocity})},
\end{itemize}

\subsection{Energy of the ejecta, transfers to the hot gas phase}
\label{energy_transfers}

As developed in subsequent sections, we introduce in our model an explicit description of the energetic content of the hot gas surrounding the galaxy. Indeed, accretion and ejecta processes generate energy. A fraction of this energy (kinetic, thermal, and luminous) leaves the galaxy and is distributed in the hot halo gas phase. This energy contributes to the heating of the hot gas.\\

We follow the internal energy of
\begin{itemize}
 \item{the hot gas evolving around the galaxy: $E_{Th,atm}$ (\ref{Ethatm});}
 \item{the hot accreted gas: $E_{Th,acc}$ (\ref{Ethacc});} 
 \item{the wind generated by the galaxy: $E_{Th,wind}$ (\ref{Ethwind}).}
\end{itemize}
With these three terms, it is possible to follow the evolution of the mean temperature of the hot gas phase $\overline{T}$ and estimate the mean temperature $\overline{T}_{wind}$ of the wind produced by the galaxy.

In this section, we separately discuss the three terms that are considered as three different internal energy sources. The properties of the hot gas are then computed using the following steps:
\begin{itemize}
\item[1.]{The internal energy of the accreted gas $E_{Th,acc}$ is added to the internal energy of the pre-existing hot gas $E_{Th,atm}$.}
\item[2.]{We then deduce a mean temperature $\overline{T}_{atm}$ for the pre-existing hot gas phase.}
\item[3.]{We deduce a mean temperature $\overline{T}_{wind}$ for the wind phase from its internal energy ($E_{Th,wind}$).}
\item[4.]{We deduce a mean temperature $\overline{T}_{wind}$ for the wind phase, by taking these two mean temperatures ($\overline{T}_{wind}$ and $\overline{T}_{atm}$) into account in a distribution function, $f^{\dagger}$. We finally compute the escape mass fraction $f_{esc}$ and the effective mean temperature $\overline{T}$.}  
\end{itemize}

\subsubsection{Thermal energy of the pre-existing hot phase}
\label{E_pre-existing}

At a given time t, the gas mass in the hot pre-existing atmosphere is $M_{hot}$. This gas is considered in thermal equilibrium in the dark-matter potential well. If the mean temperature of the gas\footnote{In this context, $\overline{T}$ used here is the result at the previous time step of the Eq. \ref{T_hot}.} is $\overline{T}$, and removing the mass that has condensed during the time step ($\dot{M}_{cool}\Delta t$, Eq. \ref{cooling_rate}), the total internal energy of the pre-existing hot gas phase at the end of the time step ($t+\Delta t$) is given by
  \begin{equation}
    E_{Th,atm} = \dfrac{3k_B\overline{T}}{2\mu m_p}\left[M_{hot}(t)-\underbrace{\dot{M}_{cool}\Delta t}_{condensed~mass}\right]\, .
    \label{Ethatm}
  \end{equation}
  
\subsubsection{Thermal energy in the gas accretion}
\label{E_acc_sh}

As explained in Sect.~\ref{shock_heated_part}, the hot isotropic atmosphere is also fed by cosmological hot (shock-heated) accretion. The thermal energy of the gas accreted during the previous time step $\Delta t$ is
  \begin{equation}
    E_{Th,acc} = \dfrac{3k_B T_{vir}}{2 \mu m_p}\left[\underbrace{\dot{M}_{sh}\Delta t}_{accreted~mass}\right]\, ,
    \label{Ethacc}
  \end{equation}
where $T_{vir}$ is the virial temperature of the dark-matter halo (Eq. \ref{T_vir}) and $\dot{M}_{sh}\Delta t$ (Eq. \ref{hot_cosmo_acc}) is the cosmological hot gas mass accreted during the time step $\Delta t$. 

\subsubsection{Thermal energy in the ejecta}
\label{time-everage_thermal_energy}

During a SNe explosion, a fraction $f_{Kin,sn}$ of the total energy ($E_{sn}$) is converted into kinetic energy. The remaining fraction is distributed in luminous energy (photons) and in thermal energy in the ejected gas. We can write the instantaneous non-kinetic power produced by SNe:
  \begin{equation}
      \dot{Q}_{sn} = (1-f_{Kin,sn})E_{sn}\eta_{sn}\dot{M}_{\star}\, .
    \label{Qsn}
  \end{equation}
The instantaneous non-kinetic power produced by SMBH activity follows 
  \begin{equation}
    \begin{split}
    \dot{Q}_{\BH} & = \dot{M}_{\BH,acc}c^2\left(1-f_{Kin,\BH}\right)\\
                        & =  \dot{M}_{\BH,inf}\dfrac{\eta_{\BH}}{1 + \eta_{\BH}}c^2\left(1-f_{Kin,\BH}\right)\, .
  \end{split}
  \label{Qagn}
\end{equation}
Equations \ref{Qsn} and \ref{Qagn} are used to compute the instantaneous power. We then compute the (time-)average on the evolution time step $\Delta t$ of
\begin{itemize}
  \item{the non-kinetic power, $\left<\dot{Q}_{sn}\right>$;}
  \item{the SMBH non-kinetic power, $\left<\dot{Q}_{\BH}\right>$.}
\end{itemize} 
With these two quantities, we can derive the thermal energy carried by the galactic winds during $\Delta t$ following
  \begin{equation}
    \begin{split}
    E_{Th,wind} = \Delta t\left[f_{Th,sn}\left<\dot{Q}_{sn}\right> + f_{Th,agn}\left<\dot{Q}_{\BH}\right>\right]\, ,
    \end{split}
    \label{Ethwind}
  \end{equation}
where we define $f_{Th,sn}$ and $f_{Th,\BH}$ as two free parameters describing the fraction of the non-kinetic energy distributed in thermal energy by SNe and SMBH feedback, respectively. The reference values are set to $f_{Th,sn} = 0.1$ and $f_{Th,\BH} = 10^{-3}$. These values have been chosen so as to produce a temperature distribution centred on
\begin{itemize}
\item{$10^6$ K for SN wind. Indeed, typical SN remnant emission shows temperatures close to $k_BT \in [0.1 : 0.7]$ keV, which correspond to $T\in [1 : 8] 10^{6}$ K} \citep{Koo_2002, Sasaki_2014;}
\item{$10^7$ K for SMBH wind.}
\end{itemize}

\subsubsection{Temperatures}

During a time step $\Delta t$, the hot gas phase is fed by accretion, cooling, and ejecta processes. In this context, the mean temperature of the hot atmosphere after a time step $\Delta t$ is given by
  \begin{equation}
    \begin{split}
      \overline{T}_{atm} & = \dfrac{2\mu m_p}{3k_B}\left[\dfrac{E_{Th,atm} + E_{Th,acc}}{M_{hot}(t)+\underbrace{\dot{M}_{sh}\Delta t}_{accreted~mass}-\underbrace{\dot{M}_{cool}\Delta t}_{condensed~mass}}\right]\\
      & = \dfrac{\overline{T}\left(M_{hot}(t)-\dot{M}_{cool}\Delta t\right)+ T_{vir}\dot{M}_{sh}\Delta t}{M_{hot}(t)+\dot{M}_{sh}\Delta t-\dot{M}_{cool}\Delta t}\, .
      \end{split}
      \label{Tatm}
  \end{equation}
In parallel we consider that the temperature of the wind is constant during the time step $\Delta t$. It is given by
  \begin{equation}
    \overline{T}_{wind}= \dfrac{2\mu m_p}{3k_B}\dfrac{E_{Th,wind}}{\underbrace{M_{wind}}_{ejected~mass}}\, .
    \label{Twind}
  \end{equation}

\subsection{Escape fraction and mean temperature}
\label{escape_fraction}

\subsubsection{Average velocity and mass of the galaxy wind}

On the galaxy scale, the instantaneous total wind velocity (SN and SMBH) is computed using velocity winds of each component weighted by their ejecta rate. We define 
  \begin{equation}
    \small{
    V_{wind} = \dfrac{\dot{M}_{wind,sn}V_{wind}+\dot{M}_{wind,\BH}V_{esc,\Gal}}{\dot{M}_{wind,sn}+\dot{M}_{wind,\BH}}\, .
    \label{Vwind_gal}}
  \end{equation}
This instantaneous wind velocity is only valid on the galaxy scale. As described in Eq.~\ref{adaptive_time_step} the evolution of the galaxy components (disc and bulge) is performed with an adaptative time step $\delta t$ shorter than the hot atmosphere time step $\Delta t = \sum_i\delta t_i$. To take the impact of the hot gas wind on the hot gas atmosphere into account, such as the permanently lost fraction (see Sect. \ref{escape_fraction} and Eq. \ref{f_esc}), it is necessary to estimate the average wind velocity for the time scale used for the hot phase. The wind velocity is therefore computed following
\begin{equation}
    \small{
    \ds \left<V_{wind}\right> = \dfrac{1}{\Delta t}\sum_{i=1}^N V_{wind}(i\delta t)\delta t\, .
    \label{Vwind_hot}}
  \end{equation}
The total mass contained in the wind is therefore given by
  \begin{equation}
    \small{
    \ds M_{wind} = \sum_{i=1}^N \dot{M}_{wind}(i\delta t)\delta t\, .
    \label{Mwind_hot}}
  \end{equation}

\subsubsection{Velocity distributions}

We assume that the hot atmosphere in which the wind extends is in hydrostatic equilibrium in the dark-matter halo potential well. The hot gas phase has a mean temperature $\overline{T}_{atm}$. In these conditions, the velocity probability distribution of a particle in this hot gas phase is given by the well-known Maxwell-Boltzman distribution, $f_{MB}(v,\overline{T}_{atm})dv$ (see Eq. \ref{MB_dist}). 

Concerning the wind, in the gas frame of reference, the probability velocity distribution is also given by a Maxwell-Boltzman distribution, $f_{MB}(v,\overline{T}_{wind})dv$. In the fixed halo (or galaxy) referential, if we take the mean velocity of the wind into account, the velocity probability distribution is also given by the conditional following relation
\begin{equation}
f(v,\overline{T}_{wind})dv = \left\{
  \begin{array}{ll}
   f_{MB}\left(v ',\overline{T}_{wind}\right)dv  & \mbox{: if $v' > 0$}\\
    & \\
    0 & \mbox{: otherwise,}
  \end{array}\right.
\end{equation}
with $v' = v-\left<V_{wind}\right>$.

At any given time, we can consider that the hot gas phase is composed of a pre-existing hot atmosphere in which the hot gas produced by feedback processes is injected. Therefore, the global velocity probability distribution is given by,
\begin{equation}
  \begin{split}
   f^{\dagger}(v,\overline{T}_{atm},\overline{T}_{wind})dv = & (1-f_{wind})f_{MB}(v,\overline{T}_{atm})dv\\
            &+f_{wind}f(v,\overline{T}_{wind})dv
  \label{F_dag}
  \end{split}
\end{equation}
with
\begin{equation}
   f_{wind} = \dfrac{M_{wind}}{M_{hot}+M_{wind}+\underbrace{\dot{M}_{sh}\Delta t}_{accreted~ mass}-\underbrace{\dot{M}_{cool}\Delta t}_{condensed~mass}}\,.
   \label{fwind}
\end{equation}

\subsubsection{The effective escape fraction}

At a given time some particles in the pre-existing hot gas phase and even more in the wind phase can have higher velocities than the escape velocity of the dark-matter structure $V_{esc,dm}$ (Eq. \ref{Vesc_dm}). We assume that only the dark-matter mass and therefore the dark-matter gravitational well have a direct impact on the hot gas phase confinement. \cite{Suto_1998} demonstrate that the self-gravity of the hot gas phase has an impact on the hot gas phase density profile but only at large radius ($r>r_{dm} $). The density profile of the hot gas is an important quantity in the gas cooling modelling, but if we consider as \cite{Suto_1998} that only the core part of the hot atmosphere can cool in a short time, the impact of not taking the self-gravity on cooling processes into account is small. \\

The mass fraction $f_{esc}$ that can leave the dark-matter potential well is given by 
  \begin{equation}
    f_{esc} = 1 - \int_0^{V_{esc,dm}}f^{\dagger}(v,\overline{T}_{atm},\overline{T}_{wind})dv\, .
    \label{f_esc}
  \end{equation}

\begin{figure}[h]
  \begin{center}
    \includegraphics[scale = 0.9]{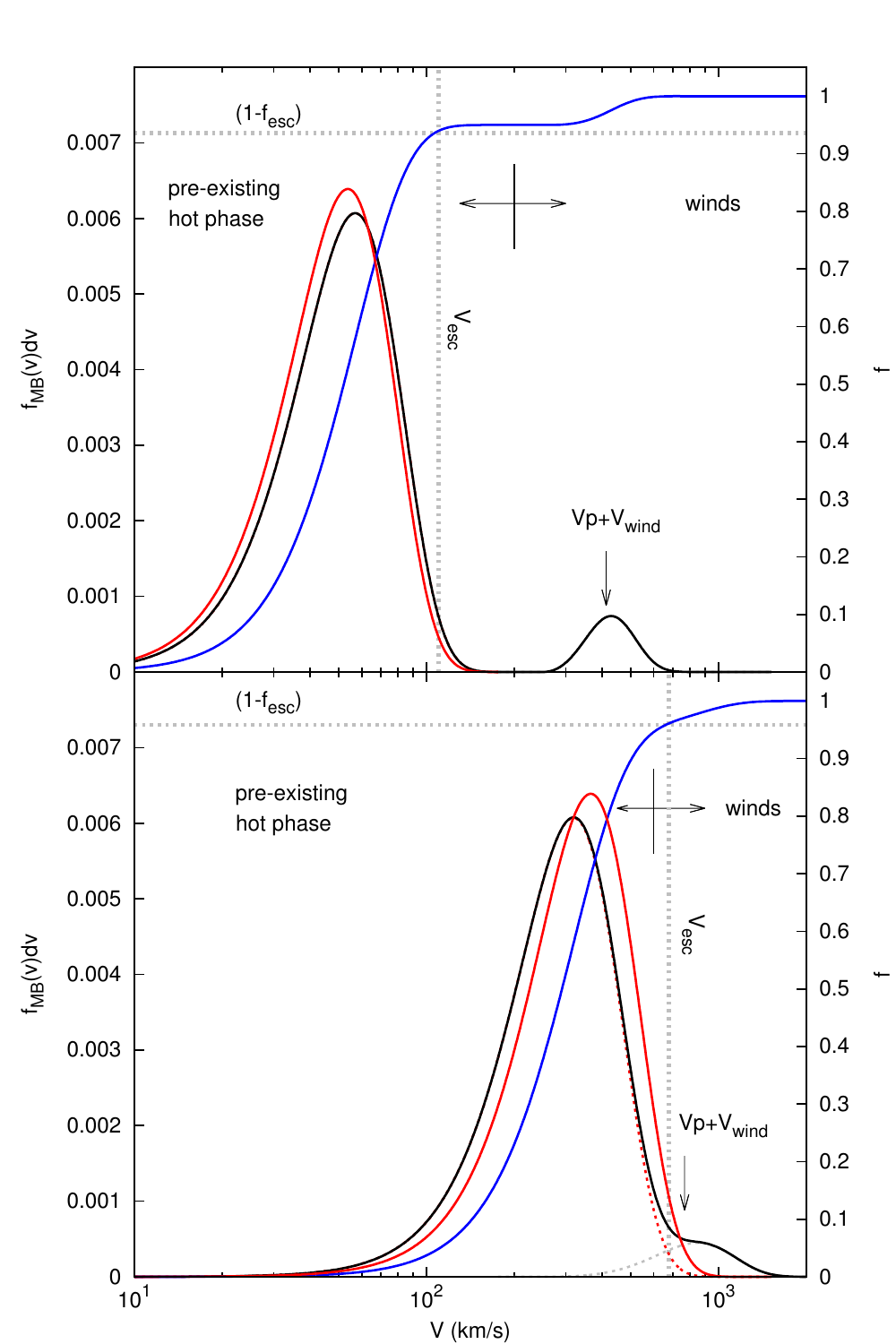}
  \caption{\ftns{Distribution functions of the velocity in the pre-existing hot gas phase and in the wind. The upper and lower panels show the distributions for two different dark-matter haloes with $M_{vir} = 10^{10}~\Msun$ and $M_{vir} = 10^{13}~\Msun$, respectively. In the two cases, the wind speed is close to $V_{wind} = 250$ km/s but the temperature is close to $\overline{T} =10^6$ K in the upper panel and $\overline{T}=10^7$ K in the lower panel. The total gas distribution, in black, is the sum of the pre-existing hot gas distribution (red dotted line) and the gas wind distribution (grey dotted line). The blue curve gives the mass fraction. The vertical dotted grey line indicates the escape velocity of the dark-matter structure. The intersection of the mass fraction and this velocity threshold allows us to determine the escape fraction. The red curve shows the new hot gas phase distribution recomputed with the new temperature $\overline{T}$ (Eq.~\ref{T_hot}) derived after accretion, cooling, and ejection processes have been taken into account during the previous time step $\Delta t$. In the low-mass case, all the mass contained in the wind leaves the structure. In the high-mass case, a fraction of the wind mass can be retained.}}
  \label{Fig_f_esc}
  \end{center}
\end{figure}

Figure~\ref{Fig_f_esc} shows the distribution of $f^{\dagger}(v,\overline{T}_{atm},\overline{T}_{wind})dv$ (Eq. \ref{F_dag}) for two different dark-matter halo masses. In the two cases the wind velocity is close to $V_{wind} = 250$ km/s but for $M_{vir} = 10^{10}~\Msun$, the temperature of the wind is close to $\overline{T} =10^6$ K, while for $M_{vir} = 10^{13}~\Msun$, it is close to $\overline{T} =10^7$ K. As expected, the fraction of the wind mass that can definitively leave the structure is larger for the lowest dark-matter halo mass. Indeed, for such small structures, all the mass contained in the wind leaves the potential well. 

After ejection and evaporation, the distribution and the average temperature are recomputed. In Fig.~\ref{Fig_f_esc} the new $f^{\dagger}(v,\overline{T}_{atm},\overline{T}_{wind})dv$ distribution is shown in red. For the halo with $M_{vir} = 10^{13}~\Msun$, the temperature has increased because the hot wind has participated in heating the hot isotropic atmosphere.

\subsubsection{Mass transfer and mass loss}
\label{hot_mass_transfers}

The mass $M_{IGM}=f_{esc}M_{hot}$ leaves the hot halo owing to its high velocity. If the galaxy that generates the outflows is hosted by an S-H (satellite), we recompute the ejection processes, following the previous description, in the host M-H referential. Indeed, the halo finder algorithm creates links between S-H and their host M-H. We compute the escape velocity of M-H, and deduce the fraction of ejected mass that can leave its dark-matter potential well. The difference between the mass lost by S-H and M-H is added to the hot reservoir of M-H. The fraction of the mass that can leave M-H is definitively removed from the hot reservoir and added to a passive $IGM$ reservoir (see Fig.~\ref{baryon_halo_transfer}).

After mass ejection, the average temperature of the gas is modified. This new mean temperature is then given by
  \begin{equation}
    \overline{T} = \dfrac{2\mu m_p}{3k_B(1-f_{esc})}\int_0^{V_{esc,dm}}v^2f^{\dagger}(v,\overline{T}_{atm},\overline{T}_{wind})dv\, .
    \label{T_hot}
  \end{equation}

\subsection{Some remarks concerning the SMBH feedback efficiency}

SMBH feedback has been implemented in different models \citep[e.g.][]{Croton_2006,Cattaneo_2006,Bower_2006,Malbon_2007,Somerville_2008,Ostriker_2010,Bower_2012}. For example, in \cite{Croton_2006} (their Eqs. 10 and 11) and/or \cite{Somerville_2008} (their Eq. 21) SMBH activity directly affects the cooling of the hot gas phase. Indeed in these two models, a fraction of the power generated by the accretion is considered as a heating gas rate $\dot{M}_{heat}$. This heating rate is then used to compute an effective cooling rate, $\dot{M}_{cool,eff} = MAX(0,\dot{M}_{cool,eff} - \dot{M}_{heat})$. This approach leads to a very efficient SMBH feedback in which a large fraction of the power generated by the accretion is directly used to reduce the cooling process.

In our approach, SMBH feedback contributes to gas ejection (Eq.~\ref{agn_tot_ejecta_rate}), but the impact on the cooling rate is only given by a possible increase in the hot gas phase temperature due to wind (thermal energy in the wind, Eqs.~\ref{Qagn} and \ref{Ethwind}). In this context, even if the SMBH activity produces high-temperature wind ($\simeq 10^{7}$ K), our SMBH feedback implementation leads to less efficiency and therefore has a weaker impact on the cooling process.

\subsection{The hot gas density profile}
\label{hot_profile}

To describe the hot isotropic atmosphere located around the galaxy and to take the impact of the cooling process into account, we need to define a density profile. This section is dedicated to describing of i) the density profile and ii) the cooling process. All the parameters and their definitions are listed in Table~\ref{cooling_symbol}.

\begin{table}[t]
  \begin{center}
    \footnotesize{
      \begin{tabular*}{0.5\textwidth}{@{\extracolsep{\fill}}cr}
        \hline
        Symbol & Definition \\
        \hline     
        $M_{hot}$                   & Hot atmosphere mass [Eqs. \ref{hot_mass_evolution}, \ref{Mhot_r}, Fig. \ref{baryon_halo_transfer}]\\
        $\rho_g(r)$               & Hot atmosphere radial density profile [Eq. \ref{Mhot_r}]\\
        $T_0$                       & Hot atmosphere central temperature [Eq. \ref{T0}]\\
        $\rho_g^{id}(r)$          & Radial density profile (ideal gas) [Eq. \ref{HEC_id}]\\
        $P^{id}(r)$                  & Radial pressure profile (ideal gas) [Eq. \ref{HEC_id}]\\
        $\rho_{g,0}$               & Hot gas central density [Eqs. \ref{rhog0}, \ref{rho_gas_core_id}, \ref{rho_gas_core_poly}]\\ 
        $\mathcal{F}^{id}(x)$ & Geometrical function (density profile) [Eqs. \ref{Fidx}, \ref{rhog0}]\\
        $\Phi(x)$                  & Geometrical function (density profile)  [Eqs. \ref{Phix}, \ref{Fidx}]\\
        $c_{dm}$                    & Dark-matter halo concentration [Eq. \ref{dm_mass_rvir}]\\
        $P^{poly}(r)$               & Radial pressure profile (polytropic gas) [Eq. \ref{HEC_poly}]\\
        $\rho_g^{poly}(r)$       & Radial density profile (polytropic gas) [Eqs. \ref{HEC_poly}, \ref{rho_poly}]\\
        $T^{poly}(r)$                 & Radial temperature profile (polytropic gas) [Eq. \ref{HEC_poly}]\\
        $\mathcal{F}^{poly}(x)$ & Geometrical function (density profile) [Eqs. \ref{F_poly}, \ref{rho_poly}]\\
        $\overline{T}^{poly} $  & (Mass-)average temperature [Eqs. \ref{eq_avg_temp_poly}, \ref{T0_Tpoly}]\\  
        $T_0^{min}$                 & Minimal central temperature [Eq. \ref{T0_min}]\\
        $t(r)$                        & Cooling time function [Eq. \ref{cooling_time_function}]\\
        $Z_g$                       & Hot gas metalicity [Eq. \ref{cooling_time_function}]\\
        $r_{cool}$                   & Cooling radius [Eq. \ref{cooling_time_function}]\\
        $t_{cool}$                   & Effective cooling time [Eq. \ref{cooling_time_function}]\\
        $\dot{M}_{cool}$         & Cooling rate [Eq. \ref{cooling_rate}]\\ 
        \hline
      \end{tabular*}}
  \end{center}  
  \caption{\footnotesize{Symbols and their definition for the hot atmosphere and cooling process. Symbols are listed in order of appearance.}}
  \label{cooling_symbol}
\end{table}

We first assume that the hot stable mass $M_{hot}$ (in hydrostatic equilibrium) is enclosed in the virial radius $r_{vir}$ of the dark-matter structure. To model the mass distribution of the hot gas we assume a spherical geometry and thus define
  \begin{equation}
    M_{hot} = 4\pi \int_0^{r_{vir}}\rho_g(r)r^2dr \ .
    \label{Mhot_r}
  \end{equation}
The density distribution $\rho_g(r)$ of the hot atmosphere is deduced from the hydrostatic equilibrium condition (HEC) computed in a potential dominated by the dark-matter,
\begin{center}
  \begin{equation}
    \dfrac{1}{\rho_g(r)}\dfrac{dP(r)}{dr}  = -\dfrac{G M_{h}(<r)}{r^2}\, ,
    \label{HEC_id}
  \end{equation}
\end{center}
where $G$ is the gravitational constant, $P(r)$ the pressure radial profile and $M_{h}(<r)$ the dark-matter mass enclosed in the radius $r$. In our model it is possible to use two different density profiles deduced from an ideal or a polytropic gas. In the reference model, we use the isothermal gas case. For completeness we describe the polytropic gas case in Appendix \ref{poly_gas_case}.

\subsubsection{The isothermal ideal gas case}
\label{ideal_gas_case}

For an ideal (\textit{id}) gas, the equation of state links the pressure, the density, and the temperature by the well-known following relation
  \begin{equation}
    P^{id}(r)=\frac{k_b T_0}{\mu m_p}\rho_g^{id}(r) \ ,
    \label{T0}
  \end{equation}
where $k_B$ is the Boltzman constant, $\mu$ the mean molecular weight of the gas, $m_p$ the proton mass, $P^{id}(r)$ the radial pressure profile, $\rho_g^{id}(r)$ the density profile, and $T_0$ the central temperature of the gas. In this simple case of an ideal gas we assume, that the radial profile of the temperature $T(r)$ is set to a constant value and that this constant value is equal to the central value $T_0$. Here, $T_0$ is equal to the average value $\overline{T}_{atm}$.
The HEC in the dark-matter halo potential \citep{Suto_1998, Makino_1998, Komatsu_2001, Capelo_2012} gives 
  \begin{equation}
    \dfrac{dln(\rho_g^{id})}{dr} = -\dfrac{G\mu_g m_p}{k_BT_0}\frac{M_{h}(<r)}{r^2} = -\dfrac{4\pi G\mu m_p\rho_{dm}r_{dm}}{k_BT_0}\frac{\phi(x)}{x^2} \ ,
  \end{equation}
where $\rho_{dm}$ and $r_{dm}$ are the core density and the characteristic radius of the dark-matter halo, respectively and $\phi(x)$ is the geometrical function given in Eq.~\ref{phix}.

For an NFW profile \citep{Navarro_1995, Navarro_1996, Navarro_1997}, the hot gas profile that follows the HEC is \cite{Suto_1998, Capelo_2012}
  \begin{equation}
    \rho_g^{id}(r) = \rho_{g,0}\mathcal{F}^{id}(x)~~\mbox{with}~~x=\frac{r}{r_{dm}}\, ,
    \label{rhog0}
  \end{equation}
with $\mathcal{F}^{id}(x)$, a geometrical function defined as 
  \begin{equation}
    \mathcal{F}^{id}(x) = exp\left[-\dfrac{4\pi G\mu m_p\rho_{dm}r_{dm}^2}{k_bT_0}\Phi(x)\right]\, ,
    \label{Fidx}
  \end{equation}
where  $\Phi(x)$ is another geometrical function based on the previous one $\phi(x)$ (Eq. \ref{phix}) and defined by 
  \begin{equation}
    \Phi(x)=\ds\int_{0}^{x}\dfrac{\phi(u)}{u^2}du = 1 - \dfrac{ln(1+x)}{x}\, .
    \label{Phix}
  \end{equation}

In the case of an ideal gas, the geometrical distribution of the hot gas is completely set by the dark-matter properties. The last parameter that remains to be determined is the gas core density $\rho_{g,0}$. It is extracted from the total mass contained in the hot halo (Eq.~\ref{Mhot_r}) as
  \begin{equation}
    \rho_{g,0} = M_{hot}\left[4\pi r_{dm}^3\int_0^{c_{dm}} \mathcal{F}^{id}(x)x^2dx \right]^{-1}\, .
    \label{rho_gas_core_id}
  \end{equation}

\subsection{Cooling process}
\label{cooling}

The cooling and condensation of the hot atmosphere is computed using the classical model initially proposed by \cite{White_1991}. We recall the cooling time function here, which gives the cooling time of the gas mass enclosed in a radius $r$ and which follows a given density $\rho_g(r)$ and temperature profile $T(r)$,   
  \begin{equation}
    t(r) = 1.04\dfrac{\mu m_p T(r)}{\rho_g(r)\Lambda[T(r),Z_g]}\, .
    \label{cooling_time_function}
  \end{equation}
The cooling time function also depends on the mean molecular weight of the gas $\mu$, the proton mass $m_p$, and on the cooling function $\Lambda$ that depends on the temperature $T(r)$ and metallicity $Z_g$ of the hot gas. We use the cooling function $\Lambda$ from \cite{Sutherland_1993}, for temperature between $10^4$ K and $10^{8.5}$ K and metallicity between $10^{-30}~Z_{\odot}$ and $10^{0.5}~Z_{\odot}$. 

\subsubsection{Cooling time, radius, and rate}

To compute the cooling rate, and therefore the condensed mass of gas that can fall into the galaxy, we need to define the cooling radius, $r_{cool}$ which the radial extension of the hot atmosphere that can condense during a given time $t_{cool}$. The cooling radius $r_{cool}$ is the solution of the following equation
  \begin{equation}
    t(r_{cool}) = t_{cool}
    \label{cooling_radius_equation}\, ,
  \end{equation}
where $t(r)$ is the cooling time function (Eq.~\ref{cooling_time_function}).

Different definitions for $t_{cool}$ exist in the literature \citep[][e.g.]{Cole_2000, Hatton_2003}, and thus for $r_{cool}$ (see \cite{DeLucia_2010} for a discussion). In our model, we link $t_{cool}$ to a cooling clock:
\begin{itemize}
  \item{The cooling clock starts when the hot reservoir $M_{hot}$ receives some mass. The clock runs as long as the hot phase contains gas. Therefore, $t_{cool}$ increases with the evolving time of the hot atmosphere.}
  \item{The cooling clock is set to 0 when a major merger event occurs. We consider a major merger event when the mass ratio of the two hot gas reservoirs ($M_{hot}^1$ and $M_{hot}^2$)
\begin{equation}
  MIN(M_{hot}^1,M_{hot}^2)/MAX(M_{hot}^1,M_{hot}^2) >1/3\, .
\end{equation}}
  \item{After a minor merger, the cooling clock is set to the cooling time $t_{cool}$ of the most massive progenitor.\footnote{The most massive structure is defined as the structure with the higher mass of hot gas $M_{hot}$, which is not necessarily the highest $M_{vir}$ dark-matter structure.}}
\end{itemize}

The cooling rate of the hot atmosphere is then computed following
  \begin{equation}
    \dot{M}_{cool} = \dfrac{\varepsilon_{cool}}{2t_{dyn}}\int_0^{MIN(r_{cool},r_{vir})}\rho_g(r)r^2dr\, .
    \label{cooling_rate}
  \end{equation}
The mass enclosed in the cooling radius $r_{cool}$ falls into the galaxy with a rate close to the free-fall rate.  We assume that the hot atmosphere extends up to the virial radius, so that the cooling radius cannot be larger than this virial radius. The free efficiency parameter is set to $\varepsilon_{cool}=1.0$.

%
%

\section{Galaxy components}
\label{galaxy}

\begin{figure}[h]
  \begin{center}
    \includegraphics[scale = 0.12]{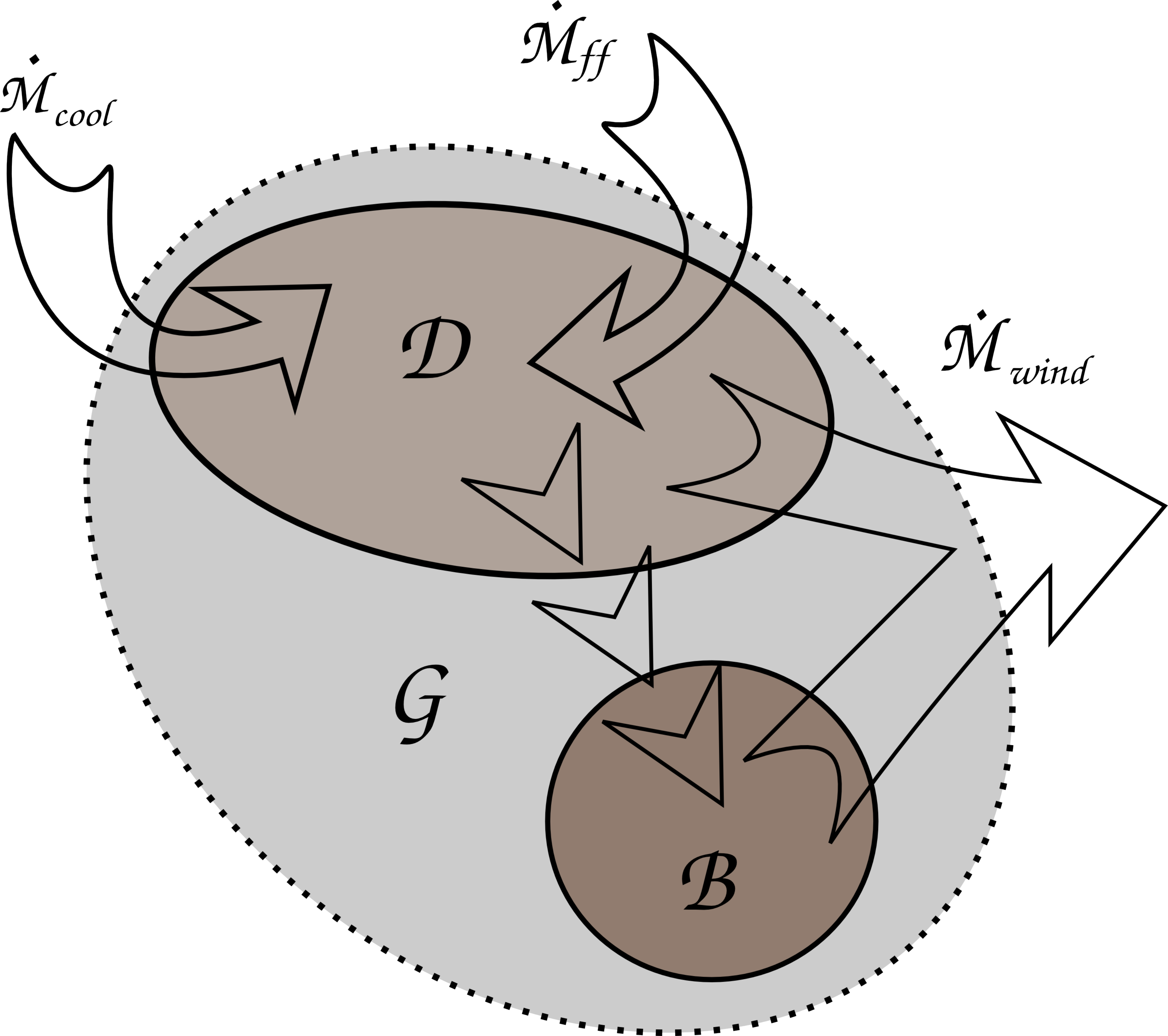}
  \caption{\ftns{\textbf{Exchanges on the galaxy scale}. The galaxy scale is considered as an intermediate scale. A galaxy can be made of two different parts, a disc $\Disc$ and a bulge $\Bulge$. The fresh gas ($\dot{M}_{ff}$ and $\dot{M_{cool}}$), coming from the cold and the hot reservoirs, only feeds the disc component. The bulge component is created during a merger or by the clump migration process. SNe and the activity of the SMBH produce ejecta ($M_{wind}$). These ejecta are considered at an upper scale (see Fig.~\ref{baryon_halo_transfer}). The gas exchanges associated with the disc and the bulge are described in Figs. \ref{disc_transfer} and \ref{bulge_agn_transfer}.}}
  \label{gal_transfer}
  \end{center}
\end{figure}

We present the relations and definitions linked to galaxy formation and evolution in the next sections. 
An \textit{e}GalICS galaxy (Fig.~\ref{gal_transfer}) is composed of two distinct but interacting components: a disc $\Disc$ (Fig.~ \ref{disc_transfer}) and a bulge or bulge-like component $\Bulge$ (Fig.~\ref{bulge_agn_transfer}).

\subsection{Accretion on the galaxy}

At the first steps of galaxy formation, the cold gas (from filamentary collimated structures and/or the isotropic cooling flow) falls into the center of the dark-matter halo. We assume that this cold gas initially forms a thin exponential disc. The gas acquires angular momentum during the mass transfer \citep{Peebles_1969}. After its formation, the disc is supported by its angular momentum. This paradigm is based on the prescription given by \cite{Blumenthal_1986} and \cite{Mo_1998} and has been frequently used in SAMs \citep[e.g.][]{Cole_1991, Cole_2000, Hatton_2003,Somerville_2008}. During the secular evolution of the galaxy, some mass may be transferred to a pseudo-bulge component by disc instabilities (see Sects. \ref{clumps_formation} and \ref{clumps_transfert} for more information). 
After a major merger event (see Sect.~\ref{major_or_micro_merger_events}), the remnant galaxy has a spheroidal morphology (see Sect.~\ref{bulge}). The gas contained in the remnant galaxy and the freshly accreted gas, which will fall into the centre of the halo in the next time step, will then form a new disc. \\

\begin{table}[t]
  \begin{center}
    \footnotesize{
      \begin{tabular*}{0.5\textwidth}{@{\extracolsep{\fill}}cr}
        \hline
        Symbol & Definition \\
        \hline    
        $\Gal$           & Common symbol for galaxy properties\\         
        $V_{circ,\Gal}$   & Circular velocity [Eq. \ref{gal_circular_velocity}]\\
        $ V_{esc,\Gal}$  & Escape velocity  [Eq. \ref{gal_escape_velocity}]\\
        $r_{\Gal,50}$     & Half-mass radius [Eq. \ref{gal_escape_velocity}]\\
        \hline
      \end{tabular*}}
  \end{center}  
  \caption{\footnotesize{Galaxy symbols and their definition. Symbols are listed in order of appearance.}}
  \label{galaxy_symbol}
\end{table}

\subsection{The circular and escape velocity of the galaxy}

The radial profile of the circular velocity of the galaxy is deduced from the dynamic of all its components, the dark-matter halo $\DM$, disc $\Disc$, and bulge $\Bulge$. We obtain 
  \begin{equation}
    V_{circ,\Gal}^2 = V_{circ,\DM}^2 + V_{circ,\Disc}^2 + V_{circ,\Bulge}^2 \, .
    \label{gal_circular_velocity}
  \end{equation}
All the terms will be defined in their corresponding description section (Eqs.~\ref{Vcirc_dm}, \ref{Vcirc_d}, and \ref{Vcirc_b}).

The escape velocity of the galaxy is computed following
  \begin{equation}
    V_{esc,\Gal}^2 = \dfrac{2M_{\Gal}(r<r_{\Gal,50})+M_{vir}(r<r_{\Gal,50})]}{r_{\Gal,50}} \, ,
    \label{gal_escape_velocity}
  \end{equation}
where $M_{\Gal}(r<r_{\Gal,50})$ and $M_{vir}(r<r_{\Gal,50})$ are the galaxy and dark-matter mass enclosed in the galaxy half-mass radius $r_{\Gal,50}$, respectively.

\subsection{The disc $\Disc$}
\label{disc}

Table~\ref{disc_symbol} summarised the parameters of the disc. In the classical scenario adopted here, the disc component is the first structure formed in a new galaxy. We use the standard approach as proposed by \cite{Cole_2000, Hatton_2003}, among others.

\begin{table}[t]
  \begin{center}
    \footnotesize{
      \begin{tabular*}{0.5\textwidth}{@{\extracolsep{\fill}}cr}
        \hline
        Symbol & Definition \\
        \hline    
        $\Disc$               & Common symbol for disc properties\\    
        $M_{\Disc}(r)$         & Disc mass enclosed in a radius r [Eq. \ref{disc_profile}]\\
        $M_{\Disc}$            & Total mass of the disc [Eq. \ref{disc_profile}] \\
        $x_{\Disc}$             & Dimensionless radius [Eq. \ref{disc_profile}]\\ 
        $r_{\Disc}$              & Exponential characteristic radius [Eq. \ref{disc_profile}]\\   
        $M_{\Disc,\sfg}$       & Gas mass in the disc [Eq. \ref{disc_object}]\\ 
        $M_{\Disc,\star}$      & Stellar mass in the disc [Eq. \ref{disc_object}]\\ 
        $V_{circ,\Disc}$        & Circular velocity [Eqs. \ref{Vcirc_d},  \ref{gal_circular_velocity}]\\
        $\sigma_{v,\Disc}$  & Average velocity dispersion [Eqs. \ref{E_vdisp},\ref{t_disc}]\\
        $E_{\sigma_v}$        &  Energy: gas velocity dispersion [Eq. \ref{E_vdisp}]\\
        $E_{V}$                 &  Energy: gas rotation [Eq. \ref{E_Vrot}]\\
        $E_{inf}$                & Rotational energy of the infall accreted disc [Eq. \ref{infall_energy}]\\
        $E_{sn,\sigma_v} $   & Turbulent energy injected by sn [Eq. \ref{Esn_sv}]\\
        $f_{disp}$               & \textbf{Energy dissipation factor = 0.05} [Eq. \ref{dEsv}]\\
        $t_{dyn,\Disc}$         & Disc dynamical time [Eq. \ref{t_disc}]\\
        $\sigma_r$           & Average radial velocity dispersion [Eq. \ref{t_disc}]\\
        $Q$                     & Toomre parameter [Eq. \ref{Toomre}]\\
        $\kappa$             & Epicyclic frequency [Eqs. \ref{kappa}, \ref{Toomre}]\\
        $\Sigma_{\Disc,\sfg}$ & Average gas surface density [Eq. \ref{Toomre}]\\
        $r_{\Disc,90}$          & Disc radius that encloses 90\% of the mass [Eq. \ref{Toomre}]\\
        $\Omega$           & Angular velocity [Eq. \ref{kappa}] \\
        $Q_{crit}$               & \textbf{Critical Toomre parameter = 1.0} [Eq. \ref{Toomre}]\\
        \hline
      \end{tabular*}}
  \end{center}  
  \caption{\footnotesize{Disc symbols and their definition. Symbols are listed in order of appearance. Model parameters are in boldfaced text.}}
  \label{disc_symbol}
\end{table}
We assume that the disc component is infinitely thin. Its mass radial distribution is given by
  \begin{equation}
    M_{\Disc}(r) = M_{\Disc}\left[1-exp(-x_{\Disc})\left(1+x_{\Disc}\right)\right]\, ,
    \label{disc_profile}
  \end{equation}
where $x_{\Disc}$ is the dimensionless radius $x=r/r_{\Disc}$. 

As explained previously, we assume that the gas accretion is only supported by the disc. At a given time $n$, the disc is defined by its total mass $M_{\Disc}^n$ and its exponential radius $r_{\Disc}^n$. During a time step $\Delta t = t_{n+1} - t_n$, we assume that the new accreted mass acquires angular momentum and therefore forms a very thin disc structure with a mass $(\dot{M}_{ff} + \dot{M}_{cool})\Delta t$ and an exponential radius deduced from the dark-matter halo virial radius $r_{vir}$ using the well-know spin parameter $\lambda$, $r_{acc} = \dfrac{\lambda}{\sqrt{2}}r_{vir}$ \citep{Blumenthal_1986, Sellwood_2005, Maccio_2008, Cole_2008, Antonuccio-Delogu_2010, Munoz-Cuartas_2011}

The assembly of the pre-existing disc with the accreted mass leads to a secular evolution of the disc exponential radius that follows
\begin{equation}
  r_{\Disc}^2 = \dfrac{M_{\Disc}^{n+1}}{\underbrace{\dfrac{M_{\Disc}^{n} - \overbrace{\dot{M}_{wind}\Delta t}^{ejected~mass}}{r_{\Disc}^n}}_{pre-existing~disc} + \underbrace{\dfrac{(\dot{M}_{ff} + \dot{M}_{cool})\Delta t}{\dfrac{\lambda}{\sqrt{2}}r_{vir}}}_{accretion}}\, .
  \label{rd_evolution}
\end{equation} 

In the model, the disc is defined as an object that contains two different parts: 
\begin{equation}
\Disc = \left\{
  \begin{array}{ll}
   M_{\Disc,\sfg}  & \mbox{: A star-forming (cold) gas component} \\
    & \\
    M_{\Disc,\star} & \mbox{: A stellar population.}
  \end{array}\right.
\label{disc_object}
\end{equation}

\begin{figure}[h]
  \begin{center}
    \includegraphics[scale = 0.18]{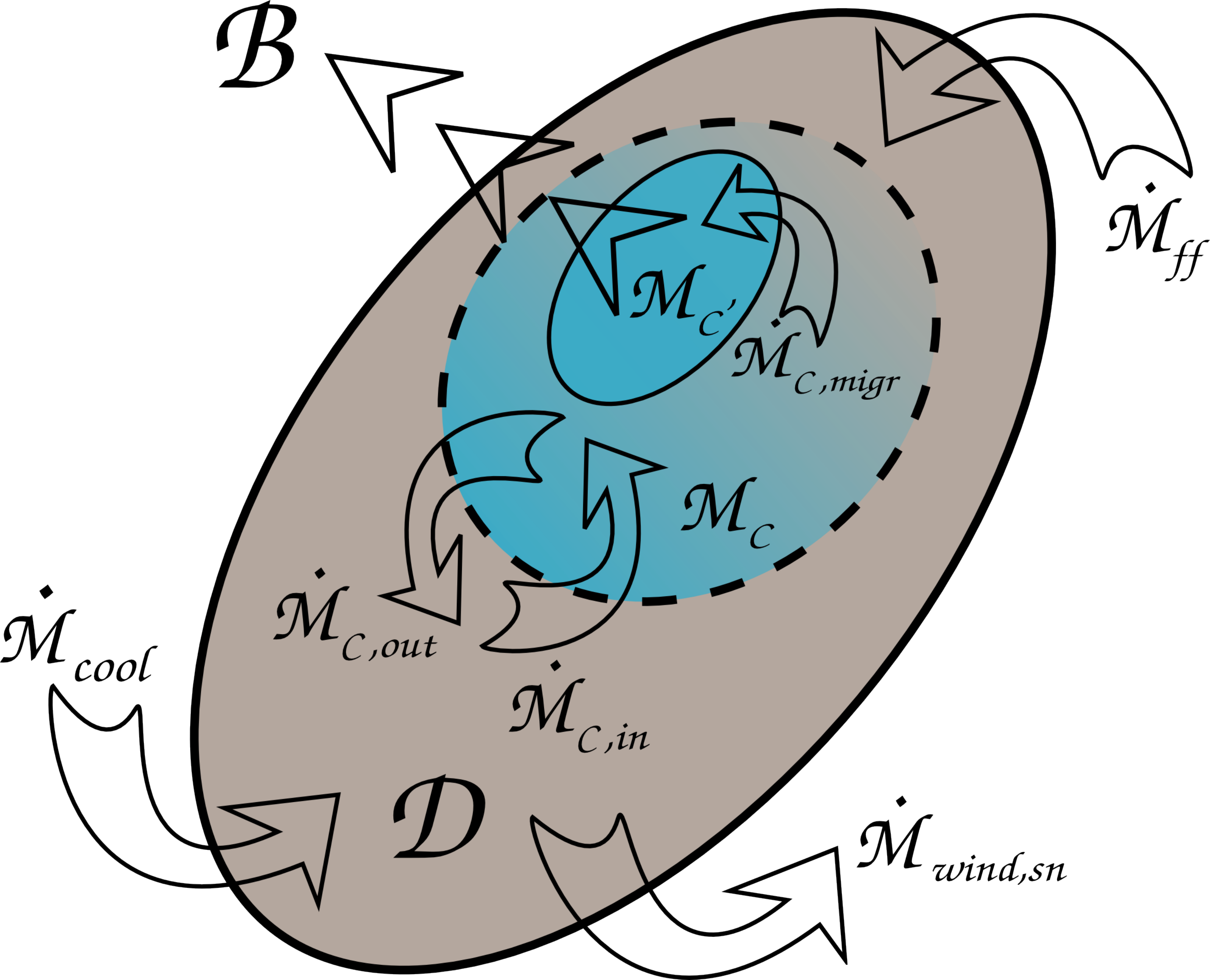}
  \caption{\ftns{\textbf{Exchanges on the disc scale}. The disc is the main component. It is fed by the fresh gas coming from the large scales ($\dot{M}_{ff}$ and $\dot{M}_{cool}$). SNe associated with the stellar population of the disc generate ejecta $\dot{M}_{wind,sn}$. In a disc, dynamics and gravity are in balance. Under some conditions, giant clumps can be formed ($\dot{M}_{\Clumps,in}$) and/or disrupted ($\dot{M}_{\Clumps,out}$). At any given time, a given fraction $M_{\Clumps}$ of the total disc mass is in those clumps. The clumps migrate ($\dot{M}_{\Clumps,migr}$) and feed a pseudo-bulge component ($\Bulge$).}}
  \label{disc_transfer}
  \end{center}
\end{figure}

The radial profile of the disc circular velocity is given by \citep{Freeman_1970}
  \begin{equation}
    V_{circ,\Disc}^2(r) = \dfrac{G M_{\Disc}}{2r_{\Disc}^3}r^2\left[I_0K_0-I_1K_1\right]_{\frac{r}{2r_{\Disc}}} \, ,
    \label{Vcirc_d}
  \end{equation}
where $I_r$ and $K_r$ are modified Bessel functions of rank $r$.

\subsubsection{Disc velocity dispersion and dynamical time}
\label{disc_vel_disp}
In this section, we use a new method based on energy conservation that is used to monitor the average velocity dispersion of the gas in the disc component.  At a given time $n$, we assume that the energy dispersion of the gas $\sigma_{v,\Disc}$ is given by
  \begin{equation}
    E_{\sigma_v}^n = \frac{1}{2}M_{\sfg,\Disc}\sigma_{v,\Disc}^2\, ,
    \label{E_vdisp}
  \end{equation}
where $M_{\sfg,\Disc}$ is the total mass of the gas in the disc and $\sigma_{v,\Disc}$ the average velocity dispersion in the disc component. At time $n$ the kinetic energy contained in the disc rotation is
  \begin{equation}
    E_{V}^n =\dfrac{M_{\Disc}^n}{2r_{\Disc}^2}\int_0^{r_max = 11r_{\Disc}}exp\left(-\frac{r}{r_{\Disc}}\right)V_{circ,\Gal}^2(r)rdr\, ,
    \label{E_Vrot}
  \end{equation}
where $V_{circ,\Gal}(r)$ is the circular velocity of the galaxy (Eq. \ref{gal_circular_velocity}). Here the integral is computed from 0 to $11~r_{\Disc}$, which encloses 99.99\% of the disc mass.

As proposed by \cite{Khochfar_2009} or \cite{Ocvirk_2008}, we assume that the velocity dispersion is generated by gas infall and SNe energy injection. Since the infall mass is structured through a very thin disc in our model, assuming that the dynamics of the gas is given by the total potential well gives the following definition for the kinetic energy transported by gas fuelling
  \begin{equation}
    E_{inf} = \left(\dot{M}_{cool}+\dot{M}_{ff}\right)\Delta t \int_0^{11r_{\Disc}}exp\left(-\frac{r}{r_{\Disc}}\right)V_{circ,\Gal}^2(r)rdr \, ,
    \label{infall_energy}
  \end{equation}
where $\dot{M}_{cool}$ and $\dot{M}_{ff}$ are the cooling rate and the free-fall rate fuelling the galaxy, respectively. 

In addition we take the SNe kinetic energy injection in the gas into account,
  \begin{equation}
    E_{sn,\sigma_v} = (1 - \varepsilon_{\Disc})\eta_{sn}(1-f_{Kin,sn})E_{sn}\dot{M}_{\star,\Disc}\Delta t\, ,
    \label{Esn_sv}
  \end{equation}
where
\begin{itemize}
\item{$(1 - \varepsilon_{\Disc})$ is the fraction of SNe kinetic energy used to feed disc velocity dispersion;}
\item{$\eta_{sn}$ is the number of SNe produced per unit of stellar mass formed ($[\eta_{sn}] = \Msun^{-1}$). This parameter is linked to the initial mass function IMF;}
\item{$f_{Kin,sn}E_{sn}$ is the SNe energy converted in kinetic energy. We use $E_{sn} = 10^{44}$~Joules and $f_{Kin,sn} = 0.3$ \citep{Kahn_1975,Aguirre_2001};}
\item{$\dot{M}_{\star,\Disc}$  is the star formation rate in the disc.}
\end{itemize}

To compute the evolution of the average velocity dispersion, we assume that the variation of the rotational energy between times $n$ and $n+1$,  ($\Delta E_{V} = E_{V}^{n+1} - E_{V}^n$) is supported by gas infall energy ($E_{inf} \ge \Delta E_{V}$) and that the variation of the energy dispersion ($\Delta E_{\sigma_v,} = E_{\sigma_v}^{n+1} - E_{\sigma_v}^{n}$) is then given by 
  \begin{equation}
    \Delta E_{\sigma_v,\Disc} = (1-f_{disp})\left[E_{sn,\sigma_v,\Disc} + (E_{inf}-\Delta E_{V})\right] \, ,
    \label{dEsv}
  \end{equation}
where  $f_{disp}=0.05$ is a dissipation factor.
The mean velocity dispersion in the disc is therefore given by
  \begin{equation}
    \sigma_{v,\Disc} = \sqrt{\dfrac{2(E_{\sigma_v}^n +\Delta E_{\sigma_v})}{M_{\sfg,\Disc}}} \, .
    \label{sigma_v}
  \end{equation}

\begin{figure}[h]
  \begin{center}
    \includegraphics[scale = 0.9]{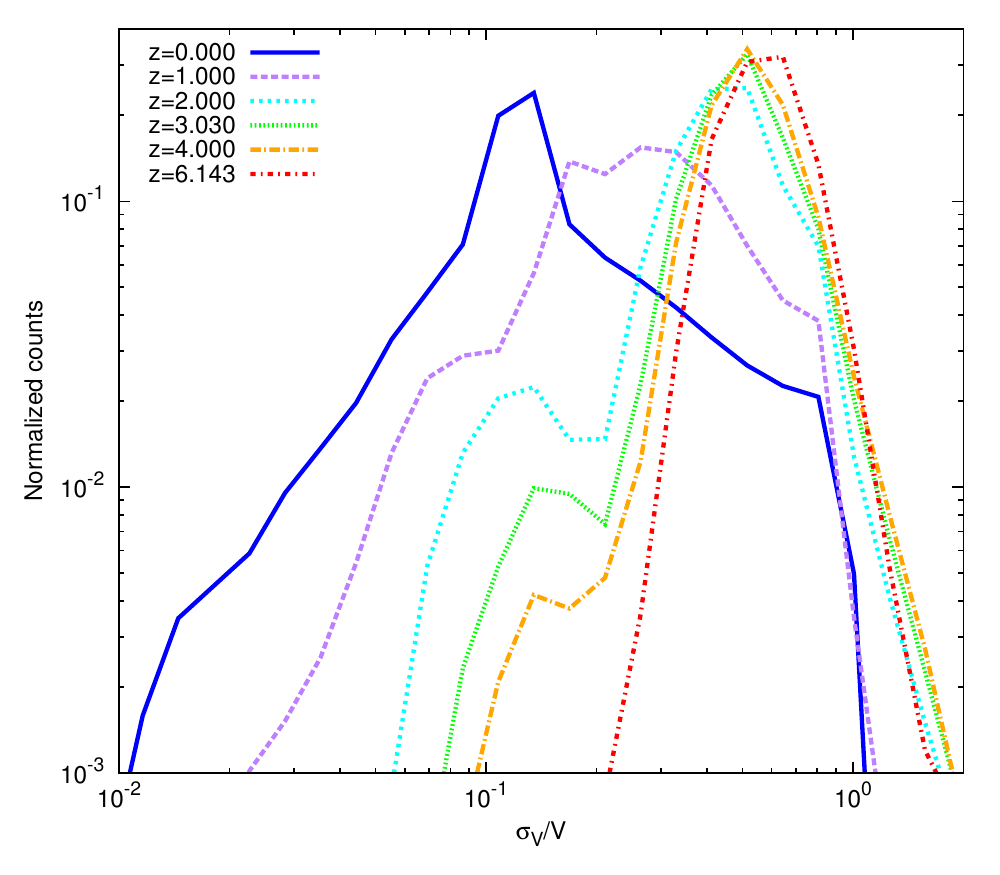}
  \caption{\ftns{Distribution of the $\sigma_{r}/V_{circ,\Gal}(r=r_{\Gal,50})$ ratio for $0<z<6$ (coloured lines). The mean ratio decreases from 0.6 for high-$z$ structures to 0.1 for local structures. The unstable (high sigma disc) population formed at high $z$ disappears progressively and stable disc structures appear.}}
  \label{sVoV}
  \end{center}
\end{figure}

Figure~\ref{sVoV} shows the evolution with redshift of the normalized distribution of $\sigma_{r}/V_{circ,\Gal}(r=r_{\Gal,50})$. We have assumed an isotropic three-dimensional velocity dispersion, $\sigma_r = \sigma_{v,\Disc} /\sqrt{3}$. While high-redshift discs present an average ratio close to $0.6$, local discs have ratios close to $0.1$. The disturbed disc population disappears progressively with time and a more stable population is generated.\\

The disc dynamical time is set to the minimal time between the disc's full rotational time and local velocity dispersion. We use 
  \begin{equation}
    t_{dyn,\Disc} = MIN(2r_{\Gal,50}\sigma_r^{-1}~;~2\pi V_{circ,\Gal}^{-1}(r=r_{\Gal,50}))\, .
    \label{t_disc}
  \end{equation}

\subsection{Disc instability, bulge-like component formation}

For more than one decade, observations \citep[e.g.][]{Cowie_1995,Van_den_Bergh_1996,Elmegreen_2005,Genzel_2008,Bournaud_2008} and hydrodynamic simulations \citep[e.g.][]{Bournaud_2007,Ceverino_2010,Ceverino_2012} have shown that there are of gas-rich, turbulent discs. These kinds of discs are unstable, and undergo gravitational fragmentation that forms giant clumps. Such clumps interact and migrate to the centre of the galaxy where they form a bulge-like component \citep[e.g.][]{Elmegreen_2009,Dekel_2009b}.

Based on these works, we assume in our model that mass overdensities and low velocity dispersion (Eq.~\ref{sigma_v}) lead to the formation and the migration of giant clumps. We present an analytic self-consistent model of disc instabilities here. All parameters used in this section are described in Table~\ref{clumps_symbol}.

\subsubsection{Origin of instabilities}

Following \cite{Toomre_1963,Toomre_1964}, a disc becomes unstable if its gas surface density ($\Sigma_{\Disc,\sfg}$) is high and its circular velocity and/or mean velocity dispersion is low. The disc stability is controlled by the following parameter
  \begin{equation}
    Q = \dfrac{\sigma_r\kappa}{\pi G \Sigma_{\Disc,\sfg}}=\dfrac{r_{\Disc,90}^2 \sigma_r\kappa }{0.9 G M_{\Disc,\sfg}}\, ,
    \label{Toomre}
  \end{equation}
where
 \begin{equation}
    \kappa = \dfrac{1}{M_{\Disc,\sfg}}\int_0^{11r_{\Disc}}\left[4\Omega^2\left(1+\dfrac{r}{2\Omega}\dfrac{d\Omega}{dr}\right)\right]^{1/2}\Sigma_{\Disc,\sfg}(r)rdr\, ,
    \label{kappa}
  \end{equation}
is the mass-weighted epicyclic frequency with $\Omega$ the angular velocity and $\Sigma_{\Disc,\sfg}(r)$ the gas mass surface density.

In Eq.~\ref{Toomre}, $\sigma_r$ is the mean radial velocity dispersion. As for the disc dynamical time computation (Eq.~\ref{t_disc}), we assume an isotropic three-dimensional velocity dispersion, $\sigma_r = \sigma_{v,\Disc} /\sqrt{3}$. The gaseous component of the disc $M_{\Disc,\sfg}$ is considered as stable or marginally stable if $Q > Q_{crit} = 1.0 $. We use this stability criterium to evaluate the mass that will form clumps by gravitational instabilities. 

\subsubsection{The clump component, $\Clumps$}
\label{clumps}

We assume that disc instabilities are linked to the formation of giant clumps. These giant clumps are formed into the disc and migrate to the centre of the disc to form a pseudo-bulge component (see Fig.~\ref{disc_transfer}).  As previously, all parameters used in the description of disc instabilities linked to clumps are listed in the dedicated Table~\ref{clumps_symbol}.\\

\begin{table}[t]
  \begin{center}
    \footnotesize{
      \begin{tabular*}{0.5\textwidth}{@{\extracolsep{\fill}}cr}
        \hline
        Symbol & Definition \\
        \hline    
        $\Clumps$               & Common symbol for clumps properties\\  
        $M_{\Clumps}$             & Total mass in clumps\\
        $M_{\Clumps,\sfg} $      & Gas mass in clumps\\
        $M_{\Clumps,\star}$      & Stellar mass in clumps\\
        $Q'$                          & \textit{Reduced}-Toomre parameter [Eq. \ref{unstable_mass}]\\
        $M_{\Disc,\sfg,}^s$       & Stable mass of the disc [Eq. \ref{unstable_mass}]\\
        $M_{\Disc,\sfg,}^u$       & Unstable mass of the disc [Eqs. \ref{unstable_mass}, \ref{unstable_mass_2}]\\
        $\dot{M}_{\Clumps,in}$ & Clump formation rate [Eq. \ref{clumps_in}, Fig. \ref{disc_transfer}]\\
        $t_{dyn,\Clumps}$         & Dynamical time [Eq. \ref{t_clumps}]\\ 
        $r_{\Clumps}$              & Typical size of clumps [Eqs. \ref{r_clumps}, \ref{t_clumps}]\\
        $M_{\Clumps,ind}$        & Mass of clumps [Eqs. \ref{m_clumps}, \ref{t_clumps}]\\
        $\delta$                   & Disc mass fraction [Eq. \ref{delta}]\\   
        $\dot{M}_{\Clumps,migr}$ & Clumps migration rate [Eq. \ref{clumps_migr}, Fig. \ref{disc_transfer}]\\  
        $M_{\Clumps'}$             & Mass of clumps in transfer [Eq. \ref{clumps_migr}]\\
        $t_{migr,\Clumps}$         & Clumps migration time scale  [Eq. \ref{t_clumps_trans}]\\
        $\dot{M}_{\Clumps,out}$ & Clumps disruption rate [Eq. \ref{clumps_out}, Fig. \ref{disc_transfer}]\\ 
        $t_{disr,\Clumps}$          & Clumps disruption time scale [Eqs. \ref{t_clumps_disr}, \ref{clumps_out}]\\
        \hline
      \end{tabular*}}
  \end{center}  
  \caption{\footnotesize{Clumps symbols and their definition. Symbols are listed in order of appearance.}}
  \label{clumps_symbol}
\end{table}

In our model, the clumpy phase ($\Clumps$) is not considered as a new separated component but is considered as an inherent part of the disc. The amount of gas in clumps is therefore given by a fraction of the disc mass. As for the disc component, clumps contain a cold star-forming gas $M_{\Clumps,\sfg}$ and a stellar population $M_{\Clumps,\star}$.
We monitor the stellar and the gaseous mass in clumps by taking their formation (Sect.~\ref{clumps_formation}), transfer (Sect. ~\ref{clumps_transfert}), and disruption (Sect.~\ref{clumps_disruption}) into account. Star formation in the disc is divided into a clumpy and a homogeneous component. 

\subsubsection{Clump formation}
\label{clumps_formation}

\begin{figure*}[t]
  \begin{center}
    \includegraphics[scale = 0.9]{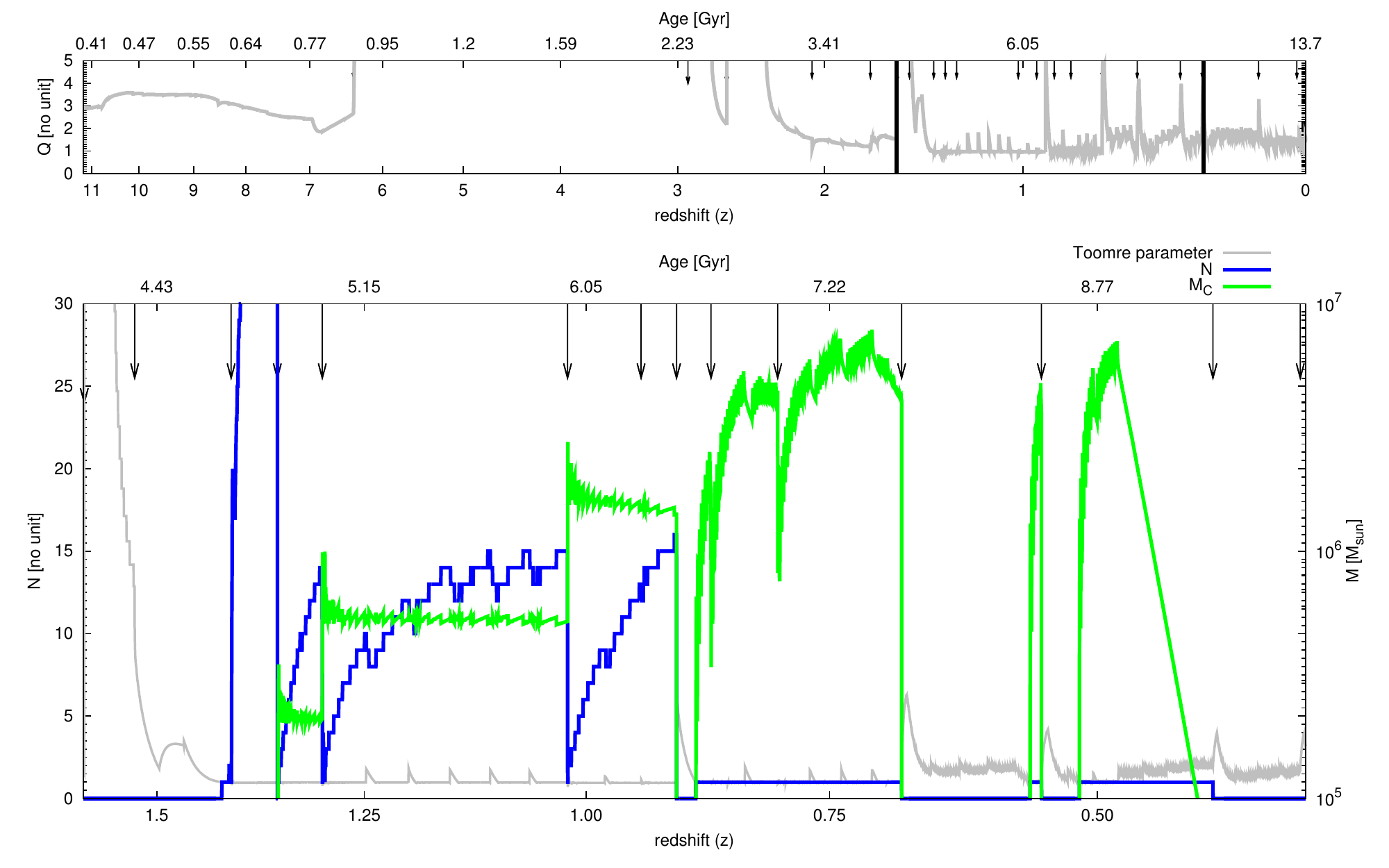}
  \caption{\ftns{Evolution of the disc-stability criterion (Toomre parameter $Q$) and clump properties. In the upper panel is shown the evolution of the Toomre parameter with $z$ (Eq.~\ref{Toomre}, grey line). On this timeline, vertical arrows indicate a merger event (with short and long arrows for minor and major mergers, respectively).  For $11<z<1.5$, the Toomre parameter is larger than the critical value $Q > Q_{crit} =1$, and therefore the disc is stable, but for $1.5<z<0.25$ (highlighted by the two vertical black lines), the disc becomes unstable. During the unstable transition, bounded by two merger events, many clumps are formed and migrate. The lower panel shows a zoom of this unstable period. The blue and green lines show the number of clumps (left y-axis) and the average individual mass of these clumps (right y-axis), respectively. After a merger event, disc instabilities lead to the formation of a large number of clumps (42) with small individual masses ($M_{\Clumps,ind} < 10^5\Msun$). Then, during a series of merger events, clumps are destroyed and formed again, but with higher individual masses. Between two merger events, the number of clumps increases while the individual mass $M_{\Clumps,iind}$ remains stable. A decrease in one or two clumps is linked to the migration process. It is interesting to note that the migration process creates a temporary stable disc ($Q > Q_{crit} =1$). In the second part of the unstable period, only one clump is formed but with a much higher mass. Finally, this single clump is disrupted and the disc becomes stable again.}}
  \label{clumps_time_line}
  \end{center}
\end{figure*}

We consider that disc instabilities (and therefore clumps formation) are only due to an excess of mass. When the homogeneous (non-clumpy) gas fraction of the disc ($M_{\Disc,\sfg} - M_{\Clumps,\sfg})$ is unstable ($Q < Q_{crit}$), we divide the gaseous mass of the disc in two parts, a stable $M_{\Disc,\sfg}^s$ and an unstable part $M_{\Disc,\sfg}^{u}$. We have
  \begin{equation}
    \begin{split}
      \dfrac{1}{Q'}  & = \dfrac{0.9 G (M_{\Disc,\sfg}^s + M_{\Disc,\sfg}^{u})}{r_{\Disc,90}^2\sigma_r\kappa}\\
      & = \dfrac{1}{Q_{crit}} + \dfrac{0.9 G M_{\Disc,\sfg}^{u}}{r_{\Disc,90}^2\sigma_r\kappa}\, ,
    \end{split}
    \label{unstable_mass}
  \end{equation}
with $Q'$ is a reduced-Toomre parameter that takes into account only the fraction of the disc homogeneous gas mass (the gaseous mass of the clumps being subtracted in the gas surface density computation). Within this framework, at a given time, a new amount of unstable gas feeds the clumpy part, $\Clumps$ (see Fig. \ref{disc_transfer}). The unstable gas mass is given by 
  \begin{equation}
      M_{\Disc,\sfg}^{u} = \dfrac{r_{\Disc,90}^2\sigma_r\kappa}{0.9G}\left(\dfrac{1}{Q'} - \dfrac{1}{Q_{crit}}\right) \, .
\label{unstable_mass_2}
  \end{equation}
Knowing the new unstable mass in the disc, we can compute the instantaneous formation rate of the clump component ($\dot{M}_{\Clumps,in}$, see Fig.~\ref{disc_transfer}) following
  \begin{equation}
      \dot{M}_{\Clumps,in} =  \dfrac{M_{\Disc,\sfg}^{u}}{t_{dyn,\Disc}}\, .
      \label{clumps_in}
  \end{equation}
  Clumps can form stars. The dynamical time for star formation in clumps
  \begin{equation}
      t_{dyn,\Clumps} = \sqrt{\dfrac{r_{\Clumps}^3}{Gf_{\sfg}M_{\Clumps,ind}}}~~\mbox{with}~~f_{\sfg}=\dfrac{M_{\Clumps,\sfg}}{M_{\Clumps,\sfg}+M_{\Clumps,\star}}\, ,
      \label{t_clumps}
  \end{equation}
where $r_{\Clumps}$ and $M_{\Clumps,ind}$ are the characteristic radius and mass of one clump, which we extract from \citealt{Dekel_2009b} (their Eqs. 8 and 9),
  \begin{equation}
    r_{\Clumps} = \dfrac{\pi}{6}\delta r_{\Disc,90}\, ,
    \label{r_clumps}
  \end{equation}
and
  \begin{equation}
    M_{\Clumps,ind} = \dfrac{\pi^2}{36}\delta M_{\Disc}\, .
    \label{m_clumps}
  \end{equation}
In these relations, $\delta$ is the disc mass fraction defined as 
  \begin{equation}
    \delta = \dfrac{M_{\Disc}(<r_{\Disc,90})}{M_{vir}(<r_{\Disc,90}) + M_{\Disc}(<r_{\Disc,90}) + M_{\Bulge}(<r_{\Disc,90})}\, ,
    \label{delta}
  \end{equation}
where $M_{vir}(<r_{\Disc,90})$, $M_{\Disc}(<r_{\Disc,90})$, and $M_{\Bulge}(<r_{\Disc,90})$ are the mass, enclosed in $r_{\Disc,90}$ (radius for 90\% of the disc mass), of the dark-matter halo, disc, and bulge component, respectively. As explained in the following section, the individual mass $M_{\Clumps,ind}$ is used as a threshold mass that allows us to compute the clump transfer to the pseudo-bulge component.  

\subsubsection{Clumps transfer}
\label{clumps_transfert}

To take the migrations into account, we define $M_{\Clumps '}$ as the mass of clumps that is transferred to the pseudo-bulge. The transfer rate, $\dot{M}_{\Clumps,migr}$, writes as
  \begin{equation}
      \dot{M}_{\Clumps,migr} =  \dfrac{M_{\Clumps} -M_{\Clumps'}}{t_{migr,\Clumps}}\, ,
      \label{clumps_migr}
  \end{equation}
where $M_{\Clumps}$ is the total mass in clumps, and $M_{\Clumps '}$ the mass of the clumps that is already in a transfer process. We add $t_{migr,\Clumps}$ as the typical clump transfer time scale. Like other clump properties, we compute this characteristic time following \citealt{Dekel_2009b} (their Eq. 19)
  \begin{equation}
    t_{migr,\Clumps} = 2.1\left(\dfrac{Q}{\delta}\right)^2t_{dyn,\Disc}\, .
    \label{t_clumps_trans}
  \end{equation}
When the transferred mass reservoir $M_{\Clumps '}$ becomes larger than the reference mass $M_{\Clumps,ind}$, the stellar and gas mass contained in this reservoir ($\Clumps '$) is subtracted from the disc and added to the pseudo-bulge component.  The transfer is considered as a micro-merger event. When the \textit{micro}-merger is finished, we reset the transferred mass, and recompute the characteristic individual mass of clumps $M_{\Clumps,ind}$. This new characteristic mass will be the new threshold mass for the next clump migration process.

\subsubsection{Clumps disruption}
\label{clumps_disruption}

At a given time, a disc $\Disc$ may have giant clumps ($\Clumps$). If, at the same time $Q > Q_{crit}$ then we compute the disruption rate 
  \begin{equation}
      \dot{M}_{\Clumps,out} = \dfrac{M_{\Clumps}}{t_{disr,\Clumps}}\, ,
      \label{clumps_out}
  \end{equation}
where $t_{\Clumps}^{disr}$ is the disruption time computed as proposed by \citealt{Dekel_2009b} (their Eq. 14):
  \begin{equation}
    t_{disr,\Clumps} = 1.4\dfrac{t_{dyn,\Disc}}{Q}\, .
    \label{t_clumps_disr}
  \end{equation}

In Fig.~\ref{clumps_time_line} we show the evolution of the Toomre stability parameter for one disc chosen arbitrarily for $z<11$. We also show the evolution of the number and individual mass of clumps for the unstable period. We see that for $0.8<z<1.5$ the disc instability generates numerous clumps that are destroyed after each merger event and systematically formed again but with higher individual mass. During this period of instabilities, the decrease in the number of clumps has to be linked to the migration process. After $z=0.8$ only one giant clumps can be formed, but with a higher mass. For $z<0.5$ the disc is stable ($Q > Q_{crit} = 1$), the clumps disruption process is therefore turned on and the clumps mass decrease. After the last merger event, the disc does cannot form any clumps owing to the lack of gas.

\subsubsection{Summary of mass transfers}
The evolution of the mass transfers of the different components of the disc is governed by the following equations:
\begin{equation}
\Disc = \left\{
  \begin{array}{ll}
   \dot{M}_{\Disc,\sfg}  = & + \dot{M}_{ff} + \dot{M}_{cool}  + \dot{M}_{wind,\star} \\
    & - \dot{M}_{\star,\Disc} - \dot{M}_{wind,sn,\Disc} - \left[M_{\Clumps ',\sfg}\right]_{M_{\Clumps'} > M_{\Clumps,ind}} \\
    &\\
    \dot{M}_{\Disc,\star} = & + \dot{M}_{\star,\Disc} - \dot{M}_{wind,\star}  - \left[M_{\Clumps ',\star}\right]_{M_{\Clumps'} > M_{\Clumps,ind}}
  \end{array}\right.\, .
\end{equation}

\subsection{Pseudo-bulge or spheroidal galaxy $\Bulge$}
\label{bulge}

\begin{table}[h]
  \begin{center}
    \footnotesize{
      \begin{tabular*}{0.5\textwidth}{@{\extracolsep{\fill}}cr}
        \hline
        Symbol & Definition \\
        \hline    
        $\Bulge$                & Common symbol for bulge properties\\  
        $M_{\Bulge,\sfg} $      & Gas mass in the bulge\\
        $M_{\Bulge,\star}$      & Stellar mass in the bulge\\
        $M_{\Bulge}(r)$          & Bulge mass enclosed in the radius r [Eq. \ref{bulge_profile}]\\
        $M_{\Bulge}$             & Total bulge mass [Eq. \ref{bulge_profile}]\\
        $r_{\Bulge}$              & Bulge characteristic radius [Eqs. \ref{bulge_profile}, \ref{merger_bulge_radius}]\\
        $V_{circ,\Bulge}$        & Circular velocity [Eqs. \ref{Vcirc_b}, \ref{gal_circular_velocity}]\\
        $V_{esc,\Bulge}$        & Escape velocity [Eq. \ref{Vesc_b}]\\
        $t_{dyn,\Bulge} $       & Dynamical time [Eq. \ref{t_bulge}]\\ 
        $\sigma_{v,\Bulge}$  & Velocity dispersion [Eq. \ref{t_bulge}]\\ 
        $r_{\Bulge,50}$         & Half-mass bulge radius [Eqs. \ref{t_bulge}, \ref{merger_bulge_half_mass_radius}]\\
        \hline
      \end{tabular*}}
  \end{center}  
  \caption{\footnotesize{Bulge symbols and their definition. Symbols are listed in order of appearance.}}
  \label{bulge_symbol}
\end{table}

\begin{figure}[h]
  \begin{center}
    \includegraphics[scale = 0.22]{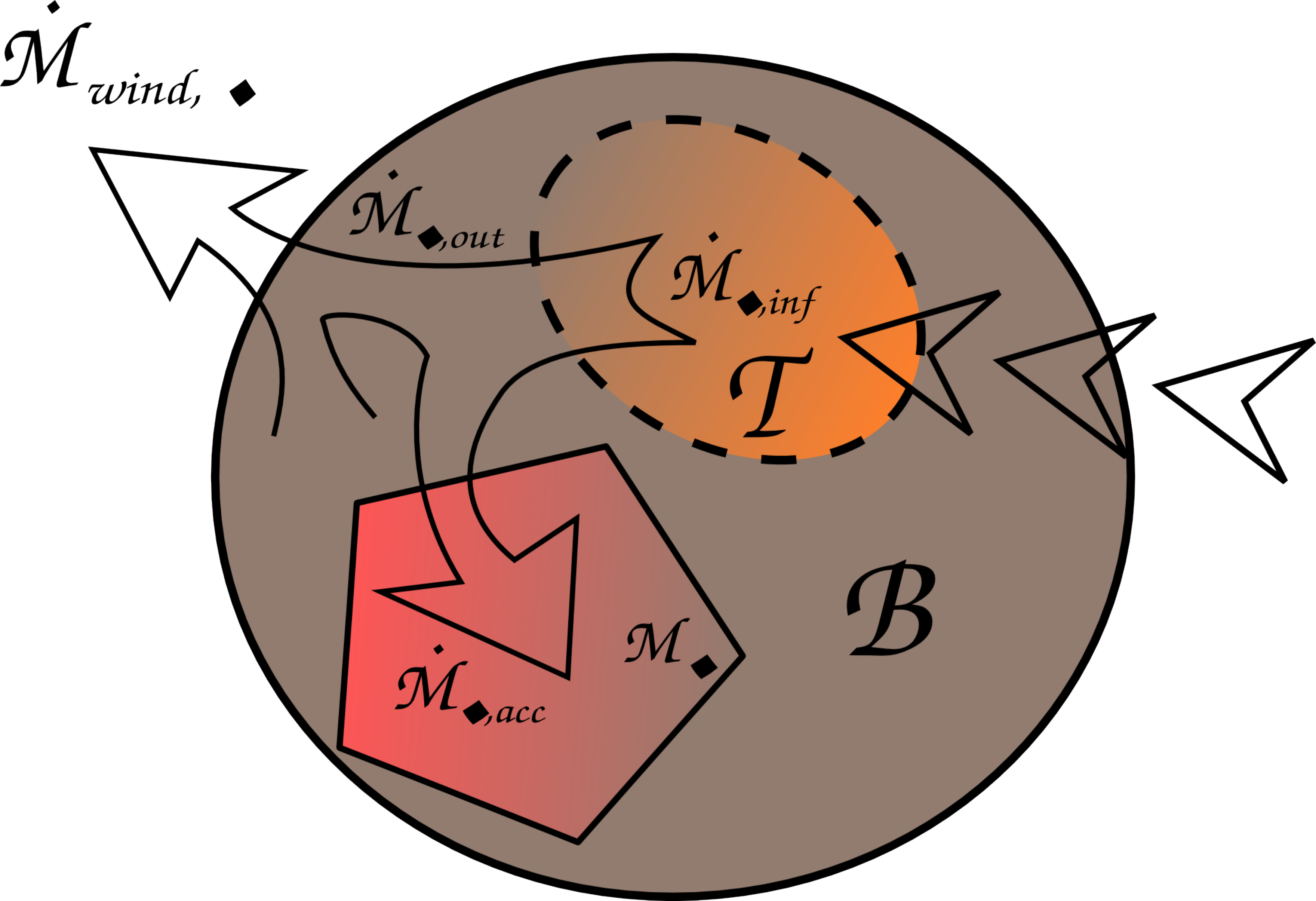}
  \caption{\ftns{\textbf{Exchanges on the bulge scale}. The bulge is considered as a passive component not fed by fresh accreted gas. A bulge ($\Bulge$) is formed during a merger or by clumps migrations from the disc ($\Disc$). An SMBH can evolve in the bulge. It is fed by a gas reservoir (the torus, $\Torus$) with an accretion rate $\dot{M}_{acc}$. The torus is fed during a merger or by the clump migration process. The activity of the SMBH generates ejecta ($\dot{M}_{\BH,out}$).}}
  \label{bulge_agn_transfer}
  \end{center}
\end{figure}

In our model, we assume the same properties for a post-major merger galaxy (with spheroidal morphology) and for a pseudo-bulge formed by giant-clumps migration. The mass transfers of these components $\Bulge$ are illustrated in Fig.~\ref{bulge_agn_transfer}, and the parameters are listed in Table~\ref{bulge_symbol}.\\

As for the disc, we treat the (pseudo-)bulge as a numerical object composed of a cold gas reservoir ($M_{\Bulge,\sfg}$) and a stellar population ($M_{\Bulge,\star}$). In a pseudo bulge the cold gas reservoir is fed by clump migration. This gas can be used to form new stars. In a spheroidal galaxy, formed during a merger event, the gas reservoir is only fed by the stellar cycle (see Fig. \ref{stellar_cycle}). In this context, the amount of gas is very low and therefore no star formation is possible. At each merger, the gas contained in the bulge component is added to the disc component of the remnant galaxy.\\ 

The mass distribution of the spheroidal galaxy or the pseudo-bulge is described by a \cite{Hernquist_1990} model,
  \begin{equation}
   M_{\Bulge}(r) = M_{\Bulge} \dfrac{r^2}{(r+r_{\Bulge})^2}\, ,
    \label{bulge_profile}
  \end{equation}
where $r_{\Bulge}$ is the characteristic radius of the bulge. 

\subsubsection{Bulge circular and escape velocity}

The circular velocity of the bulge or \textit{pseudo}-bulge is given by \citep{Hernquist_1990} 
  \begin{equation}
    V_{circ,\Bulge}^2(r) = \dfrac{G M_{\Bulge,\BH}(r) r}{(r_{\Bulge}+r)^2}\, ,
    \label{Vcirc_b}
  \end{equation}
with
  \begin{equation}
    M_{\Bulge,\BH}(r) = M_{\Bulge}\dfrac{r^2}{(r+r_{\Bulge})^2} + M_{\BH}\, ,
    \label{Vesc_b}
  \end{equation}
where $M_{\Bulge}$ is the total mass (gas + stars) of the bulge, $M_{\BH}$ the mass of the SMBH, and $r_{\Bulge}$ the characteristic radius. We give also here the escape velocity of the bulge,
  \begin{equation}
    V_{esc,\Bulge}^2(r) = \dfrac{2G M_{\Bulge,\BH}}{(r_{\Bulge}+r)}\, .
    \label{Vesc_b}
  \end{equation}

\subsubsection{Bulge dynamical time and velocity dispersion}
As commonly used in SAMs, we define the dynamical time of the bulge using its velocity dispersion ($\sigma_{v,\Bulge}$) instead of its circular velocity, 
  \begin{equation}
   t_{dyn,\Bulge} = \dfrac{2r_{\Bulge,50}}{\sigma_{v,\Bulge}}\, ,
    \label{t_bulge}
  \end{equation}
where $r_{\Bulge,50}$ is the radius that enclosed 50\% of the bulge mass, and where we used the definition of the velocity dispersion proposed by \citealt{Hernquist_1990} (their Eq. 10), computed at the half-mass radius. \\

\subsubsection{Summary of the mass transfers}
The physical processes acting in the bulges are linked to clump migration and gas ejection ($\dot{M}_{wind,agn}$ due to SMBH activity). As in the disc, the gas in the bulge is converted into stars and these stars feed the gas reservoir ($\dot{M}_{wind,\star}$, Sect.~\ref{stellar_component} and Fig.~\ref{stellar_cycle}). We summarize the mass transfer rates with the following equations: 
\begin{equation}
\Bulge = \left\{
  \begin{array}{ll}
   \dot{M}_{\Bulge,\sfg}  = & \left[M_{\Clumps ',\sfg}\right]_{M_{\Clumps '} > M_{\Clumps,ind}} + \dot{M}_{wind,\star}\\
   & - \dot{M}_{\star,\Bulge} - \dot{M}_{wind,sn,\Bulge} - \dot{M}_{wind,\BH}\\
    &\\
    \dot{M}_{\Bulge,\star} = & \left[M_{\Clumps ',\star}\right]_{M_{\Clumps '} > M_{\Clumps,ind}} + \dot{M}_{\star,\Bulge} - \dot{M}_{wind,\star}.
  \end{array}\right.
\end{equation}

\section{The SMBH component}
\label{agn_component}

\begin{table}[t]
  \begin{center}
    \footnotesize{
      \begin{tabular*}{0.5\textwidth}{@{\extracolsep{\fill}}cr}
        \hline
        Symbol & Definition \\
        \hline    
        $M_{\Bulge,\BH}$                 & \textbf{Bulge min mass to host a SMBH = $10^6~\Msun$} \\
        $M_{\BH,min}$                     & \textbf{SMBH min initial mass  = $10^3~\Msun$} \\
        $\dot{M}_{BH,inf} $             & SMBH total infall rate  [Eqs. \ref{agn_mass_conservation}, \ref{agn_tot_ejecta_rate}, \ref{agn_inf}, Fig. \ref{bulge_agn_transfer}]\\ 
        $\dot{M}_{Bondi}$              & Bondi accretion rate  [Eqs. \ref{agn_inf}, \ref{Bondi}]\\
        $M_{\BH}$                         & SMBH mass [Eqs. \ref{Bondi}, \ref{Edd_acc_rate}]\\
        $\dot{M}_{Edd}$                & Eddington limited accretion rate [Eqs. \ref{agn_inf}, \ref{Edd_acc_rate}]\\
        $\dot{M}_{\Torus}$             & Accretion rate linked to the torus [Eqs. \ref{agn_inf_torus}, \ref{agn_inf}]\\ 
        $M_{\Torus}$                      & Gas mass in the torus [Eq. \ref{agn_inf_torus}]\\  
        $\delta_{\Torus} = 10^{-4}$ & \textbf{Gas fraction accreted in a merger = $10^{-4}$}\\
        $L_{\BH,bol} $                    & agn bolometric luminosity [Eq. \ref{agn_Lum}]\\
        \hline
      \end{tabular*}}
  \end{center}  
  \caption{\footnotesize{SMBH symbols and their definition. Symbols are listed in order of appearance. Model parameters are in boldfaced text.}}
  \label{SMBH_symbol}
\end{table}

An SMBH can be formed and evolve in the centre of the galaxy. This section describes the hypothesis behind the formation of an SMBH and the accretion mechanisms that govern its evolution. The parameters used in this section are listed in a dedicated table (Table~\ref{SMBH_symbol}).\\

We decided to form the SMBH by converting a fraction of the bulge mass into an SMBH when the bulge mass ($M_{\Bulge}$) becomes greater than a threshold $M_{\Bulge,\BH} = 10^6 \Msun$. We assume that the formation process of an SMBH is only possible during a merger. Indeed, we assume that the SMBH formation needs powerful processes. Therefore, after a merger, if the total mass of the bulge is higher than $M_{\Bulge,\BH}$ an SMBH with a mass $M_{\BH,min}=10^3M_{sun}$ is created. We subtract the equivalent mass in their respective proportions from the gas phase and from the stellar population. As can be expected, the hypothesis of a constant seed mass as a threshold to create an SMBH generates a scatter in the $M_{\BH}$ versus $M_{\Bulge}$ relation. This scatter is similar to what is obtained by \cite{Croton_2006}.\\

The SMBH evolves in the centre of the bulge and grows by accretion processes. The total accretion rate is given by
  \begin{equation}
    \dot{M}_{\BH,inf} = MIN\left(MAX\left(\dot{M}_{\Torus},\dot{M}_{Bondi}\right),\dot{M}_{Edd}\right)\, ,
    \label{agn_inf}
  \end{equation}
where $\dot{M}_{\Torus}$ is an accretion rate linked to the evolution of a gaseous torus formed around the black hole and defined as
  \begin{equation}
    \dot{M}_{\Torus}=\dfrac{M_{\Torus}}{2t_{dyn,\Bulge}}\,.
    \label{agn_inf_torus}
  \end{equation}
In this last equation, $M_{\Torus}$ is the total mass of gas in the torus and $t_{dyn,\Bulge}$ the dynamical time of the bulge. During the galaxy evolution processes the torus $M_{\Torus}$ is fed by
\begin{itemize}
  \item{the gas coming from clumps migration ($M_{\Clumps ',\sfg}$),}
  \item{a fraction of the bulge gas ($\delta_{\Torus} = 10^{-4}$)\footnote{This value has been chosen to reproduce the trend of the $M_{\BH}$ vs. $M_{\Bulge}$ relation}, instantaneously added during a merger event.}
\end{itemize}
In Eq.~\ref{agn_inf}, $\dot{M}_{Bondi}$ is the classical \cite{Bondi_1952} accretion rate. We use the same calculation as proposed by \cite{Somerville_2008},
  \begin{equation}
    \dot{M}_{Bondi} = \dfrac{3\pi G \mu m_p}{4}\dfrac{k_b T_{\BH}}{\Lambda(T_{\BH},Z)}M_{\BH}\,,
\label{Bondi}
  \end{equation}
where $T_{\BH}\simeq 10^7~K$ is the mean temperature in the accretion torus, $\Lambda$ the cooling function, and $M_{\BH}$ the SMBH mass. 
At any time, the accretion onto the SMBH is possible only if the torus contains gas ($M_{\Torus} > 0$) and only in the Eddington accretion rate limit, 
  \begin{equation}
    \dot{M}_{Edd} = \dfrac{4\pi G m_p}{\left[1-\left(\frac{1}{1+\eta_{\BH}}\right)\left(1+f_{Kin,\BH}\right)\right]\sigma_Tc}M_{\BH}\,.
\label{Edd_acc_rate}
  \end{equation}

We recall that the complete description of ejecta processes driven by the SMBH is given in Sects.~\ref{agn_ejecta} and \ref{time-everage_thermal_energy}.

\begin{figure}[h]
  \begin{center}
    \includegraphics[scale = 0.9]{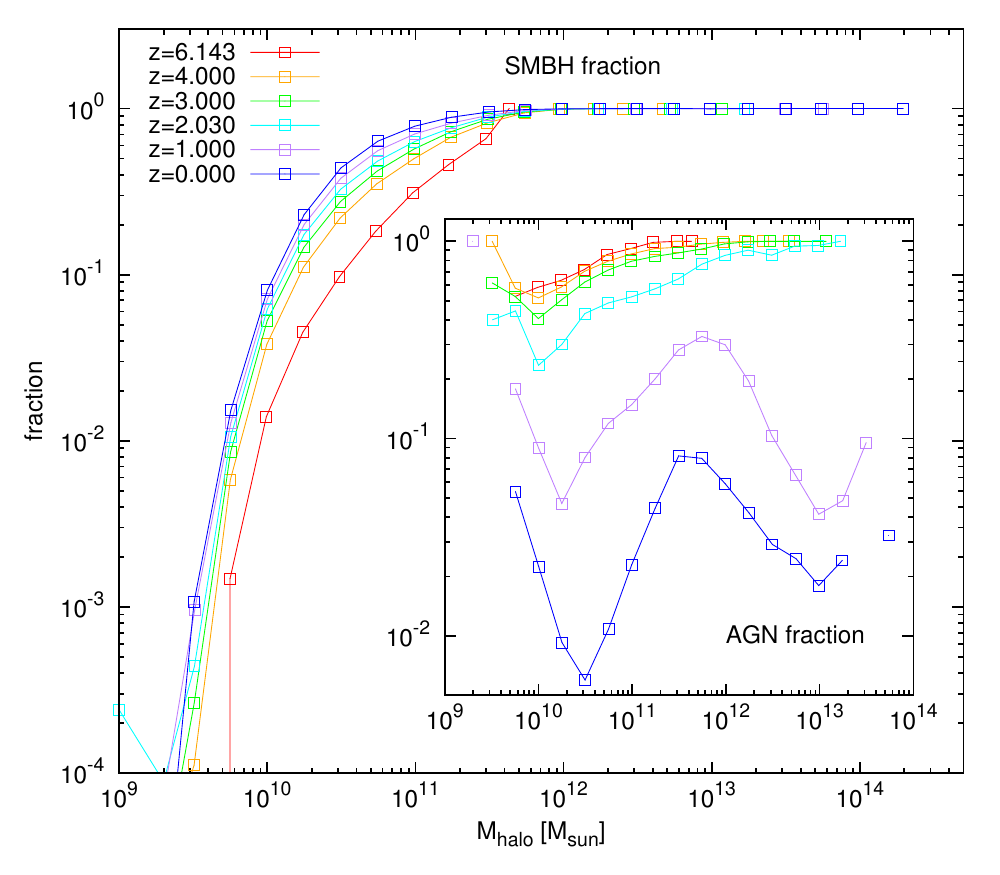}
  \caption{\tiny{Fraction of structures hosting an SMBH in the centre of their galaxy as a function of the dark-matter halo mass for a set of redshift (coloured lines). The fraction increases with the dark-matter halo mass to reach a value of 1 for haloes more massive than $\simeq 4\times 10^{11}~\Msun$. The insert panel shows the fraction of active galaxy nuclei as a function of the dark-matter halo mass. An SMBH is considered to be active if its bolometric luminosity is higher than 5\% of its Eddington luminosity.}}
  \label{BH_AGN_farction}
  \end{center}
\end{figure}

The instantaneous bolometric luminosity produced by the SMBH is deduced from Eq.~\ref{Qagn} and is set to
  \begin{equation}
    L_{\BH,bol} = \left(1-f_{Th,\BH}\right)\dot{Q}_{\BH}\,.
    \label{agn_Lum}
  \end{equation}

We show in Fig.~\ref{BH_AGN_farction} the evolution of the fraction of dark-matter haloes that hosts an SMBH, as a function of the halo mass. The fraction increases with $M_{h}$. In the model presented here, all structures with dark-matter halo mass larger than $\simeq 4\times 10^{11}\Msun$ host an SMBH, regardless of the redshift. 

We also show the fraction of active SMBH in Fig.~\ref{BH_AGN_farction}. An SMBH is considered as active if $L_{\BH,bol} > 0.05 L_{Edd}$. For redshift $z>2$ this fraction is large then $>40\%$ but it decreases for structures evolving at lower redshift (z$\le$1). This decrease is mainly visible at the highest and lowest mass. Haloes with intermediate mass ($10^{11}<M_{vir}<10^{12}~\Msun$) host most of the active SMBH. 

\section{The stellar component}
\label{stellar_component}

In the disc, bulge, and clumps, the gas ($M_{\sfg,\Disc}$, $M_{\sfg,\Bulge}$, $M_{\sfg,\Clumps}$, respectively) is converted into stars ($M_{\star,\Disc}$, $M_{\star,\Bulge}$, $M_{\star,\Clumps}$, respectively). We adopt the same definition for the star formation rate in all these components,
  \begin{equation}
    \dot{M}_{\star,i} = \varepsilon_{\star,i}\dfrac{M_{\sfg,i}}{t_{dyn,i}}\,.
    \label{sfr}
  \end{equation}

In this formulation, $\varepsilon_{\star,i}$ is a free efficiency parameter, depending on the component. We adopt $\varepsilon_{\star,i} = 0.02$ in discs and bulge and $0.2$ in clumps. These values allow to be reproduced the \cite{Kennicutt_1998} law on both galactic and clump scales. Here $M_{\sfg,i}$ is the total mass of star-forming, gas and $t_{dyn,i}$ is the dynamical time (Eqs.~\ref{t_disc}, \ref{t_bulge}, and \ref{t_clumps} for the disc, bulge and, clumps, respectively). 

\begin{figure}[h]
  \begin{center}
    \includegraphics[scale = 0.40]{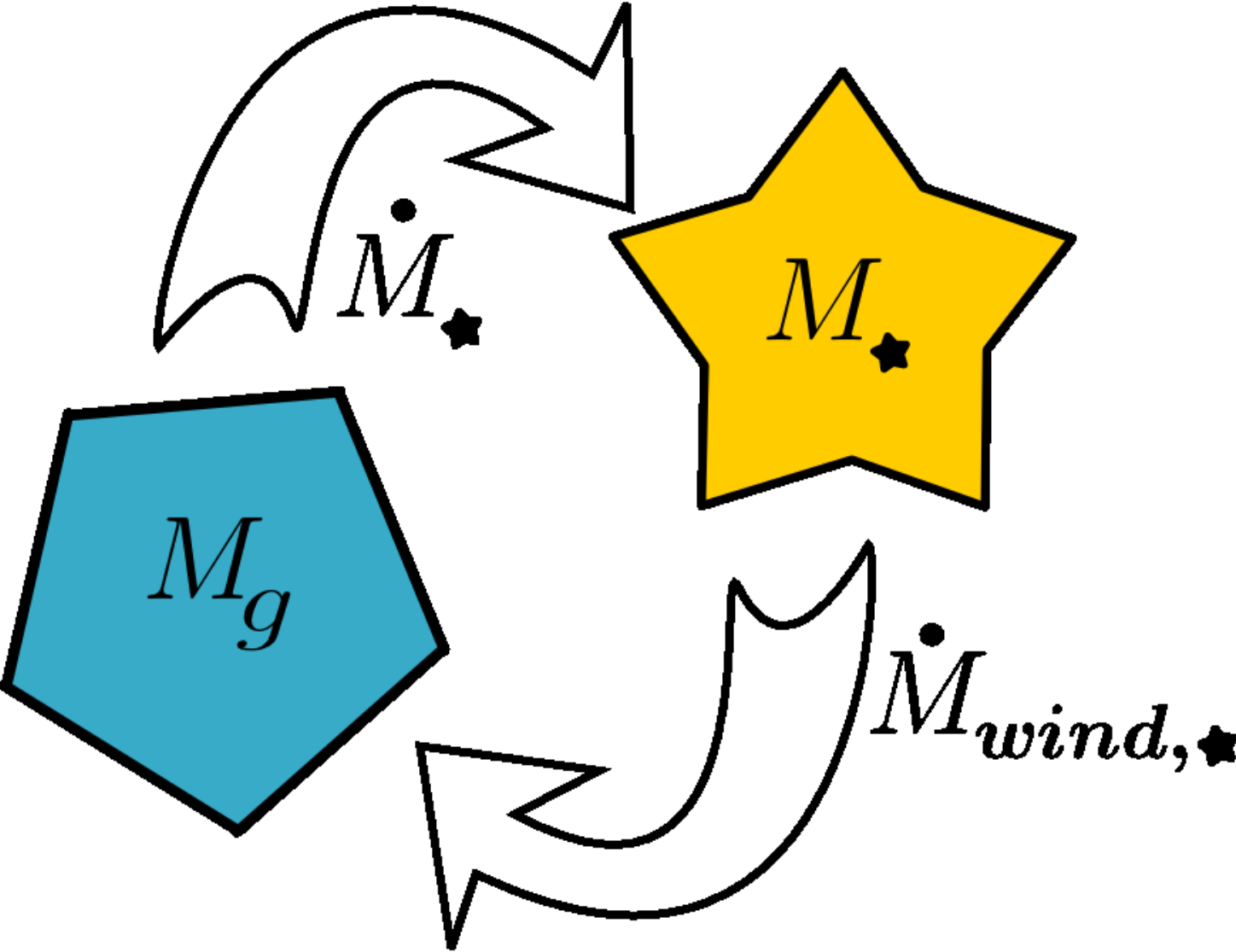}
  \caption{\ftns{Cartoon showing the stellar cycle. Stars are formed from cold gas in the galaxy components ($\dot{M}_{\star}$). Stars evolve and die. We follow the evolution of the ejected mass produced by the stellar population $\dot{M}_{wind,\star}$.}}
  \label{stellar_cycle}
  \end{center}
\end{figure}

If the disc contains some clumps, the star formation is divided into two parts: i) the homogeneous gas part in the disc ($M_{\sfg,\Disc}$, Sect. \ref{clumps_formation} and Eq.~\ref{unstable_mass}) and ii) the clumpy part ($M_{\sfg,\Clumps}$). The overall disc star formation rate is computed by considering the fraction of mass in each of the two components
  \begin{equation}
    \dot{M}_{\star,\Disc} + \dot{M}_{\star,\Clumps} = \varepsilon_{\star,\Disc}\dfrac{{M}_{\sfg,\Disc}}{t_{dyn,\Disc}} + \varepsilon_{\star,\Clumps}\dfrac{{M}_{\sfg,\Clumps}}{t_{dyn,\Clumps}}\, .
    \label{sfr_disc_all}
  \end{equation}

As shown in Fig.~\ref{stellar_cycle} the evolution of stellar population is closely related to the gas reservoirs. At the end of their life, stars re-injects gas into the galaxy's interstellar medium. This gas is enriched in metals. As in its predecessor \citep{Hatton_2003}, the metals enrichment is followed using the \cite{Devriendt_1999} prescription.

%
%

\section{Mergers}
\label{merger_events}

\subsection{Major or minor merger?}

We define the merger type (major or minor) using the following mass ratio, 
  \begin{equation}
    \eta_{merger} = \dfrac{MIN(M_{1/2,1}~;~M_{1/2,2})}{MAX(M_{1/2,1}~;~M_{1/2,2})}\,,
    \label{eta_merger}
  \end{equation}
where $M_{1/2} = M_{\Gal}(<r_{\Gal,50}) + 2M_{vir}(<r_{\Gal,50}) $ is the total mass (galaxy and dark-matter halo), enclosed in the galaxy half-mass radius $r_{\Gal,50}$. 
We consider a minor merger event when $\eta_{merger} < 0.25$. Table \ref{merger_symbol} gives symbols and their definition associated with mergers. 

\begin{table}[h]
  \begin{center}
    \footnotesize{
      \begin{tabular*}{0.5\textwidth}{@{\extracolsep{\fill}}cr}
        \hline
        Symbol & Definition \\
        \hline    
        $\eta_{merger}$                   & Mass ratio [Eq. \ref{eta_merger}]\\
        $E_{them,i} $                        & Thermal energy of a hot gas phase [Eqs. \ref{Etherm_merger}, \ref{Thot_merger}]\\
        $E_{grav} $                          & Gravitational interaction energy [Eqs. \ref{Egrav_merger}, \ref{Thot_merger}]\\
        $\varepsilon_{boost}(\tau)$ & Merger boost time function [Eq. \ref{boost_factor}]\\
        $\eta_{gas}$                        & Gas fraction in the remnant system [Eq. \ref{boost_factor}]\\
        $\tau_{merger}$                   & Characteristic merger time scale [Eq. \ref{boost_factor}]\\
        \hline
      \end{tabular*}}
  \end{center}  
  \caption{\footnotesize{Merger event symbols and their definition. Symbols are listed in order of appearance.}}
  \label{merger_symbol}
\end{table} 

\subsection{Minor mergers}
\label{minor_merger}

During a minor merger, we keep the disc and the bulge. The stellar populations of each disc (bulge), are added to form the remnant disc (bulge). In contrast, cold gas reservoirs are added in the remnant disc. The size of the remnant disc (bulge) is set to the larger disc (bulge) progenitor size. The velocities (dispersion and circular) are recomputed with the properties of the remnant galaxy and its dark-matter halo. \\

We consider that the minor merger of two discs has a strong impact on the evolution and formation of clumps. We decided to destroy all clumps in the remnant disc. New clumps have to be formed in the new disc after the merger, by taking the post-merger dark-matter and disc properties into account.

\subsection{Major mergers}
\label{major_or_micro_merger_events}

In the case of a major merger ($\eta_{merger} > 0.25$), the galaxy structure and dynamic are completely modified. The stellar population of the remnant galaxy presents an elliptical form modelled by the same \cite{Hernquist_1990} mass distribution than the bulge component. In parallel the gas components of the two progenitors are added and a new disc is formed. The characteristics of this new disc are computed in the context of the post-merger dark-matter halo ($r_{vir}, \lambda$).

At the merger time, the half-mass bulge radius $r_{\Bulge,50}$ is computed using \citep{Hatton_2003}, 
  \begin{equation}
    r_{\Bulge,50} = \dfrac{\left(M_{1/2,1} + M_{1/2,2}\right)}{\frac{M_{1/2,1}^2}{r_{\Gal,1}} + \frac{M_{1/2,2}^2}{r_{\Gal,2}} - 2.5\frac{M_{1/2,1} M_{1/2,2}}{(r_{\Gal,1}+r_{\Gal,2})}}\,,
    \label{merger_bulge_half_mass_radius}
  \end{equation}
where $M_{1/2,i}$ has been introduced in Eq.~\ref{eta_merger}. The characteristic bulge radius ($r_{\Bulge}$) is then deduced following
  \begin{equation}
    r_{\Bulge,50} = r_{\Bulge}\left(1 + \sqrt{2}\right)\,.
    \label{merger_bulge_radius}
  \end{equation}
In the case of a micro-merger event, when a clump migrates to the centre, we use the same framework as in Eq. \ref{eta_merger}, $M_{1/2,1}$ being in this case the bulge plus dark-matter halo mass enclosed in the bulge half mass radius, and $M_{1/2,2}\simeq\Clumps_{ind}$ the transferred clump mass.

\subsection{SMBH: evolution and/or formation}

For the SMBH component, if the two progenitors contain an SMBH, their masses are added, and a fraction $\delta_{\Torus} = 10^{-4}$ of the gas contained in the remnant galaxy is added to the SMBH accretion torus ($M_{\Torus}$). If only one of the two bulge components contains an SMBH, the remnant galaxy keeps this SMBH component. If the progenitors do not contain any SMBH, but if the remnant bulge mass becomes higher than the mass threshold, an SMBH is created with a mass $M_{\BH,min} = 10^3~\Msun$.

\subsection{Merger boost factor}
\label{boostfactor}

Mergers are violent events. During a short time-scale galaxy evolution processes are strongly modified and secular evolution laws (efficiencies) are no longer valid. To take these modifications into account we define a merger boost factor following
  \begin{equation}
    \varepsilon_{boost}(\tau) = 100\eta_{merger}\eta_{gas}exp\left(-\dfrac{\tau}{\tau_{merger}}\right)\,. 
\label{boost_factor}
  \end{equation}
In this definition, $\eta_{merger}$ is the progenitor mass ratio (Eq.~\ref{eta_merger}), $\eta_{gas}$ the gas fraction in the post-merger structure,  $\tau$ the time elapsed since the last merger event, and $\tau_{merger} = 0.05~Gyr $ is the characteristic merger time scale. This factor is used to modify the efficiencies just after a merger event. We use $MAX\left[1,\varepsilon_{boost}(\tau)\right]$ as a correction factor to the star formation law (Eq.~\ref{sfr}), for each component of the galaxy (disc and bulge) and for the SMBH torus accretion rate (Eq.~\ref{agn_inf_torus}). The formulation used here allows simulating a time-dependent gas compression (decreasing with time), and takes the gas content of the two progenitors into account. The more gas they contain, the higher the gas compression.

\subsection{The hot gas phase}

During a merger, the hot gas masses of the two haloes are added, and the equilibrium temperature is computed with the thermal energy of each hot atmosphere,
\begin{equation}
  E_{them,i} = \dfrac{3}{2}k_B\overline{T}_i\dfrac{M_{hot,i}}{\mu m_p}\,,
\label{Etherm_merger}
\end{equation}
which is added to the gravitational interaction energy,
\begin{equation}
  E_{grav} = G\dfrac{M_{hot,1}M_{hot,2}}{r_{hot,1}+r_{hot,2}}\,,
  \label{Egrav_merger}
\end{equation}
where $r_{hot,i}$ is the half-mass radius. 
The equilibrium temperature is then given by
\begin{equation}
  \overline{T}=\dfrac{2\mu m_p}{3k_B}\dfrac{E_{them,1} + E_{them,2} + E_{grav}}{M_{hot,1}+M_{hot,2}}\,.
\label{Thot_merger}
\end{equation}

%
%

\section{Conclusions}
\label{Overview_perspectives}
 
We have presented in this paper the physical mechanisms acting on the formation and evolution of a galaxy in a dark-matter halo in a semi-analytical framework. The model follows the dark-matter structuration through merger trees. The growth of dark-matter haloes was followed using a set of properties, including the virial mass and radius and density profile parameters. The dark-matter accretion rate is computed using background particles, which were identified as particles that have never been assigned to any other halo at a previous time step. These particles allowed following the baryonic accretion. Using a photoionisation model, the baryonic accretion rate was deduced from the smooth dark-matter accretion rate, built with the background particles. The baryonic accretion is then divided into two parts: 
\begin{itemize}
  \item{A cold mode, that dominates the growth of small structures. It relies on the fact that the cooling time of the gas accreted by low-mass haloes ($M_{vir} < 5\times 10^{10}~\Msun$) is shorter than their dynamical time.}
    \item{A hot mode: for massive haloes ($M_{vir} > 5\times 10^{10}~\Msun$), the accreted hot gas has to cool before collapsing.}
\end{itemize}
In the model, the evolution of the hot atmosphere, in equilibrium in the dark-matter potential well, is based on a detailed monitoring of its average temperature and density profile. These two quantities were followed by using energy conservation laws based on SN feedback and hot-gas accretion. In addition, the coupling between the pre-existing hot-gas phase and the wind phase generated by the galaxy, allowed us to compute the gas escape fraction. The cooling rate, computed using the standard \cite{White_1991} prescription, relies on this explicit monitoring of the hot-gas phase properties. 

Coming from the cooling process or directly from the cold accretion mode, the cold gas forms a disc at the centre of the dark-matter halo. The gas in the disc can form stars. In the fiducial approach, we used the Kennicutt law based on the total gas mass and the dynamical time of the disc. 

A feature of this new model is the treatment of instabilities in the disc. Based on \cite{Dekel_2009b}, the model we developed accounts for the formation, migration, and disruption of giant clumps. The dynamic of the disc is followed using the rotational profile that is deduced from the entire mass distribution (dark-matter halo, disc, and bulge). To this rotational profile is added a model to follow the average gas velocity dispersion. The evolution through cosmic time of the ratio $\sigma_{r}/V_{cric,\Gal}$ indicates that the population of highly disturbed discs ($\sigma_{r}/V_{cric,\Gal} \simeq 0.6$), formed at  high $z$, decreases progressively with time, and gives birth to a more quiet disc population ($\sigma_{r}/V_{cric,\Gal} \le 0.1$).  The mean velocity dispersion, added to the gas surface density computation, allows the stability behaviour of the disc to be derived. Using a method based on the Toomre parameter, we compute at each time-step the unstable mass, which is stored in clumps. The clumpy mass then follows a migration process. The mass that is transferred participates to the formation of a pseudo-bulge component. 

This pseudo-bulge, or bulge formed during mergers, can host the formation and the evolution of an SMBH. The accretion of mass onto the SMBH generates energy that is at the origin of mass ejection through wind. Together with the mass ejected due to SNe, the SMBH activity participates in the heating of the hot atmosphere around the galaxy. SN and SMBH are the two main process that allow the star formation to be regulated. Indeed gas ejection induces gas depletion in the galaxy and the heating of the hot atmosphere reduces the cooling rate.\\     

In our model all the up-to-date mechanisms used to regulate the star formation process were considered. Some variations of the fiducial recipes are discussed in detail in Cousin et al. (2014). We showed in this paper that even using strong efficiencies for the photoionisation and feedback processes, the star formation activity remains too efficient in low-mass galaxies. We thus introduce an ad hoc modification of the standard paradigm, based on the presence of a no-star-forming gas component, and an artificial concentration of the star-forming gas in galaxy discs. The no-star-forming gas reservoir generates a delay between the accretion of the gas and star formation. This model agrees much better with the observations of the stellar mass function in the low-mass range than the reference model, and agrees quite well with a large set of observations, including the redshift evolution of the specific star formation rate.\\

SAMs are currently the best models for following a large number of galaxies in their dark-matter halo. Their predictions, deduced from large scale dark-matter simulations, allow interpreting the recent surveys of galaxy population. However, even if the decoupling between dark-matter evolution (merger tree) and the baryonic physics allows exploring a large set of prescriptions in a short computation time, SAMs cannot follow all the complexity of the galaxy evolution. This applies particularly to the computation of the star formation rate. SAMs generally assume that the gas distribution follows an exponential profile, in a infinitely thin disc. All the gas content in the disc is available to form stars. In contrast, real galaxies host a complex gas structuration; high-$z$ structures are generally gas-rich discs and show a finite scale-height. Galaxy discs are a multi-fluid system, in which multi-scale interactions (2D, 3D \citealt{Shi_2011}) generate instabilities that participate in the ISM structuration \citep[e.g.][]{Jog_1984, Hoffmann_2012}. A better description of these ISM structuration mechanisms is essential for better understanding the star formation regulation processes. In a future work, we plan to follow the gas structuration in the disc from the large ($\lambda>>h)$ to the small scales ($\lambda << h$), assuming a finite scale-height ($h$) and using the turbulent cascade description. Assuming that star formation occurs only on the smallest scales, we could estimate the star-forming gas fraction and give a better description of the star formation activity. 

\begin{acknowledgements}
We thank Nicole Nesvabda, Pierre Guillard, and Nicolas Bouch\'e for useful discussions of the SNe and SMBH feedback. We also thank J\'er\'emy Blaizot and Andrea Cattaneo for very insightful discussions.
\end{acknowledgements}

\bibliographystyle{aa} 
\bibliography{aa2}

\begin{thebibliography}{127}
\expandafter\ifx\csname natexlab\endcsname\relax\def\natexlab#1{#1}\fi

\bibitem[{{Aguirre} {et~al.}(2001){Aguirre}, {Hernquist}, {Schaye}, {Weinberg},
  {Katz}, \& {Gardner}}]{Aguirre_2001}
{Aguirre}, A., {Hernquist}, L., {Schaye}, J., {et~al.} 2001, \apj, 560, 599

\bibitem[{{Antonuccio-Delogu} {et~al.}(2010){Antonuccio-Delogu}, {Dobrotka},
  {Becciani}, {Cielo}, {Giocoli}, {Macci{\`o}}, \&
  {Romeo-Velon{\'a}}}]{Antonuccio-Delogu_2010}
{Antonuccio-Delogu}, V., {Dobrotka}, A., {Becciani}, U., {et~al.} 2010, \mnras,
  407, 1338

\bibitem[{{Aubert} {et~al.}(2004){Aubert}, {Pichon}, \&
  {Colombi}}]{Aubert_2004}
{Aubert}, D., {Pichon}, C., \& {Colombi}, S. 2004, \mnras, 352, 376

\bibitem[{{Babul} \& {Rees}(1992)}]{Babul_1992}
{Babul}, A. \& {Rees}, M.~J. 1992, \mnras, 255, 346

\bibitem[{{Baugh}(2006)}]{Baugh_2006}
{Baugh}, C.~M. 2006, Reports on Progress in Physics, 69, 3101

\bibitem[{{Behroozi} {et~al.}(2012{\natexlab{a}}){Behroozi}, {Wechsler}, \&
  {Conroy}}]{Behroozi_2012b}
{Behroozi}, P.~S., {Wechsler}, R.~H., \& {Conroy}, C. 2012{\natexlab{a}}, ArXiv
  e-prints

\bibitem[{{Behroozi} {et~al.}(2012{\natexlab{b}}){Behroozi}, {Wechsler}, \&
  {Conroy}}]{Behroozi_2012a}
{Behroozi}, P.~S., {Wechsler}, R.~H., \& {Conroy}, C. 2012{\natexlab{b}}, ArXiv
  e-prints

\bibitem[{{Benson} \& {Bower}(2011)}]{Benson_2011}
{Benson}, A.~J. \& {Bower}, R. 2011, \mnras, 410, 2653

\bibitem[{{Benson} {et~al.}(2003){Benson}, {Bower}, {Frenk}, {Lacey}, {Baugh},
  \& {Cole}}]{Benson_2003}
{Benson}, A.~J., {Bower}, R.~G., {Frenk}, C.~S., {et~al.} 2003, \apj, 599, 38

\bibitem[{{Benson} {et~al.}(2002){Benson}, {Lacey}, {Baugh}, {Cole}, \&
  {Frenk}}]{Benson_2002}
{Benson}, A.~J., {Lacey}, C.~G., {Baugh}, C.~M., {Cole}, S., \& {Frenk}, C.~S.
  2002, \mnras, 333, 156

\bibitem[{{Bertone} {et~al.}(2005){Bertone}, {Stoehr}, \&
  {White}}]{Bertone_2005}
{Bertone}, S., {Stoehr}, F., \& {White}, S.~D.~M. 2005, \mnras, 359, 1201

\bibitem[{{B{\'e}thermin} {et~al.}(2012){B{\'e}thermin}, {Dor{\'e}}, \&
  {Lagache}}]{Bethermin_2012a}
{B{\'e}thermin}, M., {Dor{\'e}}, O., \& {Lagache}, G. 2012, \aap, 537, L5

\bibitem[{{Blumenthal} {et~al.}(1986){Blumenthal}, {Faber}, {Flores}, \&
  {Primack}}]{Blumenthal_1986}
{Blumenthal}, G.~R., {Faber}, S.~M., {Flores}, R., \& {Primack}, J.~R. 1986,
  \apj, 301, 27

\bibitem[{{Bondi}(1952)}]{Bondi_1952}
{Bondi}, H. 1952, \mnras, 112, 195

\bibitem[{{Bournaud} {et~al.}(2008){Bournaud}, {Daddi}, {Elmegreen},
  {Elmegreen}, {Nesvadba}, {Vanzella}, {Di Matteo}, {Le Tiran}, {Lehnert}, \&
  {Elbaz}}]{Bournaud_2008}
{Bournaud}, F., {Daddi}, E., {Elmegreen}, B.~G., {et~al.} 2008, \aap, 486, 741

\bibitem[{{Bournaud} {et~al.}(2007){Bournaud}, {Elmegreen}, \&
  {Elmegreen}}]{Bournaud_2007}
{Bournaud}, F., {Elmegreen}, B.~G., \& {Elmegreen}, D.~M. 2007, \apj, 670, 237

\bibitem[{{Bower} {et~al.}(2012){Bower}, {Benson}, \& {Crain}}]{Bower_2012}
{Bower}, R.~G., {Benson}, A.~J., \& {Crain}, R.~A. 2012, \mnras, 422, 2816

\bibitem[{{Bower} {et~al.}(2006){Bower}, {Benson}, {Malbon}, {Helly}, {Frenk},
  {Baugh}, {Cole}, \& {Lacey}}]{Bower_2006}
{Bower}, R.~G., {Benson}, A.~J., {Malbon}, R., {et~al.} 2006, \mnras, 370, 645

\bibitem[{{Bullock} {et~al.}(2000){Bullock}, {Kravtsov}, \&
  {Weinberg}}]{Bullock_2000}
{Bullock}, J.~S., {Kravtsov}, A.~V., \& {Weinberg}, D.~H. 2000, \apj, 539, 517

\bibitem[{{Capelo} {et~al.}(2012){Capelo}, {Coppi}, \&
  {Natarajan}}]{Capelo_2012}
{Capelo}, P.~R., {Coppi}, P.~S., \& {Natarajan}, P. 2012, \mnras, 422, 686

\bibitem[{{Caputi} {et~al.}(2011){Caputi}, {Cirasuolo}, {Dunlop}, {McLure},
  {Farrah}, \& {Almaini}}]{Caputi_2011}
{Caputi}, K.~I., {Cirasuolo}, M., {Dunlop}, J.~S., {et~al.} 2011, \mnras, 413,
  162

\bibitem[{{Cattaneo} {et~al.}(2006){Cattaneo}, {Dekel}, {Devriendt},
  {Guiderdoni}, \& {Blaizot}}]{Cattaneo_2006}
{Cattaneo}, A., {Dekel}, A., {Devriendt}, J., {Guiderdoni}, B., \& {Blaizot},
  J. 2006, \mnras, 370, 1651

\bibitem[{{Ceverino} {et~al.}(2010){Ceverino}, {Dekel}, \&
  {Bournaud}}]{Ceverino_2010}
{Ceverino}, D., {Dekel}, A., \& {Bournaud}, F. 2010, \mnras, 404, 2151

\bibitem[{{Ceverino} {et~al.}(2012){Ceverino}, {Dekel}, {Mandelker},
  {Bournaud}, {Burkert}, {Genzel}, \& {Primack}}]{Ceverino_2012}
{Ceverino}, D., {Dekel}, A., {Mandelker}, N., {et~al.} 2012, \mnras, 420, 3490

\bibitem[{{Cole}(1991)}]{Cole_1991}
{Cole}, S. 1991, \apj, 367, 45

\bibitem[{{Cole} {et~al.}(1994){Cole}, {Aragon-Salamanca}, {Frenk}, {Navarro},
  \& {Zepf}}]{Cole_1994}
{Cole}, S., {Aragon-Salamanca}, A., {Frenk}, C.~S., {Navarro}, J.~F., \&
  {Zepf}, S.~E. 1994, \mnras, 271, 781

\bibitem[{{Cole} {et~al.}(2008){Cole}, {Helly}, {Frenk}, \&
  {Parkinson}}]{Cole_2008}
{Cole}, S., {Helly}, J., {Frenk}, C.~S., \& {Parkinson}, H. 2008, \mnras, 383,
  546

\bibitem[{{Cole} {et~al.}(2000){Cole}, {Lacey}, {Baugh}, \&
  {Frenk}}]{Cole_2000}
{Cole}, S., {Lacey}, C.~G., {Baugh}, C.~M., \& {Frenk}, C.~S. 2000, \mnras,
  319, 168

\bibitem[{{Couchman} \& {Rees}(1986)}]{Couchman_1986}
{Couchman}, H.~M.~P. \& {Rees}, M.~J. 1986, \mnras, 221, 53

\bibitem[{{Cowie} {et~al.}(1995){Cowie}, {Hu}, \& {Songaila}}]{Cowie_1995}
{Cowie}, L.~L., {Hu}, E.~M., \& {Songaila}, A. 1995, \aj, 110, 1576

\bibitem[{{Croton} {et~al.}(2006){Croton}, {Springel}, {White}, {De Lucia},
  {Frenk}, {Gao}, {Jenkins}, {Kauffmann}, {Navarro}, \&
  {Yoshida}}]{Croton_2006}
{Croton}, D.~J., {Springel}, V., {White}, S.~D.~M., {et~al.} 2006, \mnras, 365,
  11

\bibitem[{{De Lucia} {et~al.}(2010){De Lucia}, {Boylan-Kolchin}, {Benson},
  {Fontanot}, \& {Monaco}}]{DeLucia_2010}
{De Lucia}, G., {Boylan-Kolchin}, M., {Benson}, A.~J., {Fontanot}, F., \&
  {Monaco}, P. 2010, \mnras, 406, 1533

\bibitem[{{Dekel} {et~al.}(2009{\natexlab{a}}){Dekel}, {Birnboim}, {Engel},
  {Freundlich}, {Goerdt}, {Mumcuoglu}, {Neistein}, {Pichon}, {Teyssier}, \&
  {Zinger}}]{Dekel_2009a}
{Dekel}, A., {Birnboim}, Y., {Engel}, G., {et~al.} 2009{\natexlab{a}}, Nature,
  457, 451

\bibitem[{{Dekel} {et~al.}(2009{\natexlab{b}}){Dekel}, {Sari}, \&
  {Ceverino}}]{Dekel_2009b}
{Dekel}, A., {Sari}, R., \& {Ceverino}, D. 2009{\natexlab{b}}, \apj, 703, 785

\bibitem[{{Dekel} \& {Silk}(1986)}]{Dekel_1986}
{Dekel}, A. \& {Silk}, J. 1986, \apj, 303, 39

\bibitem[{{Devriendt} {et~al.}(1999){Devriendt}, {Guiderdoni}, \&
  {Sadat}}]{Devriendt_1999}
{Devriendt}, J.~E.~G., {Guiderdoni}, B., \& {Sadat}, R. 1999, \aap, 350, 381

\bibitem[{{Doroshkevich} {et~al.}(1967){Doroshkevich}, {Zel'dovich}, \&
  {Novikov}}]{Doroshkevich_1967}
{Doroshkevich}, A.~G., {Zel'dovich}, Y.~B., \& {Novikov}, I.~D. 1967, \sovast,
  11, 233

\bibitem[{{Efstathiou}(1992)}]{Efstathiou_1992}
{Efstathiou}, G. 1992, \mnras, 256, 43P

\bibitem[{{Efstathiou}(2000)}]{Efstathiou_2000}
{Efstathiou}, G. 2000, \mnras, 317, 697

\bibitem[{{Eke} {et~al.}(2005){Eke}, {Baugh}, {Cole}, {Frenk}, {King}, \&
  {Peacock}}]{Eke_2005}
{Eke}, V.~R., {Baugh}, C.~M., {Cole}, S., {et~al.} 2005, \mnras, 362, 1233

\bibitem[{{Eke} {et~al.}(2004){Eke}, {Frenk}, {Baugh}, {Cole}, {Norberg},
  {Peacock}, {Baldry}, {Bland-Hawthorn}, {Bridges}, {Cannon}, {Colless},
  {Collins}, {Couch}, {Dalton}, {de Propris}, {Driver}, {Efstathiou}, {Ellis},
  {Glazebrook}, {Jackson}, {Lahav}, {Lewis}, {Lumsden}, {Maddox}, {Madgwick},
  {Peterson}, {Sutherland}, \& {Taylor}}]{Eke_2004}
{Eke}, V.~R., {Frenk}, C.~S., {Baugh}, C.~M., {et~al.} 2004, \mnras, 355, 769

\bibitem[{{Elmegreen}(2009)}]{Elmegreen_2009}
{Elmegreen}, B.~G. 2009, in Astronomical Society of the Pacific Conference
  Series, Vol. 419, Galaxy Evolution: Emerging Insights and Future Challenges,
  ed. S.~{Jogee}, I.~{Marinova}, L.~{Hao}, \& G.~A. {Blanc}, 23

\bibitem[{{Elmegreen} \& {Elmegreen}(2005)}]{Elmegreen_2005}
{Elmegreen}, B.~G. \& {Elmegreen}, D.~M. 2005, \apj, 627, 632

\bibitem[{{Emonts} {et~al.}(2005){Emonts}, {Morganti}, {Tadhunter},
  {Oosterloo}, {Holt}, \& {van der Hulst}}]{Emonts_2005}
{Emonts}, B.~H.~C., {Morganti}, R., {Tadhunter}, C.~N., {et~al.} 2005, \mnras,
  362, 931

\bibitem[{{Fakhouri} \& {Ma}(2010)}]{Fakhouri_2010a}
{Fakhouri}, O. \& {Ma}, C.-P. 2010, \mnras, 401, 2245

\bibitem[{{Fakhouri} {et~al.}(2010){Fakhouri}, {Ma}, \&
  {Boylan-Kolchin}}]{Fakhouri_2010b}
{Fakhouri}, O., {Ma}, C.-P., \& {Boylan-Kolchin}, M. 2010, \mnras, 406, 2267

\bibitem[{{Faucher-Gigu{\`e}re} {et~al.}(2011){Faucher-Gigu{\`e}re}, {Kere{\v
  s}}, \& {Ma}}]{Faucher-Giguere_2011}
{Faucher-Gigu{\`e}re}, C.-A., {Kere{\v s}}, D., \& {Ma}, C.-P. 2011, \mnras,
  417, 2982

\bibitem[{{Freeman}(1970)}]{Freeman_1970}
{Freeman}, K.~C. 1970, \apj, 160, 811

\bibitem[{{Frye} {et~al.}(2002){Frye}, {Broadhurst}, \&
  {Ben{\'{\i}}tez}}]{Frye_2002}
{Frye}, B., {Broadhurst}, T., \& {Ben{\'{\i}}tez}, N. 2002, \apj, 568, 558

\bibitem[{{Genel} {et~al.}(2010){Genel}, {Bouch{\'e}}, {Naab}, {Sternberg}, \&
  {Genzel}}]{Genel_2010}
{Genel}, S., {Bouch{\'e}}, N., {Naab}, T., {Sternberg}, A., \& {Genzel}, R.
  2010, \apj, 719, 229

\bibitem[{{Genzel} {et~al.}(2008){Genzel}, {Burkert}, {Bouch{\'e}}, {Cresci},
  {F{\"o}rster Schreiber}, {Shapley}, {Shapiro}, {Tacconi}, {Buschkamp},
  {Cimatti}, {Daddi}, {Davies}, {Eisenhauer}, {Erb}, {Genel}, {Gerhard},
  {Hicks}, {Lutz}, {Naab}, {Ott}, {Rabien}, {Renzini}, {Steidel}, {Sternberg},
  \& {Lilly}}]{Genzel_2008}
{Genzel}, R., {Burkert}, A., {Bouch{\'e}}, N., {et~al.} 2008, \apj, 687, 59

\bibitem[{{Gnedin}(2000)}]{Gnedin_2000}
{Gnedin}, N.~Y. 2000, \apj, 542, 535

\bibitem[{{Guillard} {et~al.}(2012){Guillard}, {Ogle}, {Emonts}, {Appleton},
  {Morganti}, {Tadhunter}, {Oosterloo}, {Evans}, \& {Evans}}]{Guillard_2012}
{Guillard}, P., {Ogle}, P.~M., {Emonts}, B.~H.~C., {et~al.} 2012, \apj, 747, 95

\bibitem[{{Guo} {et~al.}(2011){Guo}, {White}, {Boylan-Kolchin}, {De Lucia},
  {Kauffmann}, {Lemson}, {Li}, {Springel}, \& {Weinmann}}]{Guo_2011}
{Guo}, Q., {White}, S., {Boylan-Kolchin}, M., {et~al.} 2011, \mnras, 413, 101

\bibitem[{{Guo} {et~al.}(2010){Guo}, {White}, {Li}, \&
  {Boylan-Kolchin}}]{Guo_2010}
{Guo}, Q., {White}, S., {Li}, C., \& {Boylan-Kolchin}, M. 2010, \mnras, 404,
  1111

\bibitem[{{Hatton} {et~al.}(2003){Hatton}, {Devriendt}, {Ninin}, {Bouchet},
  {Guiderdoni}, \& {Vibert}}]{Hatton_2003}
{Hatton}, S., {Devriendt}, J.~E.~G., {Ninin}, S., {et~al.} 2003, \mnras, 343,
  75

\bibitem[{{Heckman} {et~al.}(2000){Heckman}, {Lehnert}, {Strickland}, \&
  {Armus}}]{Heckman_2000}
{Heckman}, T.~M., {Lehnert}, M.~D., {Strickland}, D.~K., \& {Armus}, L. 2000,
  \apjs, 129, 493

\bibitem[{{Helly} {et~al.}(2003){Helly}, {Cole}, {Frenk}, {Baugh}, {Benson}, \&
  {Lacey}}]{Helly_2003}
{Helly}, J.~C., {Cole}, S., {Frenk}, C.~S., {et~al.} 2003, \mnras, 338, 903

\bibitem[{{Henriques} {et~al.}(2013){Henriques}, {White}, {Thomas}, {Angulo},
  {Guo}, {Lemson}, \& {Springel}}]{Henriques_2013}
{Henriques}, B.~M.~B., {White}, S.~D.~M., {Thomas}, P.~A., {et~al.} 2013,
  \mnras, 431, 3373

\bibitem[{{Hernquist}(1990)}]{Hernquist_1990}
{Hernquist}, L. 1990, \apj, 356, 359

\bibitem[{{Hoeft} {et~al.}(2006){Hoeft}, {Yepes}, {Gottl{\"o}ber}, \&
  {Springel}}]{Hoeft_2006}
{Hoeft}, M., {Yepes}, G., {Gottl{\"o}ber}, S., \& {Springel}, V. 2006, \mnras,
  371, 401

\bibitem[{{Hoffmann} \& {Romeo}(2012)}]{Hoffmann_2012}
{Hoffmann}, V. \& {Romeo}, A.~B. 2012, \mnras, 425, 1511

\bibitem[{{Ikeuchi}(1986)}]{Ikeuchi_1986}
{Ikeuchi}, S. 1986, \apss, 118, 509

\bibitem[{{Ilbert} {et~al.}(2013){Ilbert}, {McCracken}, {Le Fevre}, {Capak},
  {Dunlop}, {Arnouts}, {Aussel}, {Caputi}, {Comparat}, {Guo}, {Hudelot},
  {Kartaltepe}, {Kneib}, {Krogager}, {Le Floc'h}, {Lilly}, {Mellier},
  {Milvang-Jensen}, {Moutard}, {Onodera}, {Renzini}, {Richard}, {Salvato},
  {Sanders}, {Scoville}, {Silverman}, {Taniguchi}, {Tasca}, {Thomas}, {Toft},
  {Tresse}, {Vergani}, {Wolk}, \& {Zirm}}]{Ilbert_2013}
{Ilbert}, O., {McCracken}, H.~J., {Le Fevre}, O., {et~al.} 2013, ArXiv e-prints

\bibitem[{{Ilbert} {et~al.}(2010){Ilbert}, {Salvato}, {Le Floc'h}, {Aussel},
  {Capak}, {McCracken}, {Mobasher}, {Kartaltepe}, {Scoville}, {Sanders},
  {Arnouts}, {Bundy}, {Cassata}, {Kneib}, {Koekemoer}, {Le F{\`e}vre}, {Lilly},
  {Surace}, {Taniguchi}, {Tasca}, {Thompson}, {Tresse}, {Zamojski}, {Zamorani},
  \& {Zucca}}]{Ilbert_2010}
{Ilbert}, O., {Salvato}, M., {Le Floc'h}, E., {et~al.} 2010, \apj, 709, 644

\bibitem[{{Jog} \& {Solomon}(1984)}]{Jog_1984}
{Jog}, C.~J. \& {Solomon}, P.~M. 1984, \apj, 276, 114

\bibitem[{{Kahn}(1975)}]{Kahn_1975}
{Kahn}, F.~D. 1975, in International Cosmic Ray Conference, Vol.~11,
  International Cosmic Ray Conference, 3566

\bibitem[{{Kauffmann} {et~al.}(1999){Kauffmann}, {Colberg}, {Diaferio}, \&
  {White}}]{Kauffmann_1999}
{Kauffmann}, G., {Colberg}, J.~M., {Diaferio}, A., \& {White}, S.~D.~M. 1999,
  \mnras, 303, 188

\bibitem[{{Kauffmann} {et~al.}(1993){Kauffmann}, {White}, \&
  {Guiderdoni}}]{Kauffmann_1993}
{Kauffmann}, G., {White}, S.~D.~M., \& {Guiderdoni}, B. 1993, \mnras, 264, 201

\bibitem[{{Kennicutt}(1998)}]{Kennicutt_1998}
{Kennicutt}, Jr., R.~C. 1998, \apj, 498, 541

\bibitem[{{Kere{\v s}} {et~al.}(2005){Kere{\v s}}, {Katz}, {Weinberg}, \&
  {Dav{\'e}}}]{Keres_2005}
{Kere{\v s}}, D., {Katz}, N., {Weinberg}, D.~H., \& {Dav{\'e}}, R. 2005,
  \mnras, 363, 2

\bibitem[{{Khochfar} \& {Silk}(2009)}]{Khochfar_2009}
{Khochfar}, S. \& {Silk}, J. 2009, \apjl, 700, L21

\bibitem[{{Komatsu} \& {Seljak}(2001)}]{Komatsu_2001}
{Komatsu}, E. \& {Seljak}, U. 2001, \mnras, 327, 1353

\bibitem[{{Koo} {et~al.}(2002){Koo}, {Lee}, \& {Seward}}]{Koo_2002}
{Koo}, B.-C., {Lee}, J.-J., \& {Seward}, F.~D. 2002, \aj, 123, 1629

\bibitem[{{Kravtsov} {et~al.}(2004){Kravtsov}, {Gnedin}, \&
  {Klypin}}]{Kravtsov_2004}
{Kravtsov}, A.~V., {Gnedin}, O.~Y., \& {Klypin}, A.~A. 2004, \apj, 609, 482

\bibitem[{{Lacey} \& {Cole}(1994)}]{Lacey_1994}
{Lacey}, C. \& {Cole}, S. 1994, \mnras, 271, 676

\bibitem[{{Larson}(1974)}]{Larson_1974}
{Larson}, R.~B. 1974, \mnras, 169, 229

\bibitem[{{Leauthaud} {et~al.}(2012){Leauthaud}, {Tinker}, {Bundy}, {Behroozi},
  {Massey}, {Rhodes}, {George}, {Kneib}, {Benson}, {Wechsler}, {Busha},
  {Capak}, {Cort{\^e}s}, {Ilbert}, {Koekemoer}, {Le F{\`e}vre}, {Lilly},
  {McCracken}, {Salvato}, {Schrabback}, {Scoville}, {Smith}, \&
  {Taylor}}]{Leauthaud_2012}
{Leauthaud}, A., {Tinker}, J., {Bundy}, K., {et~al.} 2012, \apj, 744, 159

\bibitem[{{Lehnert} {et~al.}(2011){Lehnert}, {Tasse}, {Nesvadba}, {Best}, \&
  {van Driel}}]{Lehnert_2011}
{Lehnert}, M.~D., {Tasse}, C., {Nesvadba}, N.~P.~H., {Best}, P.~N., \& {van
  Driel}, W. 2011, \aap, 532, L3

\bibitem[{{Lu} {et~al.}(2011){Lu}, {Kere{\v s}}, {Katz}, {Mo}, {Fardal}, \&
  {Weinberg}}]{Lu_2011}
{Lu}, Y., {Kere{\v s}}, D., {Katz}, N., {et~al.} 2011, \mnras, 416, 660

\bibitem[{{Macci{\`o}} {et~al.}(2008){Macci{\`o}}, {Dutton}, \& {van den
  Bosch}}]{Maccio_2008}
{Macci{\`o}}, A.~V., {Dutton}, A.~A., \& {van den Bosch}, F.~C. 2008, \mnras,
  391, 1940

\bibitem[{{Makino} {et~al.}(1998){Makino}, {Sasaki}, \& {Suto}}]{Makino_1998}
{Makino}, N., {Sasaki}, S., \& {Suto}, Y. 1998, \apj, 497, 555

\bibitem[{{Malbon} {et~al.}(2007){Malbon}, {Baugh}, {Frenk}, \&
  {Lacey}}]{Malbon_2007}
{Malbon}, R.~K., {Baugh}, C.~M., {Frenk}, C.~S., \& {Lacey}, C.~G. 2007,
  \mnras, 382, 1394

\bibitem[{{Martin}(1999)}]{Martin_1999}
{Martin}, C.~L. 1999, \apj, 513, 156

\bibitem[{{Mo} {et~al.}(1998){Mo}, {Mao}, \& {White}}]{Mo_1998}
{Mo}, H.~J., {Mao}, S., \& {White}, S.~D.~M. 1998, \mnras, 295, 319

\bibitem[{{Morganti} {et~al.}(2005{\natexlab{a}}){Morganti}, {Oosterloo},
  {Tadhunter}, {van Moorsel}, \& {Emonts}}]{Morganti_2005a}
{Morganti}, R., {Oosterloo}, T.~A., {Tadhunter}, C.~N., {van Moorsel}, G., \&
  {Emonts}, B. 2005{\natexlab{a}}, \aap, 439, 521

\bibitem[{{Morganti} {et~al.}(2005{\natexlab{b}}){Morganti}, {Tadhunter}, \&
  {Oosterloo}}]{Morganti_2005b}
{Morganti}, R., {Tadhunter}, C.~N., \& {Oosterloo}, T.~A. 2005{\natexlab{b}},
  \aap, 444, L9

\bibitem[{{Mu{\~n}oz-Cuartas} {et~al.}(2011){Mu{\~n}oz-Cuartas}, {Macci{\`o}},
  {Gottl{\"o}ber}, \& {Dutton}}]{Munoz-Cuartas_2011}
{Mu{\~n}oz-Cuartas}, J.~C., {Macci{\`o}}, A.~V., {Gottl{\"o}ber}, S., \&
  {Dutton}, A.~A. 2011, \mnras, 411, 584

\bibitem[{{Navarro} {et~al.}(1995){Navarro}, {Frenk}, \&
  {White}}]{Navarro_1995}
{Navarro}, J.~F., {Frenk}, C.~S., \& {White}, S.~D.~M. 1995, \mnras, 275, 56

\bibitem[{{Navarro} {et~al.}(1996){Navarro}, {Frenk}, \&
  {White}}]{Navarro_1996}
{Navarro}, J.~F., {Frenk}, C.~S., \& {White}, S.~D.~M. 1996, \apj, 462, 563

\bibitem[{{Navarro} {et~al.}(1997){Navarro}, {Frenk}, \&
  {White}}]{Navarro_1997}
{Navarro}, J.~F., {Frenk}, C.~S., \& {White}, S.~D.~M. 1997, \apj, 490, 493

\bibitem[{{Nelson} {et~al.}(2013){Nelson}, {Vogelsberger}, {Genel}, {Sijacki},
  {Kere{\v s}}, {Springel}, \& {Hernquist}}]{Nelson_2013}
{Nelson}, D., {Vogelsberger}, M., {Genel}, S., {et~al.} 2013, \mnras, 429, 3353

\bibitem[{{Nesvadba} {et~al.}(2008){Nesvadba}, {Lehnert}, {De Breuck},
  {Gilbert}, \& {van Breugel}}]{Nesvadba_2008}
{Nesvadba}, N.~P.~H., {Lehnert}, M.~D., {De Breuck}, C., {Gilbert}, A.~M., \&
  {van Breugel}, W. 2008, \aap, 491, 407

\bibitem[{{Nesvadba} {et~al.}(2006){Nesvadba}, {Lehnert}, {Eisenhauer},
  {Gilbert}, {Tecza}, \& {Abuter}}]{Nesvadba_2006}
{Nesvadba}, N.~P.~H., {Lehnert}, M.~D., {Eisenhauer}, F., {et~al.} 2006, \apj,
  650, 693

\bibitem[{{Ocvirk} {et~al.}(2008){Ocvirk}, {Pichon}, \&
  {Teyssier}}]{Ocvirk_2008}
{Ocvirk}, P., {Pichon}, C., \& {Teyssier}, R. 2008, \mnras, 390, 1326

\bibitem[{{Okamoto} {et~al.}(2008){Okamoto}, {Gao}, \& {Theuns}}]{Okamoto_2008}
{Okamoto}, T., {Gao}, L., \& {Theuns}, T. 2008, \mnras, 390, 920

\bibitem[{{Ostriker} {et~al.}(2010){Ostriker}, {Choi}, {Ciotti}, {Novak}, \&
  {Proga}}]{Ostriker_2010}
{Ostriker}, J.~P., {Choi}, E., {Ciotti}, L., {Novak}, G.~S., \& {Proga}, D.
  2010, \apj, 722, 642

\bibitem[{{Peebles}(1969)}]{Peebles_1969}
{Peebles}, P.~J.~E. 1969, \apj, 155, 393

\bibitem[{{Press} \& {Schechter}(1974)}]{Press_1974}
{Press}, W.~H. \& {Schechter}, P. 1974, \apj, 187, 425

\bibitem[{{Proga} {et~al.}(2000){Proga}, {Stone}, \& {Kallman}}]{Proga_2000}
{Proga}, D., {Stone}, J.~M., \& {Kallman}, T.~R. 2000, \apj, 543, 686

\bibitem[{{Quinn} {et~al.}(1996){Quinn}, {Katz}, \& {Efstathiou}}]{Quinn_1996}
{Quinn}, T., {Katz}, N., \& {Efstathiou}, G. 1996, \mnras, 278, L49

\bibitem[{{Rees}(1986)}]{Rees_1986}
{Rees}, M.~J. 1986, \mnras, 218, 25P

\bibitem[{{Sellwood} \& {McGaugh}(2005)}]{Sellwood_2005}
{Sellwood}, J.~A. \& {McGaugh}, S.~S. 2005, \apj, 634, 70

\bibitem[{{Shapiro} {et~al.}(1994){Shapiro}, {Giroux}, \&
  {Babul}}]{Shapiro_1994}
{Shapiro}, P.~R., {Giroux}, M.~L., \& {Babul}, A. 1994, \apj, 427, 25

\bibitem[{{Shi} {et~al.}(2011){Shi}, {Helou}, {Yan}, {Armus}, {Wu}, {Papovich},
  \& {Stierwalt}}]{Shi_2011}
{Shi}, Y., {Helou}, G., {Yan}, L., {et~al.} 2011, \apj, 733, 87

\bibitem[{{Shu} {et~al.}(2005){Shu}, {Mo}, \& {Shu-DeMao}}]{Shu_2005}
{Shu}, C.-G., {Mo}, H.-J., \& {Shu-DeMao}. 2005, \cjaa, 5, 327

\bibitem[{{Silk}(2003)}]{Silk_2003}
{Silk}, J. 2003, \mnras, 343, 249

\bibitem[{{Somerville}(2002)}]{Somerville_2002}
{Somerville}, R.~S. 2002, \apjl, 572, L23

\bibitem[{{Somerville} {et~al.}(2012){Somerville}, {Gilmore}, {Primack}, \&
  {Dom{\'{\i}}nguez}}]{Somerville_2012}
{Somerville}, R.~S., {Gilmore}, R.~C., {Primack}, J.~R., \& {Dom{\'{\i}}nguez},
  A. 2012, \mnras, 423, 1992

\bibitem[{{Somerville} {et~al.}(2008){Somerville}, {Hopkins}, {Cox},
  {Robertson}, \& {Hernquist}}]{Somerville_2008}
{Somerville}, R.~S., {Hopkins}, P.~F., {Cox}, T.~J., {Robertson}, B.~E., \&
  {Hernquist}, L. 2008, \mnras, 391, 481

\bibitem[{{Somerville} \& {Kolatt}(1999)}]{Somerville_1999}
{Somerville}, R.~S. \& {Kolatt}, T.~S. 1999, \mnras, 305, 1

\bibitem[{{Stoll} {et~al.}(2009){Stoll}, {Mathur}, {Krongold}, \&
  {Nicastro}}]{Stoll_2009}
{Stoll}, R., {Mathur}, S., {Krongold}, Y., \& {Nicastro}, F. 2009, ArXiv
  e-prints

\bibitem[{{Sutherland} \& {Dopita}(1993)}]{Sutherland_1993}
{Sutherland}, R.~S. \& {Dopita}, M.~A. 1993, apjs, 88, 253

\bibitem[{{Suto} {et~al.}(1998){Suto}, {Sasaki}, \& {Makino}}]{Suto_1998}
{Suto}, Y., {Sasaki}, S., \& {Makino}, N. 1998, \apj, 509, 544

\bibitem[{{Thoul} \& {Weinberg}(1996)}]{Thoul_1996}
{Thoul}, A.~A. \& {Weinberg}, D.~H. 1996, \apj, 465, 608

\bibitem[{{Toomre}(1963)}]{Toomre_1963}
{Toomre}, A. 1963, \apj, 138, 385

\bibitem[{{Toomre}(1964)}]{Toomre_1964}
{Toomre}, A. 1964, \apj, 139, 1217

\bibitem[{{Tormen} {et~al.}(1997){Tormen}, {Bouchet}, \& {White}}]{Tormen_1997}
{Tormen}, G., {Bouchet}, F.~R., \& {White}, S.~D.~M. 1997, \mnras, 286, 865

\bibitem[{{Tweed} {et~al.}(2009){Tweed}, {Devriendt}, {Blaizot}, {Colombi}, \&
  {Slyz}}]{Tweed_2009}
{Tweed}, D., {Devriendt}, J., {Blaizot}, J., {Colombi}, S., \& {Slyz}, A. 2009,
  \aap, 506, 647

\bibitem[{{van Daalen} {et~al.}(2011){van Daalen}, {Schaye}, {Booth}, \& {Dalla
  Vecchia}}]{van_Daalen_2011}
{van Daalen}, M.~P., {Schaye}, J., {Booth}, C.~M., \& {Dalla Vecchia}, C. 2011,
  \mnras, 415, 3649

\bibitem[{{van de Voort} {et~al.}(2010){van de Voort}, {Schaye}, {Booth},
  {Haas}, \& {Dalla Vecchia}}]{Voort_2010}
{van de Voort}, F., {Schaye}, J., {Booth}, C.~M., {Haas}, M.~R., \& {Dalla
  Vecchia}, C. 2010, ArXiv e-prints

\bibitem[{{van den Bergh}(1996)}]{Van_den_Bergh_1996}
{van den Bergh}, S. 1996, \aj, 112, 2634

\bibitem[{{van den Bosch}(2002)}]{Van_den_Bosch_2002}
{van den Bosch}, F.~C. 2002, \mnras, 331, 98

\bibitem[{{Wechsler} {et~al.}(2002){Wechsler}, {Bullock}, {Primack},
  {Kravtsov}, \& {Dekel}}]{Wechsler_2002}
{Wechsler}, R.~H., {Bullock}, J.~S., {Primack}, J.~R., {Kravtsov}, A.~V., \&
  {Dekel}, A. 2002, \apj, 568, 52

\bibitem[{{White} \& {Frenk}(1991)}]{White_1991}
{White}, S.~D.~M. \& {Frenk}, C.~S. 1991, \apj, 379, 52

\bibitem[{{White} \& {Rees}(1978)}]{White_1978}
{White}, S.~D.~M. \& {Rees}, M.~J. 1978, \mnras, 183, 341

\bibitem[{{Zhao} {et~al.}(2009){Zhao}, {Jing}, {Mo}, \&
  {B{\"o}rner}}]{Zhao_2009}
{Zhao}, D.~H., {Jing}, Y.~P., {Mo}, H.~J., \& {B{\"o}rner}, G. 2009, \apj, 707,
  354

\end{thebibliography}

%
%

\begin{appendix}

%
%

\section{An adaptive time-step scheme}
\label{adaptive_time_step}

In a galaxy, different processes act at the same time but on different time scales. It is not efficient, for example, to compute the evolution of the cold filamentary phase that evolves with a typical dark-matter dynamical time ($t_{dyn} = 10^6$ yr) and with a time step dedicated to the ejecta rate of the galaxy ($5\times 10^3$ yr). If we use the same time step to follow these two components, we generate numerical errors on the component with the shorter dynamical time.

In this context, each part of the gas cycle (in the baryonic halo phase or in the galaxy) must evolve with a time step that is as close as possible to its dynamical time. Accordingly, we have developed an adaptive time-step scheme. Each component evolves passively by receiving and transfers mass until the evolution of an other component significantly affects its own evolution. We consider that a component is modified if the variation $\Delta M$ of its mass (gas or stars) is greater than $p\% = 10\%$. This value is low enough to accurately follow star formation and ejection processes in gas-rich galaxies up to $z\simeq 4$. Naturally, for passive galaxy (like spheroidal at low z), a higherr value could be acceptable. 

\subsection{Description}

At any given time step each component with a mass $M$ must evolve with an input and output rate ($\dot{F}_{in}$ and $\dot{F}_{out}$). With these two gas flows, the optimal time step, $\delta t^{MAX}$, for a component with a mass $M$ is therefore computed as
\begin{equation}
  \delta t^{MAX} = p\%\dfrac{M}{\left|\dot{F}_{in}-\dot{F}_{out}\right|}\, .
\end{equation} 
For example, for the cold gas component $M_{cold}$ (Fig. \ref{baryon_halo_transfer}), $\dot{F}_{in} = \dot{M}_{flt}$ (Eq. \ref{flt_formation_rate}) and $\dot{F}_{out} = \dot{M}_{ff}$ (Eq. \ref{free-fall-rate}); therefore,
\begin{equation}
  \delta t^{MAX}_{cold} = p\%\dfrac{M_{cold}}{\left|\dot{M}_{flt}-\dot{M}_{ff}\right|} \, .
\end{equation}
At the first time step, the mass is not defined and the reference is then fixed to
\begin{equation}
M_{lim,bar} = p\% <f_b>\underbrace{M_{lim,dm}}_{20\times m_p} \, .
\end{equation}

\subsection{The hot-gas phase}

The case of the hot atmosphere is more complicated. Indeed, the hot phase
\begin{itemize}
  \item{receives gas from the intergalactic medium: $\dot{M}_{sh}$ (Eq. \ref{hot_cosmo_acc});}
  \item{follows cooling process: $\dot{M}_{cool}$ (Eq. \ref{cooling_rate});}
  \item{receives gas coming from the galaxy: $\dot{M}_{wind}$ (Eq. \ref{gal_ejecta}).}
\end{itemize}
This last term is not linked to the mass of the hot phase $M_{hot}$ but depends on the star formation activity of the galaxy. To follow the evolution of the hot atmosphere with a high accuracy, and therefore compute the optimal time step $\delta t^{MAX}_{hot}$, we must take possible variations in the ejecta rate into account. 

\begin{figure}[h]
 \centering
 \includegraphics[scale = 0.42]{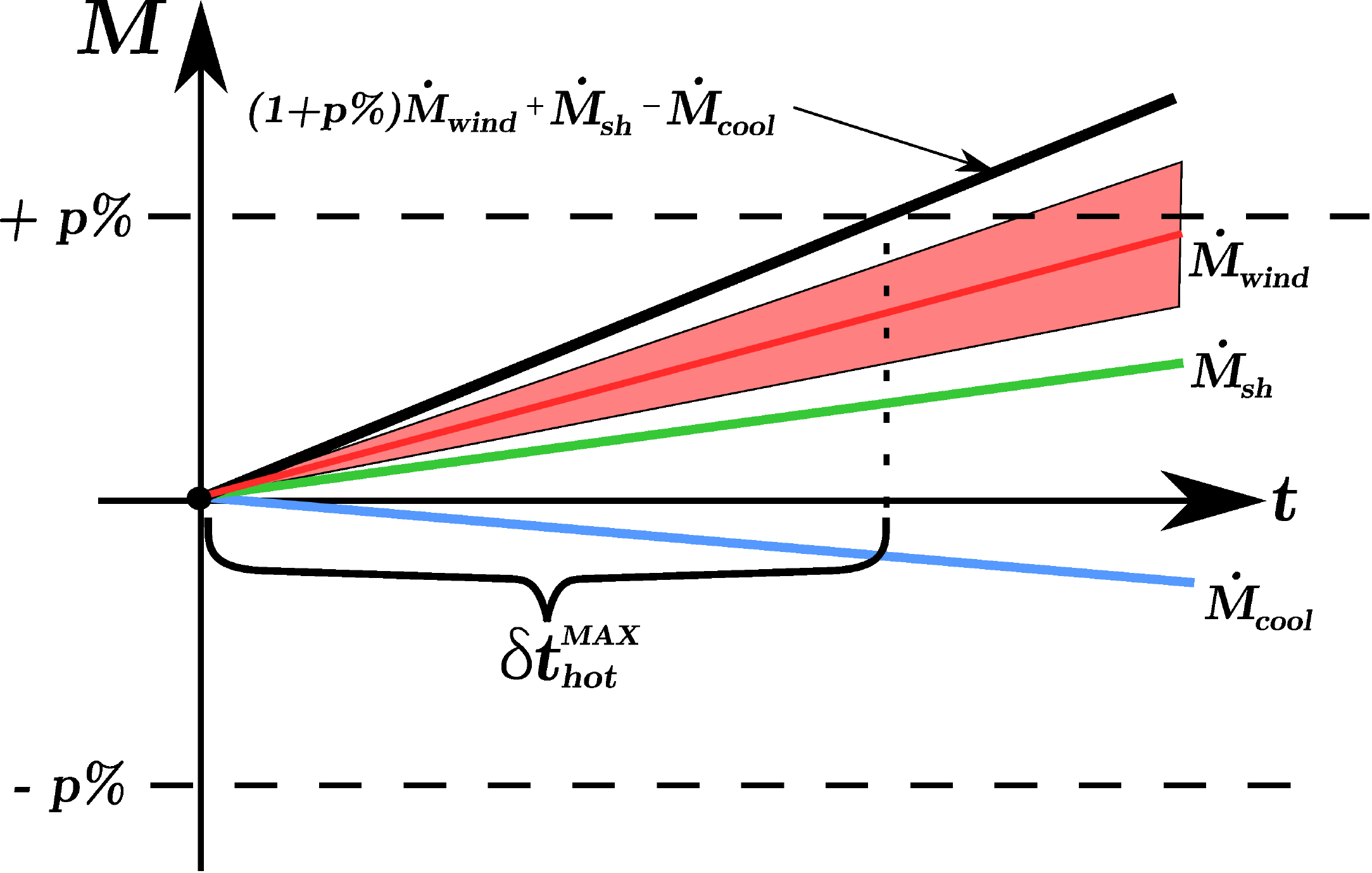}
 \caption{\tiny{Diagram illustrating the computation of the optimal time step of the hot gas phase ($\delta t^{MAX}_{hot}$). The two horizontal dash lines materialise the maximum variation, p\%. In the case shown here, the evolution of the hot gas phase is dominated by input flows: $\dot{M}_{sh} + \dot{M}_{wind} > \dot{M}_{cool}$. The optimum time step for the hot-gas phase evolution is therefore computed assuming an increase in the galaxy ejecta rate $(1+p\%)\dot{M}_{wind}$.}}
  \label{dM_to_dtmax}
\end{figure}

In this context, it is important to check whether the evolution of the hot phase is dominated by input ($\dot{M}_{sh} + \dot{M}_{wind}$) or output ($\dot{M}_{cool}$) gas flows. In the first case $\dot{M}_{in} = \dot{M}_{sh} + \dot{M}_{wind} > \dot{M}_{out} = \dot{M}_{cool}$ if we take instantaneous values of all input and output rates to compute $\delta t^{MAX}_{hot}$, an increase, in the next time step, in the galaxy ejection rate would lead to a violation, by excess, of the ``stability criterium'': $\Delta M_{hot}/M_{hot} \le p\%$. On the contrary, if the hot phase is dominated by the cooling process $\dot{M}_{in} = \dot{M}_{sh} + \dot{M}_{wind} < \dot{M}_{out} = \dot{M}_{cool}$, a decrease in the galaxy ejection rate would lead to a larger decrease in the hot gas mass than $p\%$.

We thus have
\begin{equation}
\delta t^{MAX}_{hot} = \left\{
  \begin{array}{ll}
  p\%\frac{M_{hot}}{\left|\dot{M}_{sh} + (1+p\%)\dot{M}_{wind} - \dot{M}_{cool}\right|}  & \mbox{: if $\dot{M}_{in} >\dot{M}_{out}$} \\
    & \\
  p\%\frac{M_{hot}}{\left|\dot{M}_{sh} + (1-p\%)\dot{M}_{wind} - \dot{M}_{cool}\right|}  & \mbox{: otherwise,}
  \end{array}\right.
\label{dt_hot}
\end{equation}
where we authorize, depending on the case, an increase or a decrease of p\% of the instantaneous galaxy ejection rate $\dot{M}_{wind}$. These two cases are imprinted in two limits (a lower limit: $\dot{M}_{wind}^{MIN}$ and a upper limit: $\dot{M}_{wind}^{MAX}$) beyond which the criterion of $p\%$ variation is no longer valid.

For the hot-gas phase, with this optimal time step, we can deduce the two lower and upper limits for the galaxy ejection rate. The lower limit reads as
 \begin{equation}
\dot{M}^{MIN}_{wind} = \left\{
  \begin{array}{ll}
  p\%\frac{M_{hot}}{\delta t^{MAX}_{hot}}- \dot{M}_{sh} + \dot{M}_{cool} & \mbox{: if $\dot{M}_{in} >\dot{M}_{out}$} \\
    & \\
  (1-p\%)dot{M}_{wind} & \mbox{: otherwise}
  \end{array}\right.
\label{M_wind_min}
\end{equation}
and the upper limit as
 \begin{equation}
\dot{M}^{MAX}_{wind} = \left\{
  \begin{array}{ll}
  (1+p\%)dot{M}_{wind}  & \mbox{: if $\dot{M}_{in} <\dot{M}_{out}$} \\
    & \\
  p\%\frac{M_{hot}}{\delta t^{MAX}_{hot}}- \dot{M}_{sh} + \dot{M}_{cool} & \mbox{: otherwise.}
  \end{array}\right.
\label{M_wind_max}
\end{equation}

Between these two limits, the stability criterion is followed. Figure~\ref{dM_to_dtmax} illustrates this optimal time step computation. In the case chosen for the figure, the evolution of the hot gas phase is dominated by the input flux ($\dot{M}_{sh} + \dot{M}_{wind} > \dot{M}_{cool}$). The optimal time step for the hot gas phase evolution is therefore computed by assuming an increase in the galaxy ejecta rate.

\subsection{The stop and restart mechanism}
 The two galaxy ejection rate limits are $\dot{M}^{MAX}_{wind}$ and $\dot{M}^{MIN}_{wind}$. Above $\dot{M}^{MAX}_{wind}$ and below $\dot{M}^{MIN}_{wind}$, the evolution of the hot gas phase violates the $p\%$ criterion. These two limits are transferred to the galaxy evolution procedure. During the evolution of the galaxy components, if the average\footnote{over the fraction of the time step already spent.} of the ejecta rate is above or below these two limits, the evolution process is stopped. The properties of the hot gas phase and of the galaxy are computed again at this time, and a new time step starts from this point. 

\subsection{Cold and hot phase coupling}
\label{adaptive_time_step}

In the new eGalICS algorithm, it is therefore possible to compute the optimal time step for the cold ($\delta t^{MAX}_{cold}$) and the hot phases ($\delta t^{MAX}_{hot}$). The same computation is performed for all galaxy components, such as for the disc and the bulge components, which composed a galaxy.  We present in this section the coupling that exists between the hot and cold phases.

\begin{figure}[h]
\begin{center}
 \includegraphics[scale = 0.42]{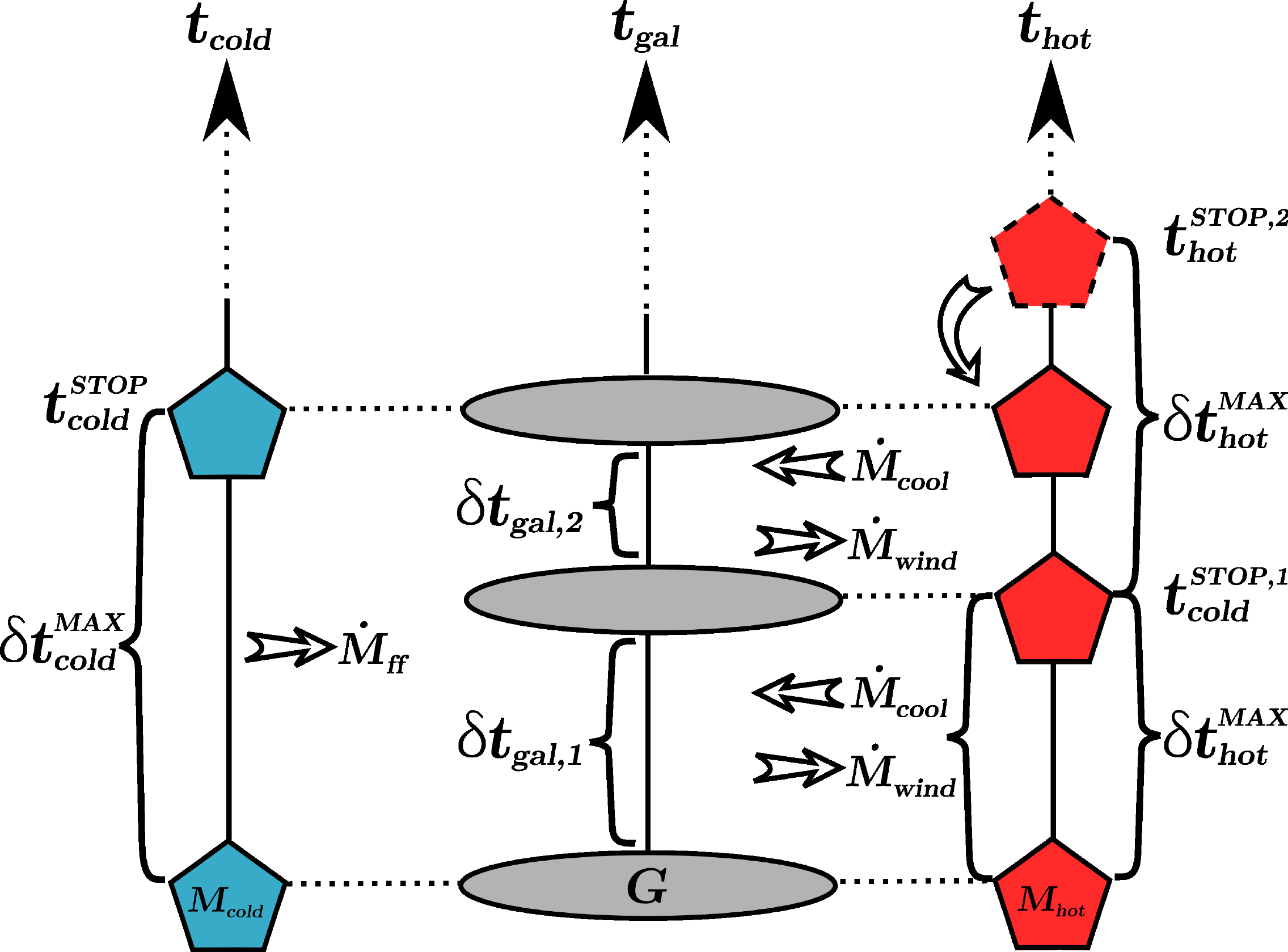}
 \caption{\tiny{Scheme presenting the coupled evolution of the cold and the hot phases, and the galaxy. During the evolution, the properties of cold and hot phases allow some barriers $t_{cold}^{STOP}$ and $t_{hot}^{STOP}$ to be established. The evolution of the galaxy is computed following these barriers.}}
  \label{cold_and_hot_coupling}
  \end{center}
\end{figure}

We assume that $\dot{M}_{ff}$ and $\dot{M}_{cool}$ are constant during $\delta t^{MAX}_{cold}$  and $\delta t^{MAX}_{hot}$, respectively. As shown in Fig. \ref{cold_and_hot_coupling} and as explained previously, the computation of $\delta t^{MAX}_{cold}$  and $\delta t^{MAX}_{hot}$ allows some barriers to be established: $t_{cold}^{STOP}$ and $t_{hot}^{STOP}$. These barriers mark the evolution of the cold and the hot phase. If the galaxy has already evolved until $t_{gal} = t_{cold} = t_{hot}$, the next galaxy time step is computed following
\begin{equation}
  \delta t_{gal} = MIN(t_{cold}^{STOP}~,\delta t^{MAX}_{hot})~- t_{gal} \,.
\end{equation}

During the first part of the evolution presented in Fig. \ref{cold_and_hot_coupling}  the time step of the galaxy is limited by the hot gas phase evolution. Indeed, the cold time step is longer ($t_{cold}^{STOP} > t_{hot}^{STOP}$). The hot gas phase and the galaxy therefore evolve on the same time grid. Later on, the galaxy evolution is limited by the cold-phase evolution ($t_{cold}^{STOP} < t_{hot}^{STOP,2}$).  Indeed,  $\dot{M}_{ff}$ has to be recomputed after $t_{cold}^{STOP}$. Owing to the strong coupling between the galaxy and the hot gas phase (ejection processes $\dot{M}_{wind}$), the evolution of these two components are synchronized at $t_{cold}^{STOP}$. In these conditions, even if the properties of the hot-gas phase should allow an evolution over a longer time step (until $t_{hot}^{STOP,2}$), its evolution is stopped. The three components are updated, their time evolution are then similar ($t_{gal} = t_{cold} = t_{hot}$), and the evolution process is started again.

\subsection{Effective time step duration}

\begin{figure}[h]
\begin{center}
 \includegraphics[scale = 0.85]{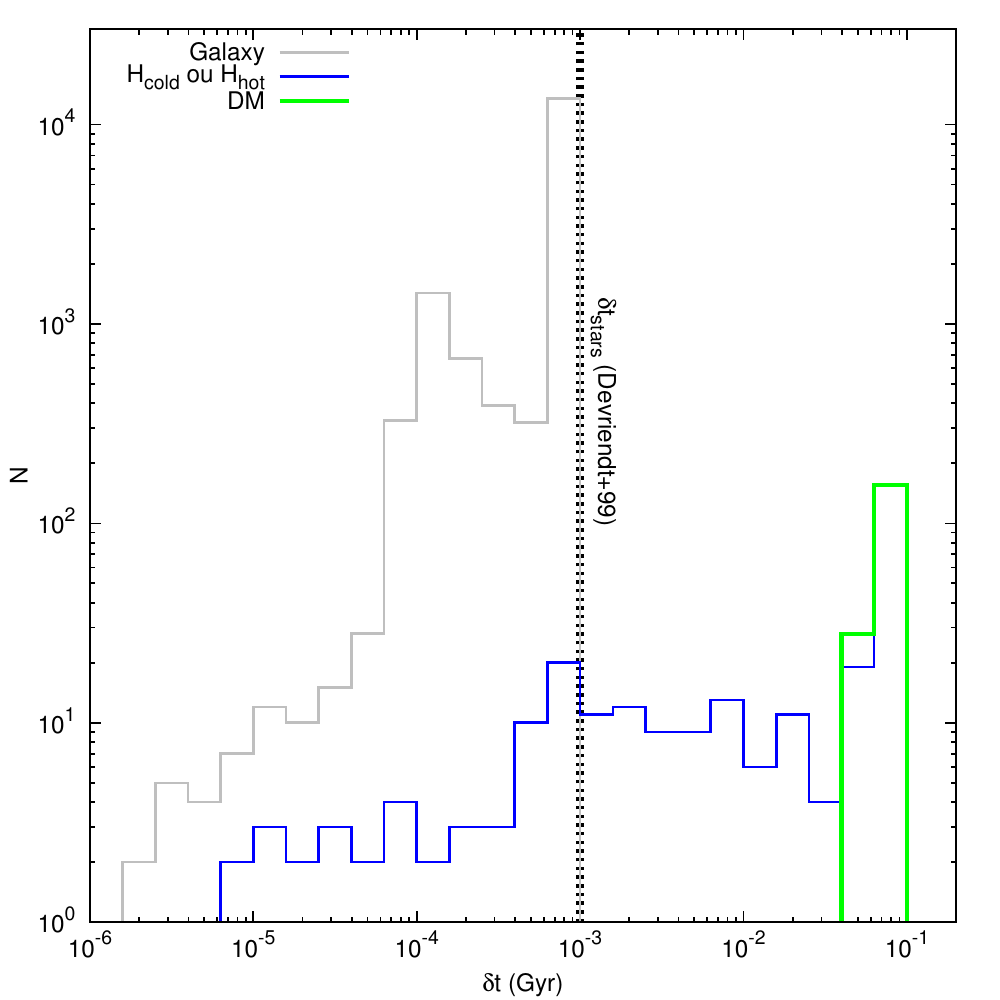}
 \caption{\tiny{Histograms of the time step duration for the galaxy components (disc or bulge in grey), the hot or cold baryonic halo phase (blue), and the dark-matter halos (green). For information, the 94 different snapshots of the N-body simulation used to follow the dark-matter component are distributed with an average time step of 0.15 Gyr.}}
  \label{dt_hist}
  \end{center}
\end{figure}

Figure~\ref{dt_hist} shows the distribution of the time step of the dark-matter component, the cold or/and hot phase and of the galaxy component (disc and bulge). The histograms are built following the analysis of a dark-matter halo and its galaxy between $z=11$ and $z=0$. The time step durations used most for the hot or cold phases are identical to those used for the dark-matter. Indeed, the bulk of the blue histogram coincides with the green histogram. Minimum values coincides with the synchronisation process ($t_{cold}^{STOP} < t_{hot}^{STOP,2}$ in Fig.~\ref{cold_and_hot_coupling}). Concerning the distribution of time steps used for the galaxy components (disc or bulge), the maximum value is linked to the stellar evolution process ($\delta t_{\star} = 10^{-3}~Gyr$, \cite{Devriendt_1999}). The second bump of the grey histogram corresponds to merger time-scale events ($\tau_{merger} = 5\times 10^{-4}~Gyr$). Lower values are again linked to the synchronisation process (clump transfers) between the disc and the bulge.

\section{The polytropic hot gas case}
\label{poly_gas_case}

For a polytropic gas (\textit{poly}), the temperature distribution is non-isothermal. The pressure, density, and temperature profiles are linked with the following relations
  \begin{equation}
    T^{poly}(r)=T_0\left[\dfrac{\rho_g^{poly}(r)}{\rho_{g,0}}\right]^{\Gamma -1} \ ,
    P^{poly} (r)=P_0\left[\dfrac{\rho_g^{poly}(r)}{\rho_{g,0}}\right]^{\Gamma} \ ,
    \label{HEC_poly}
  \end{equation}
where $T_0$, $P_0$,  and  $\rho_{g,0}$ are the central temperature, pressure, and density, respectively, and $\Gamma$ is the polytropic index of the gas.
The HEC in the dark-matter halo potential \citep{Suto_1998, Makino_1998, Komatsu_2001, Capelo_2012}  gives
  \begin{equation}
    \rho_g^{poly}(r) = \rho_{g,0}\mathcal{F}^{poly}(x,T_0)~~\mbox{with}~~x=\frac{r}{r_{dm}} \ ,
    \label{rho_poly}
  \end{equation}
with $\mathcal{F}^{poly}(x,T_0)$, a geometrical function defined as 
  \begin{equation}
    \mathcal{F}^{poly}(x,T_0) = \left[1-\dfrac{\Gamma}{\Gamma-1}\dfrac{4\pi G\mu m_p\rho_{dm}r_{dm}^2}{k_BT_0}\Phi(x)\right]^{\frac{1}{\Gamma-1}} \ ,
    \label{F_poly}
  \end{equation}
where $\Phi(x)$ is the geometrical function defined by Eq.~\ref{Phix} (and see Table \ref{cooling_symbol_2} for parameter definitions). \\

\begin{table}[h]
  \begin{center}
    \footnotesize{
      \begin{tabular*}{0.5\textwidth}{@{\extracolsep{\fill}}cr}
        \hline
        Symbol & Definition \\
        \hline     
        $M_{hot}$                   & Hot atmosphere mass [Eqs. \ref{hot_mass_evolution}, \ref{Mhot_r}, Fig. \ref{baryon_halo_transfer}]\\
        $\rho_g(r)$               & Hot atmosphere radial density profile [Eq. \ref{Mhot_r}]\\
        $T_0$                       & Hot atmosphere central temperature [Eq. \ref{T0}]\\
        $\rho_{g,0}$               & Hot gas central density [Eqs. \ref{rhog0}, \ref{rho_gas_core_id}, \ref{rho_gas_core_poly}]\\ 
        $P^{poly}(r)$               & Radial pressure profile (polytropic gas) [Eq. \ref{HEC_poly}]\\
        $\rho_g^{poly}(r)$       & Radial density profile (polytropic gas) [Eqs. \ref{HEC_poly}, \ref{rho_poly}]\\
        $T^{poly}(r)$                 & Radial temperature profile (polytropic gas) [Eq. \ref{HEC_poly}]\\
        $\mathcal{F}^{poly}(x)$ & Geometrical function (density profile) [Eqs. \ref{F_poly}, \ref{rho_poly}]\\
        $\overline{T}^{poly} $  & (Mass-)average temperature [Eqs. \ref{eq_avg_temp_poly}, \ref{T0_Tpoly}]\\  
        $T_0^{min}$                 & Minimal central temperature [Eq. \ref{T0_min}]\\
        \hline
      \end{tabular*}}
  \end{center}  
  \caption{\footnotesize{Symbols and their definition used for the polytropic gas description. Symbols are listed in order of appearance.}}
  \label{cooling_symbol_2}
\end{table}

In the polytropic gas case, $T_0$ is not equal to the mean temperature $\overline{T}_{atm}$ of the gas phase but to the mass-average temperature
  \begin{equation}
    \begin{split}
      \overline{T}^{poly}  & = M_{hot}^{-1}\int_0^{r_{vir}}\rho_g^{poly}(r)T^{poly}(r)4\pi r^2dr\\
      & = \dfrac{4\pi \rho_{g,0}T_0 r_{dm}^3}{M_{hot}}\int_0^{c_{dm}}\left[\mathcal{F}^{poly}(x,T_0)\right]^{\Gamma}x^2dx \ .
    \end{split}
    \label{eq_avg_temp_poly}
  \end{equation}
Following Eq.~\ref{rho_gas_core_id}, the core density $\rho_{g,0}$ is given by
  \begin{equation}
    \rho_{0,g} = M_{hot}\left[4\pi r_{dm}^3\int_0^{c_{dm}} \mathcal{F}^{poly}(x,T_0)x^2dx \right]^{-1} \ .
    \label{rho_gas_core_poly}
  \end{equation}
Using Eqs.~\ref{eq_avg_temp_poly} and \ref{rho_gas_core_poly}, we can compute an implicit relation between $\overline{T}$ and $T_0$
 \begin{equation}
      \ds{\overline{T}^{poly}   = T_0\dfrac{\int_0^{\frac{r_{vir}}{r_0}}\left[\mathcal{F}^{poly}(x,T_0)\right]^{\Gamma}x^2dx}{\int_0^{\frac{r_{vir}}{r_0}}\mathcal{F}^{poly}(x,T_0)x^2dx}}\, .
      \label{T0_Tpoly}
  \end{equation}

where $\mathcal{F}^{poly}(x,T_0)$ (Eq.~\ref{F_poly}) is a function of the central temperature $T_0$. Since $\overline{T}^{poly}$ is a strictly increasing function of $T_0$ for a set of dark-matter properties ($r_{dm}$ and $\rho_{dm}$), the implicit relation given above can be inverted using a dichotomy-research algorithm. We can thus compute the central temperature $T_0$ of the polytropic gas connected to the mean temperature $\overline{T}_{atm}$ deduced from energy exchanges (see Sect.~\ref{energy_transfers}).\\

The geometrical profile defined by Eq.~\ref{F_poly} must be strictly positive to ensure density (Eq.~\ref{rho_poly}) and temperature  (Eq.~\ref{HEC_poly}) consistency. This condition leads to a minimum threshold value of the central temperature $T_0^{min}$. Indeed all other parameters are linked to dark-matter halo properties and are already fixed. This critical central temperature $T_0^{min}$ depends on the geometrical function $\Phi(x)$ (Eq.~\ref{Phix}), which is a strictly increasing function of $x$. As
  \begin{equation}
\underset{x\to+\infty}{lim}\Phi(x) = 1 \ ,
  \end{equation}
then
  \begin{equation}
    T_0^{min} >  \dfrac{\Gamma}{\Gamma-1}\dfrac{4\pi G\mu m_p\rho_{dm}r_{dm}^2}{k_B}.
    \label{T0_min}
  \end{equation}
If the central temperature $T_0$ of the polytropic gas derived from the average temperature $\overline{T}_{atm}$ is lower than this minimum threshold value $T_0^{min}$, the hot atmosphere is not considered stable enough to activate the cooling processes. The gas contained in this hot phase again falls on the galaxy with a feeding rate close to the free-fall rate (as defined for the filamentary structure). We set in this case
  \begin{equation}
    \dot{M}_{cool} = \varepsilon_{ff}\dfrac{M_{hot}}{2t_{dyn}} \ ,
  \end{equation}
where $M_{hot}$ is the total mass of gas contained in the unstable hot phase, $t_0$ the dynamical time of the dark-matter structure and $\varepsilon_{ff}=1.0$ a free parameter (the same as used in Eq.\ref{free-fall-rate}).

\section{Maxwell-Boltzman distribution}

We use the following formulation of the well-known Maxwell-Boltzman velocity distribution 
\begin{center}
  \begin{equation}
    F_{MB}(v,T)dv = 4\pi\left(\dfrac{\mu m_p}{2\pi k_BT}\right)^{3/2}v^2exp\left(-\dfrac{\mu m_pv^2}{2k_B}\right)dv\, .
    \label{MB_dist}
  \end{equation}
\end{center}

\section{Data repository}
Outputs from the model are available on CDS platform \verb?\http://cdsweb.u-strasbg.fr/cgi-bin/qcat?J/A+A/? under the reference $m_1$. Data are distributed under \verb?*.fits? format and are therefore compatible with the \verb?TOPCAT? software (\verb?http://www.star.bris.ac.uk/~mbt/topcat/?). 
\end{appendix}

\end{document}